\documentclass[english,aps,prd,superscriptaddress,preprintnumbers,nofootinbib]{revtex4}
\usepackage[T1]{fontenc}
\usepackage[latin9]{inputenc}
\usepackage{verbatim}

\usepackage{empheq}
\usepackage{amssymb}
\usepackage{amsmath}
\usepackage{amsfonts}

\usepackage{ulem}

\usepackage{hyperref}

\usepackage{xcolor}
\definecolor{darkblue}{rgb}{0,0,0.7}
\definecolor{darkred}{rgb}{0.9,0,0}

\def \FLM {F_{\rm LM}}

\newcommand{\lp}{\left(}
\newcommand{\rp}{\right)}

\newcommand{\ep}{\epsilon}

\newcommand{\be}{\begin{equation}}
\newcommand{\ee}{\end{equation}}
\newcommand{\bea}{\begin{eqnarray}}
\newcommand{\eea}{\end{eqnarray}}
\newcommand{\bes}{\begin{equation}\begin{split}}

\usepackage{dcolumn}
\usepackage{graphicx}


\usepackage{babel}
\usepackage{color}
\begin{document}
\global\long\def\order#1{\mathcal{O}\left(#1\right)}
\global\long\def\d{\mathrm{d}}
\global\long\def\P{P}
\global\long\def\amp{{\mathcal M}}
\preprint{TTP20-033, P3H-20-048, OUTP-20-09P, CERN-TH-2020-152}

\def \SS {S{\hspace{-5pt}}S}
\def\FNAL{Department of Theoretical Physics, Fermilab, Batavia, IL, USA}
\def\KIT{Institute for Theoretical Particle Physics, KIT, Karlsruhe, Germany}
\def\CERN{Theoretical Physics Department, CERN, 1211 Geneva 23, Switzerland}
\def\OX{Rudolf Peierls Centre for Theoretical Physics, Clarendon Laboratory, Parks Road, 
Oxford OX1 3PU, UK}
\def\WD{Wadham College, University of Oxford, Parks Road, Oxford OX1 3PN, UK}

\title{ Mixed QCD-electroweak corrections to $W$-boson production in hadron collisions
}

\author{Arnd  Behring}            
\email[Electronic address: ]{arnd.behring@kit.edu}
\affiliation{\KIT}

\author{Federico Buccioni}            
\email[Electronic address: ]{federico.buccioni@physics.ox.ac.uk}
\affiliation{\OX}

\author{Fabrizio Caola}            
\email[Electronic address: ]{fabrizio.caola@physics.ox.ac.uk}
\affiliation{\OX}
\affiliation{\WD}

\author{Maximillian Delto}            
\email[Electronic address: ]{maximillian.delto@kit.edu}
\affiliation{\KIT}

\author{Matthieu Jaquier}            
\email[Electronic address: ]{matthieu.jaquier@kit.edu}
\affiliation{\KIT}

\author{Kirill Melnikov}            
\email[Electronic address: ]{kirill.melnikov@kit.edu}
\affiliation{\KIT}

\author{Raoul R\"ontsch }            
\email[Electronic address: ]{raoul.rontsch@cern.ch}
\affiliation{\CERN}

\begin{abstract}
  We compute  mixed QCD-electroweak  corrections to the fully-differential
  production of an on-shell $W$ boson. Decays of $W$ bosons to lepton pairs
  are included in the leading order approximation.
The required two-loop virtual corrections are computed analytically for arbitrary values of the electroweak 
gauge boson masses.
  Analytic  results for integrated subtraction terms are obtained within 
a soft-collinear subtraction scheme optimized to accommodate the structural simplicity
  of infra-red singularities of  mixed QCD-electroweak  contributions.  Numerical
  results for mixed corrections to the fiducial cross section of $pp \to W^+ \to l^+ \nu$
  and selected kinematic distributions in this process are presented. 
\end{abstract}

\maketitle

\section{Introduction} 

Studies of electroweak gauge bosons  produced in hadron collisions  played an important role
in establishing  the validity of the Standard Model of particle physics.  Given these  successes,
it is not surprising that  experiments at the LHC  continue the systematic exploration of vector-boson
production
processes \cite{cms1,cms2,atlas1,atlas2}.
Although plenty of interesting physics, 
ranging from  constraints on parton distribution functions to measurements of
the weak mixing angle to explorations of lepton universality, can be investigated by studying 
the production of single $W$ and $Z$ bosons,  the  Holy Grail of precision electroweak physics
at the LHC is the  measurement of the $W$-boson mass. Indeed, 
the  current goal is to measure the $W$ mass
with a precision of about $8~$MeV to match  the uncertainty in the value of the  $W$ mass obtained 
from precision electroweak fits \cite{Baak:2014ora,deBlas:2016ojx}.
If achieved,
it will  imply  a relative uncertainty on the (directly measured) $W$-boson mass  of about  ${\cal O}(10^{-2})$ percent, an
astounding precision.

Perturbative computations within the Standard Model
play a central role in providing a precise description of $W$ and $Z$ production processes at the LHC and are thus crucial
for the success of precision electroweak measurements. 
 Currently, 
fully-differential cross sections for dilepton production in hadron collisions are
known through next-to-next-to-leading order (NNLO) 
in perturbative QCD  \cite{nnlo1,nnlo2,nnlo3,nnlo4,nnlo5,Caola:2019nzf,nnlo7,nnlo8,nnlo9,nnlo10,nnlo11}
and through next-to-leading order (NLO)
in electroweak theory \cite{nloew1,nloew2,nloew3,nloew4,nloew5,nloew6,nloew7,nloew8,nloew9,nloew10}.
Very  recently, the total cross section for $W$ production was computed through N$^3$LO 
in perturbative QCD \cite{Duhr:2020sdp}.

Although an exact  relation between  the quality of the theoretical description of the process $pp \to W \to l \nu$ and
the precision with which the $W$ mass can be extracted is complicated and observable-dependent, it is generally
agreed that mixed QCD-electroweak corrections to this process, i.e. all effects that are suppressed
by a product of QCD and electroweak couplings ${\cal O}(\alpha_s \alpha_{EW})$
relative to the leading-order process, need to be accounted for to achieve ${\cal O}(8)~$MeV precision. 
It is also important to compute mixed  corrections
at a fully-differential level to ensure that they can be calculated for  realistic observables.

Recently, such  mixed QCD-electroweak corrections were calculated  for on-shell $Z$ production at the LHC 
in Refs.~\cite{Buccioni:2020cfi,Bonciani:2020tvf}, extending earlier results
on mixed QCD-QED corrections presented  in Refs.~\cite{qcdqed,
deFlorian:2018wcj,Cieri:2020ikq,Bonciani:2019nuy}.\footnote{Very recently, the $\mathcal O(n_f \alpha_s\alpha_{\rm EW})$ 
corrections to off-shell $W/Z$ production were computed~\cite{Dittmaier:2020vra}.}  Although the underlying mechanisms 
for  $Z$ and $W$ production  at the LHC 
are  quite similar, there are  two  outstanding issues  with extending the computation of 
QCD-electroweak  corrections to  $W$ production.  The first challenge  is that two-loop mixed QCD-electroweak
corrections are  available for  $Z$-boson production \cite{kkv} but are unknown for the $W$-boson 
case.\footnote{The form factor was computed as an expansion in $\sin^2\theta_W$ in Ref.~\cite{Bonciani:2011zz}.}
The computation
of these corrections is cumbersome,  since they  depend on several mass scales, but definitely feasible.
We present the results of this calculation in this paper. 

The second problem  that needs to be addressed  are  soft and collinear divergences.    These divergences
originate from singular soft and collinear limits of loop integrals
and real-emission amplitudes; they are known to disappear when elastic and inelastic
processes are combined. For the purpose of a fully-differential description of   $W$ production in hadron collisions, 
these divergences
need to be extracted from real-emission contributions without integration over the resolved phase space.  Several 
ways to do this in practice were developed in the context of NNLO QCD computations at hadron colliders
\cite{ant,njet1,njet2,grazi,Czakon:2010td,Czakon:2014oma,Cacciari:2015jma,Caola:2017dug}.

In this paper, we employ the so-called
nested soft-collinear subtraction scheme \cite{Caola:2017dug} that we adjust to accommodate  particularities
of mixed QCD-electroweak corrections. We note that such an adjustment was not necessary in the case of $Z$
production  since $Z$ bosons are  electrically neutral. For this reason, a simple  abelianization of 
NNLO QCD corrections to $Z$ production was sufficient \cite{deFlorian:2018wcj}.
However, since $W$ bosons are electrically charged
and, hence, interact with photons, it is not possible to 
adapt the NNLO QCD description of $W$ production to describe 
mixed QCD-electroweak  corrections. In what follows, we derive  simple
formulas that describe integrated subtraction terms required to make ${\cal O}(\alpha_s \alpha_{EW})$ corrections
to $pp \to W$ finite.  Presenting these formulas, alongside   results for the two-loop virtual corrections,
is the main goal of this paper.

We note that mixed QCD-electroweak corrections to $pp \to W \to l \nu$ can be split into three categories:
{\it i)} mixed corrections to the production process $pp \to W$, {\it ii)} QCD corrections to the production process and
electroweak corrections to the decay and {\it iii)} non-factorizable corrections that connect production
and decay processes \cite{ditt1}. The non-factorizable corrections to on-shell $W$ production 
are suppressed by powers  of $\Gamma_W/M_W$ \cite{fadin,ditt1,ditt2} and, therefore, can be neglected.
Similarly, in case of on-shell production, corrections to the production and decay stages of the
process can be defined unambiguously in a gauge-invariant way, see e.g. Ref.~\cite{ditt1}.
 NLO QCD corrections to the production and NLO electroweak corrections to the decay -- as well as mixed
 QCD-EW corrections to the decay coming from renormalization --  have already been considered
 in Refs.~\cite{ditt1,ditt2} and for this reason we do not consider them here.  The unknown contribution
 is mixed QCD-electroweak corrections to the production process $pp \to W$ since it is 
 of NNLO type.  In this paper, we focus on this contribution.

More specifically, we derive formulas for the two-loop corrections to the $q\bar q'\to W$ vertex function and
for all the subtraction terms required to compute mixed QCD-electroweak corrections to $W$-boson production at the LHC.
As an application, we provide
results for fiducial cross sections and selected kinematic distributions for the $p p \to W^+\to l^+\nu$ process. However, we do not perform detailed
phenomenological studies related to e.g. the impact of these corrections
on the $W$-mass measurement  since  such 
 studies warrant a separate publication.  We plan to return to this topic
in the near future. 

The paper is organized as follows. In Section~\ref{sect2} we provide a brief overview of the nested
soft-collinear subtraction
scheme and point out  differences between the mixed QCD-EW case studied here and the pure NNLO QCD cases considered earlier \cite{Caola:2017dug,Caola:2019nzf}.
In Section~\ref{sect3} we describe the soft limits  of scattering
amplitudes relevant for computing mixed QCD-EW corrections.
In Section~\ref{sect4} we briefly discuss the computation of NLO electroweak corrections to $pp \to W$.\footnote{The NLO QCD corrections required for computing mixed QCD-electroweak corrections can be borrowed from Refs.~\cite{Caola:2017dug,Caola:2019nzf}; for this reason, we do not discuss them here. } 
In Section~\ref{sect5} we derive all of the integrated
subtraction terms required for the full mixed QCD-EW calculation, focusing on the
most complicated $q\bar q'$ and $gq$ partonic channels. 
In Section~\ref{sect6} we present final formulas for integrated 
subtraction terms for all partonic channels.   In Section~\ref{sect7} numerical results are presented. We conclude in 
Section~\ref{sect8}.  Many intermediate  results, including the detailed discussion
of mixed QCD-electroweak corrections to the $q \bar q' \to W$ form factor, are collected in the Appendices. 


\section{An overview of the nested soft-collinear subtraction scheme and
  its modification for QCD-EW corrections to $W$ production}
\label{sect2}

In this section, we provide an overview of the soft-collinear subtraction scheme~\cite{Caola:2017dug,Caola:2019nzf}.
For the sake of definiteness,
we consider the process $pp \to W^+ \to l^+ \nu$.
It is well-known that infra-red safe observables defined for this process must
receive contributions from elastic  $pp \to W^+(l^+ \nu)$ and inelastic  $pp \to W^+(l^+ \nu) + X_f$ channels.
We note that, depending on the type of corrections that are
studied, $X_f$ stands for final states composed of  gluons, quarks and/or photons. 

It is conventional to  use
dimensional regularization to compute virtual corrections  and to regulate real-emission contributions.  In this case, 
higher-order contributions to the elastic process contain  explicit $1/\ep$ poles that originate from an 
integration over loop momenta, whereas inelastic processes  develop such $1/\ep$  poles only once 
 the integration
over phase spaces of all potentially unresolved  particles is performed.
To keep results  fully-differential,
this phase-space integration should be performed  in a way that does not affect infra-red safe observables.
A procedure that allows  one to do that defines a particular  computational 
scheme that is often referred  to as a subtraction (or a slicing) scheme. As we already mentioned, in this
paper we will use the so-called nested soft-collinear subtraction scheme introduced in Ref.~\cite{Caola:2017dug}. 

The nested subtraction scheme is based on the well-known notion of factorization of scattering amplitudes
in singular limits and the fact that, thanks to QCD color coherence, soft and collinear limits of scattering
amplitudes can be dealt with independently of each other. 
The behavior of scattering amplitudes in the singular limits is well-known; typically, they 
split into  universal functions that encapsulate singular behavior and amplitudes 
that describe lower-multiplicity processes.

The idea behind the soft-collinear subtraction scheme is to iteratively subtract such
singular limits from differential cross sections   starting from soft singularities.
The subtraction terms have to be added back 
and integrated over the unresolved  phase space. In the case of collinear singularities a similar procedure
is followed; the collinear  subtraction, however, applies to cross sections
that are {\it already} soft-subtracted.  This nested nature of the
subtraction process gives rise to a name -- the {\it nested} soft-collinear subtraction scheme.

An important challenge in constructing  subtraction terms is to ensure that the resulting limits are unambiguous. This requires
us to resolve overlapping singularities whenever they arise. In QCD overlapping singularities are present in both
soft and collinear emissions but there is no soft-collinear overlap.
To deal with soft singularities in QCD amplitudes, we {\it order} gluon energies and consider the so-called
double-soft and single-soft limits. To deal with overlapping collinear singularities, we follow Refs.~\cite{Frixione:1995ms,Czakon:2010td} and introduce partitions and sectors
that allow  us to uniquely specify  how singular collinear limits are approached. 

Although similar in spirit to the general QCD case, the calculation
of mixed QCD-EW corrections to $q \bar q' \to W^+$ is particular. There are a few reasons for that. 
The first one is that   soft singularities  in the process $q \bar q' \to W^+ g \gamma$ are not entangled. 
To understand this, we note that
when both a gluon and a photon are emitted from quark lines, the situation is QED-like and soft singularities in QED are known
to be independent from each other. If, however, a photon is emitted from a $W$-boson line and a gluon is
emitted from a quark line, the independence of  the two
soft limits is obvious. This  feature of mixed QCD-QED corrections allows us to consider soft limits of a photon and a gluon
separately, leading to simplifications of the integrated subtraction terms compared to the QCD case. Indeed, only
the product of two NLO-like  integrated soft subtractions is required, although we need them to higher order in the
$\ep$-expansion compared to a NLO calculation proper. 

Similarly, collinear limits can be simplified because photons and gluons do not interact with each other. Since two out
of the four collinear sectors described in Ref.~\cite{Caola:2017dug} for the NNLO QCD case are introduced to isolate the $g^* \to gg(q\bar q) $ splitting, the  mixed QCD-EW
case can be simplified at least inasmuch as the $Wg \gamma$ final state is concerned.  Moreover,
for $g q$ initial states, the absence of $g \to g \gamma$ splittings leads to a simplified version of the required
partition functions compared to the case discussed in Ref.~\cite{Caola:2017dug}, 
and a smaller number of singular limits that need to be considered. Contrary to the soft case, 
collinear sectors still contain genuine NNLO-like contributions that do not fully factorize into 
the product of NLO-like limits. 
Nevertheless, the features discussed above make the construction of subtraction terms
much easier than in the generic QCD case. 

As we already mentioned in the introduction,
an important difference with respect to a computation of 
NNLO QCD corrections to $pp \to W$ \cite{Caola:2019nzf} stems from the fact that $W$ bosons radiate photons.  
Since
$W$ bosons are  massive, such radiation
does not affect collinear singularities but it does change the soft ones.  Hence, formulas
for the soft  limits need to be modified compared to the QCD case.
We describe the corresponding modifications and how we deal with them in
the next section.

The final difference between the  NNLO QCD computations reported in
Refs.~\cite{Caola:2017dug,Caola:2019nzf} and the one that
we discuss in this paper is that this time 
 we perform computations in an {\it arbitrary}, i.e. not center-of-mass, 
 partonic reference  frame.
 The very fact that the soft-collinear subtraction scheme is perfectly suited  to deal with this situation
 in spite of the fact that it is not manifestly Lorentz invariant is  interesting. It  illustrates 
  the flexibility of this approach and, on a practical level, 
  it makes the treatment of soft and collinear limits very  natural and transparent.


\section{The soft limits}
\label{sect3}

As we mentioned in the previous section, an important difference between the computations of
NNLO QCD and mixed QCD-electroweak corrections is the soft limits.
In this section, we elaborate on this point and provide the required formulas to describe them.

The key feature   that we exploit to construct  soft  subtraction
terms for  mixed QCD-EW corrections  is the fact that soft-photon and soft-gluon
limits are not entangled. For this reason we can consider the two soft limits independently.
The resulting simplifications in computing integrated subtraction terms 
will become apparent when we  discuss
the NNLO computations in Section~\ref{sect4}.  In this section, we describe the soft limits relevant for the mixed case
and explain how the required eikonal integrals can be evaluated.

We focus on the most complicated process $u(p_1) + \bar d(p_2)  \to W^++ g(p_4)+\gamma(p_5)$.   We employ notations that have been used in NNLO QCD calculations
\cite{Caola:2017dug,Caola:2019nzf} and 
  denote the product of the matrix element squared of  this process and the relevant phase space factor for the
  $W$ boson (or its decay products)
 as  $F_{\rm LM}(1_u,2_{\bar d}; 4_g, 5_\gamma) $
 \be
 F_{\rm LM}(1_u,2_{\bar d}; 4_g, 5_\gamma) = 
 \mathcal N \sum_{\rm col,pol} |\mathcal A(p_1,p_2; p_W ,p_4, p_5)|^2
 (2\pi)^d\delta_d(p_1+p_2-p_W-p_4-p_5)
 \frac{\d^{d-1} p_W}{(2\pi)^{d-1}2 E_W}
 \label{eq:1}.
 \ee
A similar notation is used for lower-multiplicity processes. In Eq.(\ref{eq:1}), $\mathcal N$ stands for all the required ($d-$dimensional) initial-state color and helicity averaging
 factors, and for all the required final-state symmetry factors.  Note that $F_{\rm LM}(1_u,2_{\bar d}; 4_g, 5_\gamma) $ does
 {\it not} contain the phase-space volume elements for the potentially unresolved particles $p_4$ and $p_5$. 
    We consider the soft-gluon and the soft-photon limits separately. 
Similar to the NNLO QCD case, we describe these limits using two operators,   $S_g$ and $S_\gamma$.
The operator $S_i$ selects the most singular contribution of $|\mathcal M|^2$ in the $E_i\to 0$ limit and 
removes particle $i$ from the momentum-conserving $\delta$ function. For further details, see Refs.~\cite{Caola:2017dug,Caola:2019nzf}. 

 We  begin by considering the soft-gluon limit. In the notation that we have just reviewed, it reads 
\be
S_g F_{\rm LM}(1_u,2_{\bar d}; 4_g, 5_\gamma)  = g_s^2 \; {\rm Eik}_g(p_1,p_2;p_4) F_{\rm LM}(1_u,2_{\bar d}; 5_\gamma),
\ee
where $g_s$ is the (bare) strong coupling, and
\be
{\rm Eik}_g(p_1,p_2;p_g) = \frac{2 C_F (p_1 p_2) }{(p_1 p_g)(p_2 p_g)},
\ee
with $(p_i p_j) \equiv p_i\cdot p_j$. Also, $C_F = (N_c^2-1)/(2N_c)$ is the QCD quadratic Casimir and $N_c=3$ is the
number of colors. 
Note that this limit is independent of whether or not there is a photon in the matrix element squared;
this implies that an identical formula can be used to describe the soft gluon limit
of the process $ u(p_1) + \bar d(p_2) \to W^+ + g(p_4)$:
\be
S_g F_{\rm LM}(1_u,2_{\bar d};4_g) = g_s^2\; {\rm Eik}_g(p_1,p_2;p_4) F_{\rm LM}(1_u,2_{\bar d}).
\ee
The soft-gluon limit of different partonic channels can be trivially obtained by crossing these results.
For example,
\be
S_g F_{\rm LM}(1_\gamma,2_{\bar d};4_{\bar u},5_g) = 
g_s^2\; {\rm Eik}_g(p_2,p_4;p_5) \FLM(1_\gamma,2_{\bar d};
4_{\bar u}).
\ee

To analyze the soft-photon  limit, we write
\be
S_\gamma F_{\rm LM}(1_u,2_{\bar d}; 4_g, 5_\gamma) 
= e^2 \; {\rm Eik}_\gamma(p_1,p_2,p_W; p_5) F_{\rm LM}(1_u,2_{\bar d}; 4_g),
\label{eq3}
\ee
where $p_W = p_1 + p_2 - p_4$ is the four-momentum of the $W$ boson and $e$ is the
(bare) electric coupling. 
The QED eikonal function  reads 
\be
\begin{split} 
   {\rm Eik}_\gamma(p_1,p_2,p_W; p_\gamma) & =
   \left \{
    Q_u Q_d \frac{ 2 (p_1p_2)}{ (p_1 p_\gamma)  (p_2 p_\gamma ) }
   -  Q_W^2 \frac{p_W^2}{ (p_W p_\gamma)^2 }
   + Q_W \left (  Q_u \frac{ 2(p_W p_1)}{ (p_W p_\gamma) (p_1 p_\gamma)}
   - Q_d \frac{  2(p_W p_2)}{ (p_W p_\gamma) (p_2 p_\gamma)} \right )
\right \}.
\end{split}
\label{eq4}
\ee
In Eq.(\ref{eq4}) we used 
$Q_W = Q_u - Q_d$ to denote the electric  charge of the  $W$ boson. Note that $p_W$ 
depends on the gluon four-momentum; hence, it changes if the soft-photon and the soft-gluon
limits are considered simultaneously.

To compute the soft subtraction terms,  we  integrate the eikonal functions ${\rm Eik}_{g,\gamma}$ over
the phase spaces of a soft gluon and/or  photon. Following the NNLO QCD computations
\cite{Caola:2017dug,Caola:2019nzf}
we define phase-space
elements with an upper energy cut-off $E_{\rm  max}$
\be
[{\rm d} p] = \frac{{\rm d}^{d-1}\vec p }{(2\pi)^{d-1} 2 p_0} \; \theta(E_{\rm max} - p_0).
\ee
In the case of the soft-gluon limit,  we easily find
\be
\int [{\rm d} p_4 ]  g_s^2 \; {\rm Eik}_g(p_1,p_2;p_4)  = [\alpha_s] \frac{2 C_F (2 E_{\rm max} )^{-2\ep} }{\ep^2} \frac{\Gamma^2(1-\ep)}{\Gamma(1-2\ep)},
\label{eq7}
\ee
where
\be
[\alpha_s] = \frac{g_s^2 \Omega^{(d-2)}}{2(2\pi)^{d-1}} = 
\lp\frac{\alpha_s}{2\pi}\rp\frac{(4\pi)^\ep}{\Gamma(1-\ep)},
\ee
with $\alpha_s$ the (bare) strong coupling. 

To describe the soft-photon contribution, we need to integrate the soft-photon eikonal function
Eq.(\ref{eq4}) over the photon phase space. Since this integral is more involved than the one in the gluon case
Eq.(\ref{eq7}), it is beneficial to compute it in two special cases. 

The first case is that of a soft photon but resolved gluon.   The relevant eikonal integral
was computed  in Ref.~\cite{Alioli:2010xd} and  we borrow  it from there.  We obtain 
\be
 \int [{\rm d}p_5] e^2 {\rm Eik}_\gamma(p_1,p_2,p_W;p_5) = [\alpha] (2 E_{\rm max})^{-2\ep} \frac{\Gamma^2(1-\ep)}{\Gamma(1-2\ep)}
 J_{\gamma}(1,2,W),
\label{eq8aa}
  \ee
  where
  \be
\begin{split} 
&  J_{\gamma}(1,2,W)  =  
    \frac{Q_d^2 + Q_u^2}{\ep^2} + \frac{Q_W}{\ep} \left ( Q_W
  -  2 Q_{u} \ln \left ( \frac{\kappa_{1W}}{\sqrt{1-\beta^2}} \right )
    + 2 Q_{d} \ln \left ( \frac{\kappa_{2W}}{\sqrt{1-\beta^2}} \right ) 
    \right )
    \\
&  
-Q_W^2   \left [ \frac{1}{\beta}\ln \frac{1-\beta}{1+\beta} - \frac{1}{2} \ln^2\frac{1-\beta}{1+\beta} \right ]
-2 Q_W
\sum \limits_{i=1}^{2} 
Q_i (-1)^{i} \ln \left ( \frac{\kappa_{iW}}{1 - \beta} \right ) \ln \left ( \frac{\kappa_{iW}}{1 + \beta} \right )
\\
&
- 2 Q_W \sum \limits_{i=1}^{2} 
Q_i (-1)^i \left[  {\rm Li}_2 \left ( 1 - \frac{\kappa_{iW}}{1-\beta} \right )  +
  {\rm Li}_2 \left ( 1 - \frac{\kappa_{iW}}{1+\beta} \right ) \right ]
  +\mathcal O(\ep).
\label{eq8aaa}
\end{split} 
\ee 
In Eq.(\ref{eq8aaa})  $Q_1 = Q_u$ and $Q_2 = Q_d$,  $Q_W = Q_u - Q_d$, $\beta  = \sqrt{1 - M_W^2/E_W^2}$
and $\kappa_{iW} =(p_i p_W) /(E_i E_W)$.  Note that, similar to the QCD case, we introduced in Eq.(\ref{eq8aa})
a convenient  notation
for the (bare) fine structure constant $\alpha$
\be
[\alpha] = \frac{e^2 \Omega^{(d-2)}}{2(2\pi)^{d-1}} = 
\lp\frac{\alpha}{2\pi}\rp\frac{(4 \pi)^{\ep}}{\Gamma(1-\ep)}.
\ee

For the gluon-initiated process $g(p_1) + \bar d(p_2) \to W^+ + \bar u(p_4) + \gamma(p_5)$,
we require a similar but slightly different
integrated eikonal function. It reads
\be
    \int [{\rm d}p_5] e^2 {\rm Eik}_\gamma(p_2,p_4,p_W;p_5)
= [\alpha] (2 E_{\rm max})^{-2\ep} \frac{\Gamma^2(1-\ep)}{\Gamma(1-2\ep)} \; J_{\gamma}(2,4,W),
\label{eq14a}
  \ee
where 
  \be
\begin{split} 
& 
J_{\gamma}(2,4,W) =     \frac{Q_d^2 + Q_u^2}{\ep^2}
 + \frac{Q_W}{\ep} \left ( Q_W
  -  2 Q_{u} \ln \left ( \frac{\kappa_{4W}}{\sqrt{1-\beta^2}} \right )
    + 2 Q_{d} \ln \left ( \frac{\kappa_{2W}}{\sqrt{1-\beta^2}} \right ) 
    \right )
     - \frac{2Q_uQ_d}{\ep} \ln ( \eta_{42})
    \\
    &
   -Q_W^2
    \left [ \frac{1}{\beta}\ln \frac{1-\beta}{1+\beta}
      - \frac{1}{2} \ln^2\frac{1-\beta}{1+\beta} \right ]
    -2 Q_W
\sum \limits_{i \in \{2,4\}}^{} 
Q_i  \ln \left ( \frac{\kappa_{iW}}{1 - \beta} \right ) \ln \left ( \frac{\kappa_{iW}}{1 + \beta} \right )
\\
&  
-2 Q_W \sum \limits_{i\in \{2,4\} }^{} 
Q_i \left[  {\rm Li}_2 \left ( 1 - \frac{\kappa_{iW}}{1-\beta} \right )  +
  {\rm Li}_2 \left ( 1 - \frac{\kappa_{iW}}{1+\beta} \right ) \right ] 
 + 2 Q_d Q_u \left ( {\rm Li}_2(1-\eta_{42}) + \frac{1}{2} \ln^2 \eta_{42} \right )
 +\mathcal O(\ep)
  ,
\end{split} 
\label{eq14aa}
\ee 
and $Q_2 = Q_d$, $Q_4 = -Q_u$,  $Q_W = Q_u - Q_d$, 
$\beta  = \sqrt{1 - M_W^2/E_W^2}$
and $\kappa_{iW} = p_i p_W /(E_i E_W)$. We also used $\eta_{42} = (p_2 p_4)/(2 E_2 E_4) = 
(1-\cos\theta_{42})/2$ in Eq.(\ref{eq14aa}).

We can use the  integrated soft-photon eikonal factors shown in Eqs.(\ref{eq8aa},\ref{eq14a})  both when 
a gluon or a quark in the final
state are  resolved, so that $p_W \ne p_1 + p_2$, and when they are unresolved. However, for the latter
case one needs to evaluate the integrated photon eikonal function to higher orders in the
$\epsilon$-expansion, in the
required kinematic configuration.
It is technically more convenient to obtain this result
by {\it first} taking the required limits 
in the  eikonal function Eq.(\ref{eq4})  and integrating over the photon four-momentum after that, rather
than the other way around. 

Although soft and collinear parton emissions have a different impact on the QED eikonal function, it is easy to see 
that  we can accommodate both soft and collinear limits of
the emitted parton $p_4$ by integrating the eikonal function ${\rm Eik}_\gamma$  in an {\it arbitrary
reference
frame}  with the constraint $p_W = p_1 + p_2$.  We write such an integral as 
\be
\int [{\rm d}p_5]\;  e^2 {\rm Eik}_\gamma(p_1,p_2,p_W;p_5)|_{p_W = p_1 + p_2}
= [\alpha] (2 E_{\rm max})^{-2\ep} \frac{\Gamma^2(1-\ep)}{\Gamma(1-2\ep)} {\tilde J}_{\gamma}(E_1,E_2),
\label{eq10a}
\ee
where the function ${\tilde J}_{\gamma}(E_1,E_2)$ reads 
\be
\begin{split}
{\tilde J}_{\gamma}(E_1,E_2) & =  
  \frac{Q_u^2 + Q_d^2}{\ep^2}
  + \frac{Q_W^2}{\ep(1-2\ep)}
  + \frac{Q_W}{\ep^2}
  \left \{
  Q_u \left [  \left ( \frac{E_1}{E_2} \right )^{\ep} - 1 \right ]
  - Q_d \left [  \left ( \frac{E_2}{E_1} \right )^{\ep} - 1 \right ]
  \right \}
 \\
& 
+ \frac{Q_W}{\ep^2} \left \{
Q_u \left ( \frac{E_1}{E_2} \right )^\ep
\left [ F_{21} \left (-\ep,-2\ep,1-2\ep,1-\frac{E_2}{E_1} \right ) - 1 \right ]
\right.  \\
& \left. 
-Q_d \left ( \frac{E_2}{E_1} \right )^\ep \left [ F_{21} \left (-\ep,-2\ep,1-2\ep,1-\frac{E_1}{E_2} \right )
- 1 \right ]
\right \}  \\
& 
+ \frac{Q_W^2}{\ep(1-2\ep)} \left (
\frac{E_2}{E_1} \right )^\ep \left [ F_{21} \left (-2\ep,1-\ep,2-2\ep,1-\frac{E_1}{E_2}
\right ) - 1 \right ].
\label{eq11}
\end{split} 
\ee
We note that entries in the first line in Eq.(\ref{eq11})  are divergent contributions to ${\tilde J}_\gamma$;  
all other terms in Eq.(\ref{eq11}) have a finite  $\ep \to 0$ limit.  We also note that the  function
$ {\tilde J}_\gamma$ assumes a particularly
simple form in the partonic center-of-mass frame, $E_1 = E_2$.  We find 
\be
   {\tilde J}_\gamma(E_1, E_1) =
   \frac{Q_u^2 + Q_d^2}{\ep^2} + \frac{Q_W^2}{\ep(1-2\ep)}.
   \label{eq18}
\ee
Although, as we said earlier, we  perform all computations in an arbitrary frame, once  the poles cancellation
is achieved, we switch back to the partonic center-of-mass frame to present analytic results for the 
finite integrated
subtraction terms. The simplicity of
Eq.(\ref{eq18}) is an important reason to expect results in the center-of-mass frame  to be compact and physically transparent. 


\section{Next-to-leading order electroweak corrections}
    \label{sect4}

To introduce notations and show how the nested soft-collinear subtraction scheme is applied to a simple 
problem, we  briefly discuss the computation of NLO electroweak corrections. At this
order, we need to consider both the $q\bar q'\to W \gamma$ and the $\gamma q \to W q'$ channels. 

We first consider the $q\bar q'$ channel, and begin with the real-emission process
  $u(p_1) + \bar d(p_2) \to W^+ + \gamma(p_4) $. Using the notation introduced in Ref.~\cite{Caola:2017dug} and
  reviewed in the previous section, we write the real-emission contribution as
  \be
     2s\cdot{\rm d} \sigma^\gamma_R =  \int [{\rm d} p_4] F_{\rm LM}(1_u,2_{\bar d}; 4_\gamma ) =
     \langle F_{\rm LM}(1_u,2_{\bar d}; 4_\gamma) \rangle,
     \ee
     where $s = 2 p_1 \cdot p_2$ and $[{\rm d} p_4] = {\rm d}^{d-1} p_4/((2\pi)^{d-1} 2 E_{4}) \theta (E_{\rm max} - E_4)$. 
     We do not show  the four-momentum of the $W$ boson
in the list of arguments of the function $F_{\rm LM}$; we assume that it is always derived from energy-momentum
conservation.    We note that the phase-space integration measure
for all final-state particles but the  photon, 
as well as the delta function that ensures energy-momentum conservation,  are included in  the function $F_{\rm LM}$,
see Eq.(\ref{eq:1}).

We begin with the extraction of soft singularities and write
\be
\langle F_{\rm LM}(1_u,2_{\bar d}; 4_\gamma) \rangle =
\langle S_\gamma F_{\rm LM}(1_u,2_{\bar d}; 4_\gamma) \rangle + \langle (I -S_\gamma) F_{\rm LM}(1_u,2_{\bar d}; 4_\gamma) \rangle.
\label{eq18a}
\ee
The first term in Eq.(\ref{eq18a}) is computed using the integrated eikonal function given in Eq.(\ref{eq10a}). We find
\be
\langle S_\gamma F_{\rm LM}(1_u,2_{\bar d}; 4_\gamma) \rangle  = [\alpha]  (2 E_{\rm max})^{-2\ep}
\frac{\Gamma^2(1-\ep)}{\Gamma(1-2\ep)}\langle  {\tilde J}_\gamma(E_1,E_2)  F_{\rm LM}(1_u,2_{\bar d}) \rangle. 
\ee
The term proportional to $(I-S_\gamma)$ on the right hand side of Eq.(\ref{eq18a}) is soft-regulated, but
it still contains divergences when the  photon  becomes collinear to one of the  incoming quarks. 
To regulate them, we use the same approach we used for the QCD case~\cite{Caola:2017dug,Caola:2019nzf}.
In particular, in analogy to the soft case we introduce the collinear operators $C_{\gamma i}$ that extract  the collinear limit from $F_{\rm LM}$, see Refs.~\cite{Caola:2017dug,Caola:2019nzf} for details. 
We then write
\be
\langle (I - S_\gamma) F_{\rm LM}(1_u,2_d;4_\gamma) \rangle =
\langle {\cal O}_{\rm NLO}^{\gamma} \left [ F_{\rm LM}(1_u,2_d;4_\gamma) \right ] \rangle
+ \langle   ( I-S_\gamma)  \left ( C_{\gamma 1} + C_{\gamma 2} \right ) F_{\rm LM}(1_u,2_d;4_\gamma) \rangle,
\label{eq19a}
\ee
where ${\cal O}_{\rm NLO}^{\gamma}  = (I-S_\gamma) \left (I - C_{\gamma 1} - C_{\gamma 2} \right ) $. 
The first term on the right hand side of
Eq.(\ref{eq19a}) is fully regulated and  we do not discuss it anymore.  In the last
two terms, we need to consider the limit when the photon becomes collinear to either $p_1$ or $p_2$.
 We start with the case when the photon is 
collinear to $p_1$. The corresponding collinear limit can be directly taken from the QCD case \cite{Caola:2017dug,Caola:2019nzf}. We obtain 
\be
C_{\gamma 1}  F_{\rm LM}(1_u,2_d;4_\gamma)  = \frac{e^2 Q_u^2 }{E_4^2 \rho_{41}} \left ( 1-z \right ) {\bar P}_{qq}(z)
\frac{F_{\rm LM}(z\cdot 1_u,2_{\bar d} )}{z},
\ee
where $E_4 = (1-z) E_1$, $\rho_{ij} = p_i\cdot p_j/(E_i E_j) = 1-\cos\theta_{ij}$ and the splitting function ${\bar P}_{qq}(z)$ is defined as follows 
\be
{\bar P}_{qq}(z)  = \frac{1+z^2}{1-z} - \ep (1-z). 
\ee
Note that compared to a conventional  $q \to q+g$ splitting function, we do not include  the color factor $C_F$
in  ${\bar P}_{qq}$. The reason for that is that  one and the same  splitting function can then be used to describe 
both the $q \to q + \gamma$ and $q \to q + g$ splittings which is quite convenient. 

The next steps are identical to the QCD computation
and involve integration over the photon emission angle in the soft-regulated collinear
term~\cite{Caola:2017dug,Caola:2019nzf}. Repeating these  steps, we find 
     \be
\langle (I-S_\gamma)
C_{\gamma1}  F_{\rm LM}(1_u,2_d;4_\gamma)
= -[\alpha] \frac{Q_u^2}{\ep} \frac{\Gamma^2(1-\ep)}{\Gamma(1-2\ep)}
\left \langle (2 E_1)^{-2\ep}
\int \limits_{0}^{1} {\rm d} z P_{qq}^{\rm NLO}(z,L_1) 
\frac{F_{\rm LM} \left (z \cdot 1_u,2_{\bar d} \right )}{z} \right \rangle,
\label{eq21a}
     \ee
where $L_1 = \ln E_{\rm max}/E_1$ and 
     \be
P_{qq}^{\rm NLO}(z,L) = (1-z)^{-2\ep} {\bar P}_{qq}(z)
+ \frac{1}{\ep} \delta(1-z) e^{-2\ep L}.
\label{eq21aa}
     \ee
  The expansion of the function $P_{qq}^{\rm NLO}$ in powers of $\ep$ is given in Appendix~\ref{app:split}. 

  To obtain the final result for  the soft-regulated collinear  contribution in Eq.(\ref{eq19a}) we need
  to account for the term proportional to $C_{\gamma 2} (I - S_\gamma)$. It is easy to obtain
  it from Eq.(\ref{eq21a}) by replacing  $L_1$ with  $ L_2 = \ln E_{\rm max}/E_2$
  and $F_{\rm LM} \left (z \cdot 1_u,2_{\bar d} \right )$ with  $F_{\rm LM} \left (1_u, z \cdot 2_{\bar d} \right )$.
  Upon doing that, we find
  \be
  \begin{split} 
&    2  s\cdot {\rm d}\sigma^{\gamma}_R =
     [\alpha]  (2 E_{\rm max})^{-2\ep}
     \frac{\Gamma^2(1-\ep)}{\Gamma(1-2\ep)}
     \langle {\tilde J}_{\gamma}(E_1,E_2) \; F_{\rm LM}(1_u,2_{\bar d}) \rangle
     + \left \langle {\cal O}_{\rm NLO}^{\gamma} \left [ F_{\rm LM}(1_u,2_{\bar d};4_\gamma )  \right ] \right \rangle 
     \\
&      -
     \frac{[\alpha] }{\ep} \frac{\Gamma^2(1-\ep)}{\Gamma(1-2\ep)}
\int \limits_{0}^{1} {\rm d} z
\sum \limits_{i=1}^{2} 
Q_i^2 (2 E_i)^{-2\ep}  P_{qq}^{\rm NLO}(z,L_i)  \left \langle
F^{(i)}_{\rm LM} \left (1_u,2_{\bar d} \; | \;  z \right )
\right \rangle,
\end{split} 
\label{eq27}
\ee
where we used the notation $Q_{1,2} = Q_{u,d}$ and
\be
F^{(i)}_{\rm LM} \left (1_u,2_{\bar d} \;  | \;   z \right )
=
\left \{
\begin{array}{cc}
  F_{\rm LM} \left (z \cdot 1_u,2_{\bar d} \right )/z, & i=1,  \\
    F_{\rm LM} \left ( 1_u,z \cdot 2_{\bar d} \right )/z, & i=2. \\
\end{array}
\right.
\label{eq22a}
  \ee
We will use the notation in Eq.(\ref{eq22a}) and its natural generalizations in what follows. 
  
To obtain the final result for the NLO corrections, we have to combine the real-emission contribution
Eq.(\ref{eq27}) with virtual corrections and PDFs renormalization. The former are discussed in 
Appendix~\ref{sectA1}. We now discuss the latter. Collinear counterterms depend on 
the renormalized coupling constant $\alpha(\mu)$.\footnote{In case of NLO QCD corrections,  collinear counterterms
  depend on $\alpha_s(\mu)$.}    Since all the above results are written using 
unrenormalized couplings, we rewrite the results for the convolutions through the unrenormalized
coupling constants  as well using 
\be
[\alpha(\mu)]   = \frac{ 
  \Gamma(1 - \ep)}{\mu^{2\ep} e^{\ep \gamma_E}} \; [\alpha],
\;\;\;\;
[\alpha_s(\mu)] = \frac{
  \Gamma(1 - \ep)}{\mu^{2\ep} e^{\ep \gamma_E}} \; [\alpha_s].
\label{eq26aaa}
\ee

The collinear renormalization contribution then reads 
\be
   {\rm d} \sigma_{\rm pdf} =  \frac{[\alpha]}{\ep} \frac{ \Gamma(1 - \ep)}{  \mu^{2\ep} e^{\ep \gamma_E}}
   \int {\rm d} z \; {\bar P}_{qq}^{{\rm AP},0}(z) 
   \sum \limits_{i=1}^{2} Q_i^2 \left \langle
F^{(i)}_{\rm LM} \left (1_u,2_{\bar d} \; | \; z \right )
 \right \rangle,
\ee
where ${\bar P}_{qq}^{{\rm AP},0}$ is the (color-stripped) LO Altarelli-Parisi splitting function. Its explicit form
is given in Appendix~\ref{app:AP}.

Combining virtual and real contributions with the collinear renormalization contribution we  find the final result
for NLO electroweak corrections to the $u + \bar d \to W^+$ process. It reads
\be
\begin{split}
 2s\cdot {\rm d} & \sigma^{\rm EW}_{{\rm NLO},q\bar q'}  = 
  \left \langle F_{\rm LV}^{\rm EW,fin}(1_u,2_{\bar d})
    + {\cal O}_{\rm NLO}^{\gamma}\left [ F_{\rm LM}(1_u, 2_{\bar d};4_\gamma) \right ] \right \rangle 
    + [\alpha] \int \limits_{0}^{1} {\rm d} z  \; \sum \limits_{i=1}^{2} 
   Q_i^2 P_{qq}^{\rm fin}(z, E_i)
    \left \langle F^{(i)}_{\rm LM}(1_u, 2_{\bar d}  \; | \; z ) \right \rangle
 \\
&
    + [\alpha] \left \langle \left [
    -\frac{Q_u^2 + Q_d^2 }{\ep^2} 
    - \frac{1}{\ep} \left ( \frac{5}{2} (Q_u^2 + Q_d^2) - 2 Q_uQ_d \right )
 + Q_u Q_d \big[1-\cos (\pi \ep) \big] \left ( \frac{2}{\ep^2} + \frac{3}{\ep} \right )
    \right ] s_{12}^{-\ep} F_{\rm LM}(1_u,2_{\bar d}) \right \rangle
     \\
 &
    + [\alpha]  \frac{\Gamma^2(1 - \ep)}{\Gamma(1 - 2\ep)}
     \left \langle 
     (2 E_{\rm max} )^{-2\ep} {\tilde J}_\gamma(E_1,E_2)
     + \sum \limits_{i=1}^{2} \left  \{  
        \frac{Q_i^2 (2 E_i)^{-2\ep}}{\ep}  \left  [  \frac{3}{2}
         + \frac{1}{\ep} \lp1 - e^{-2\ep L_i} \rp  \right  ]
                           \right \}   F_{\rm LM}(1_u,2_{\bar d}) \right \rangle,
         \label{eq23a}
\end{split}
\ee
where $F_{\rm LV}^{\rm EW,fin}$ is defined in Appendix~\ref{sectA1}.
The splitting function $P^{\rm fin}_{qq}(z,E_i)$ is defined in the following way 
\be
P_{qq}^{\rm fin}(z,E_i) =
- \frac{1}{\ep} \left [
\frac{\Gamma^2(1 - \ep)}{\Gamma(1 - 2\ep)} (2 E_i)^{-2\ep}  \left (
P_{qq}^{\rm NLO}(z,0) + \frac{3}{2} \delta(1-z) \right )
   -\frac{ \Gamma(1 - \ep)}{e^{\ep \gamma_E} \mu^{2\ep} }  \bar P_{qq}^{{\rm AP},0}(z)
   \right ].
\label{eqPqqfin}
\ee

The representation of the NLO cross section  as in Eq.(\ref{eq23a}) is convenient as it allows us to compute convolutions
of these cross sections with splitting functions, required for the evaluation of mixed QCD-EW corrections,
in a straightforward way. 
It is easy to check that, upon expanding  in $\ep$,  all singularities in Eq.(\ref{eq23a})
cancel and a  finite result is obtained. In the center-of-mass frame $E_1=E_2$ and with $\mu=M_W$, 
$E_{\rm max} = E_1$,
Eq.(\ref{eq23a}) becomes
\be
\begin{split}
& 2s\cdot \d\sigma^{\rm EW}_{{\rm NLO},q\bar q'} = 
    \\
    &+
    \lp\frac{\alpha_{EW}}{2\pi}\rp\Bigg\{
    \int\limits_0^1\d z\sum_{i=1}^{2} Q_i^2 
    \mathcal P_{qq}^{\rm NLO}(z)\left \langle F^{(i)}_{\rm LM}(1_u, 2_{\bar d}  \; | \; z ) \right \rangle
+ 
    \left[\frac{\pi^2}{3}\lp Q_u^2+Q_d^2\rp + 
    \lp 2 - \frac{\pi^2}{2}\rp Q_W^2\right]
    \langle F_{\rm LM}(1_u,2_{\bar d})\rangle\Bigg\}
    \\
    &
+       \left \langle F_{\rm LV}^{\rm EW,fin}(1_u,2_{\bar d})
    + {\cal O}_{\rm NLO}^{\gamma}\left [ F_{\rm LM}(1_u, 2_{\bar d};4_\gamma) \right ] \right \rangle
         +\mathcal O(\ep).
   \label{eqNLOns}
\end{split}
\ee
In Eq.(\ref{eqNLOns}), $\alpha_{EW}$ is the renormalized coupling\footnote{Eventually, we will work in
the $G_{\mu}$ input parameter scheme. In the setup described in Section~\ref{sect7}, we obtain
$1/\alpha_{EW} = 132.338$.} and we have 
introduced
\be
\mathcal P_{qq}^{\rm NLO}(z) = 4 \left[\frac{\ln(1-z)}{1-z}\right]_+ -2(1+z)\ln(1-z)+(1-z)-
\frac{1+z^2}{1-z}\ln z.
\label{calPqq}
\ee

The extraction of singularities in the $\gamma q \to W q'$ channel proceeds in full analogy with the discussion above and, for this reason,
we do not repeat it here and limit ourselves to presenting the final result. For definiteness, we consider 
the $\gamma \bar d \to W^+ \bar u$ channel, work in the center-of-mass frame and set $\mu=M_W$, 
$E_{\rm max} = E_1=E_2$.  We obtain 
\be
2s\cdot \d\sigma^{\rm EW}_{{\rm NLO},\gamma q} = 
[\alpha]
    \int\limits_0^1\d z
	N_c Q_u^2 P_{qg}^{\rm fin}(z,E_1)
\left \langle \frac{F_{\rm LM}(z\cdot 1_u, 2_{\bar d})}{z} \right \rangle
+
    \left\langle{\cal O}_{\rm NLO}^{\bar u}\left [ F_{\rm LM}(1_\gamma, 2_{\bar d};4_{\bar u}) \right ] \right \rangle ,
\label{eq:sigmaNLOqg}
\ee
where ${\cal O}_{\rm NLO}^{\bar u} = I - C_{41}$.
We also defined
\be
P_{qg}^{\rm fin}(z,E_i) = -\frac{1}{\ep}\left[\frac{\Gamma^2(1-\ep)}{\Gamma(1-2\ep)}(2E_i)^{-2\ep}
P_{qg}^{\rm NLO}(z) - \frac{\Gamma(1-\ep)}{e^{\ep \gamma_E}\mu^{2\ep}} \bar P_{qg}^{{\rm AP},0}(z)
\right],\label{eqPfinqg}
\ee
with $\bar P_{qg}^{{\rm AP},0}(z)$ defined in Appendix~\ref{app:AP} and
\be
P_{qg}^{\rm NLO}(z) = (1-z)^{-2\ep} \big[ (1-z)^2 + z^2 - \ep \big]/(1-\ep).
\label{PqgNLOz}
\ee
We note that the factor $1/(1-\ep)$ appears because the averaging factors of hard processes, included in the definition of hard functions $F_{\rm LM},$ are different for processes with different initial states.
In case of Eq.(\ref{eq:sigmaNLOqg}), the left hand side involves
a gluon-quark cross section  where the overall factor includes an average over $(d-2)=2(1-\ep)$ gluon
polarizations; on the right hand side of Eq.(\ref{eq:sigmaNLOqg}) the cross section for quark-antiquark annihilation
appears where an average over the two quark (antiquark) polarizations is included. The mismatch between 
polarizations of gluons and quarks in the initial state leads to the factor $(1-\ep)$ that appears explicitly in our
definition of $P_{qg}^{\rm NLO}$. 

After expanding in $\ep$, we obtain
\be
\begin{split}
2s\cdot \d\sigma^{\rm EW}_{{\rm NLO},\gamma q} &= 
    \lp\frac{\alpha_{EW}}{2\pi}\rp
    \int\limits_0^1\d z
	N_c Q_u^2\mathcal P_{qg}^{\rm NLO}(z)
\left \langle \frac{F_{\rm LM}(z\cdot 1_u, 2_{\bar d})}{z} \right \rangle
+
    \left\langle{\cal O}_{\rm NLO}^{\bar u}\left [ F_{\rm LM}(1_\gamma, 2_{\bar d};4_{\bar u}) \right ] \right \rangle 
    +\mathcal O(\ep),
\end{split}
\label{eqNLOaq}
\ee
with
\be
\mathcal P_{qg}^{\rm NLO}(z) = \big[z^2+(1-z)^2\big]\ln\lp\frac{(1-z)^2}{z}\rp + 2z(1-z).
\label{calPqg}
\ee

Although Eqs.(\ref{eqNLOns},\ref{eqNLOaq}) are written for $\mu=M_W$, the full scale dependence can be easily restored using renormalization-group
arguments. 


\section{Mixed QCD-EW corrections at next-to-next-to-leading order: derivation}
\label{sect5}

The purpose of this section is to describe the computation of all the relevant  contributions to
mixed QCD-electroweak corrections. At this order, five partonic channels ($q\bar q'$,  $g q$, $\gamma q$, $\gamma g$, $q q'$) contribute.  In this section, we focus
on the first two channels. The reason for this is that the $\gamma q$ channel can be obtained with manipulations similar (but simpler) to the ones for the $g q$ channel. The $\gamma g$ and $q q'$ channels can readily be obtained
by a simple abelianization~\cite{deFlorian:2018wcj} of the QCD result~\cite{Caola:2017dug,Caola:2019nzf}. For completeness,
we will present final formulas for all the channels in the next section. 

This section is organized as follows: in Section~\ref{sect5a} we discuss in detail the double-real contribution to the $q\bar q'$ channel, the most challenging part of the calculation.
 In Section~\ref{sect6a} we briefly explain how to derive all the other required contributions
for the $q\bar q'$ channel. Finally, in Section~\ref{sect7a} we discuss the $gq$ channel.

\subsection{The $q \bar q'$ channel: double-real contribution}
\label{sect5a}

\begin{figure}
\includegraphics[width=0.5\textwidth]{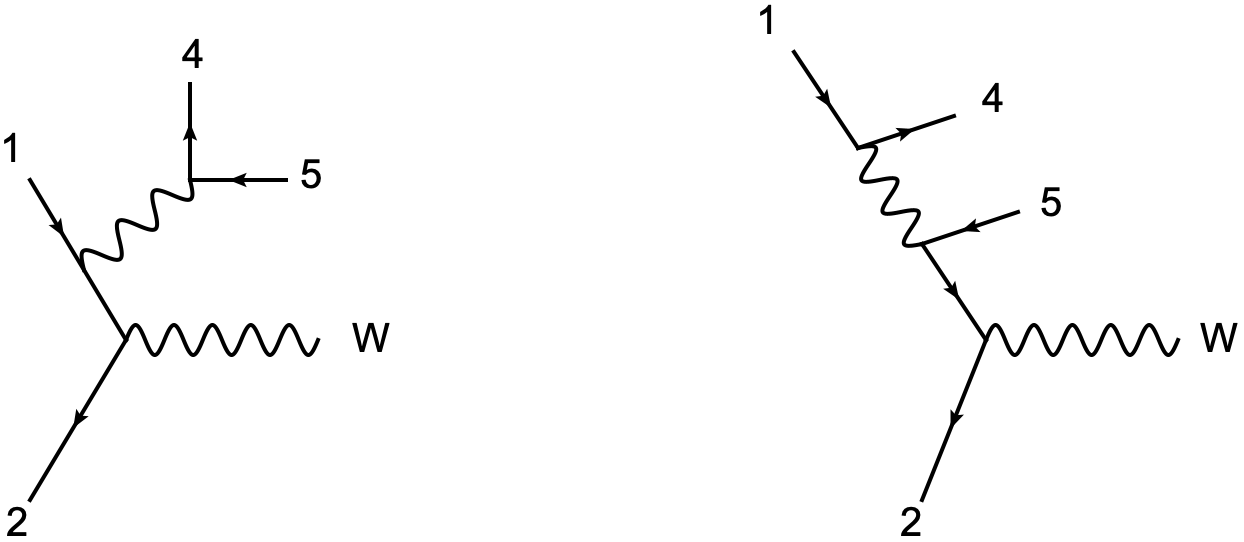}
\caption{Two-quark final state contribution to the $q\bar q'$ channel. In both Feynman graphs, the internal
vector boson can be either a gluon or a $\gamma/W/Z$. Left: representative of $s$-channel diagram. 
Right: representative of $t$-channel diagram. 
Because of color conservation, only the $s$/$t$ interference leads to a non-vanishing contribution at
$\mathcal O(\alpha_s\alpha_{EW})$.}
\label{fig:mixed_int}
\end{figure}

 To illustrate the main differences with respect to the
 earlier NNLO QCD computations ~\cite{Caola:2017dug,Caola:2019nzf},
 we consider a real-emission process $u(p_1)+ \bar d(p_2) \to W^+ f_1(p_4) f_2(p_5)$
and explain how to construct  subtraction terms and how to integrate them over unresolved phase spaces.
As mentioned earlier, we work in an arbitrary reference frame. In principle, we need to consider two options:
either $(f_1,f_2) = (g,\gamma)$ or $(f_1,f_2) = (q,\bar q)$.  
However, the singularity structure of the second contribution is
very simple: since at $\mathcal O(\alpha_s\alpha_{EW})$ the two-quark final states only contribute through a
$s$- and $t$-channel interference (see Fig.~\ref{fig:mixed_int})  the matrix element is only singular in triple collinear configurations $p_{4}||p_{5}||
p_{i}$, $i=1,2$. These configurations can be dealt with by abelianizing the corresponding QCD
result~\cite{Caola:2017dug,Caola:2019nzf}. For this reason, we do not discuss it here and focus our
attention on the $(f_1,f_2) = (g,\gamma)$ final state. 

We write  the  cross section for the partonic sub-process $u(p_1)+\bar d(p_2)\to W^+ + g(p_4)+\gamma(p_5)$ as follows 
\be
  2 s\cdot  {\rm d} \sigma_{RR}^{g \gamma} = \int [{\rm d} p_4] [{\rm d} p_5] F_{\rm LM}(1_u,2_{\bar d} \; ;  4_g, 5_\gamma)
       = \langle F_{\rm LM}(1_u,2_{\bar d}  \; ;  4_g, 5_\gamma) \rangle.
\label{eq1}
       \ee
As discussed in Section~\ref{sect2},  we do not order gluon and photon energies since there are no entangled singularities in the
kinematic limit when both of these particles become soft. 

Similar to the NNLO QCD case,  we first isolate  soft singularities. We write 
\be
\begin{split} 
\langle F_{\rm LM}(1_u,2_{\bar d};  4_g, 5_\gamma) \rangle
& =     \langle S_\gamma S_g \; F_{\rm LM}(1_u,2_{\bar d}  \; ;  4_g, 5_\gamma) \rangle
\\
&    +  \langle \left (  (I - S_g) S_\gamma
+ (I - S_\gamma)  S_g \right ) \; F_{\rm LM}(1_u,2_{\bar d}  \; ;  4_g, 5_\gamma) \rangle
   \\
   & +     \langle (I - S_\gamma) (I - S_g)  F_{\rm LM}(1_u,2_{\bar d}  \; ;  4_g, 5_\gamma )\rangle.
   \end{split} 
\label{eq2}
\ee
The different terms appearing  on the right-hand side in Eq.(\ref{eq2}) are split according to their type. The first
term $\sim S_\gamma S_g$ is the double-soft contribution where both the gluon and the photon decouple from the rest of the
process. The second group of terms proportional to $S_\gamma(I-S_g)$ and $S_g(I-S_\gamma)$ describes
cases where one of the two massless
gauge bosons (a photon or a gluon) is soft and the other one is hard (i.e. soft-regulated).
The third term on the right-hand side of Eq.(\ref{eq2})
describes a contribution where all soft singularities are  regulated. 

We can use the integrals of the eikonal functions discussed in Section~\ref{sect3}
to compute the relevant integrated soft-subtraction terms. The double-soft contribution reads
\be
\langle S_\gamma S_g F_{\rm LM}(1_u,2_{\bar d} \; ; 4_g,5_\gamma) \rangle 
= [\alpha] [\alpha_s] (2 E_{\rm max})^{-4\ep}
\left ( \frac{\Gamma^2(1-\ep)}{\Gamma(1-2\ep)} \right )^2 \frac{2 C_F}{\ep^2}\; \langle {\tilde J}_{\gamma}(E_1,E_2)
F_{\rm LM}(1_u,2_{\bar d}) \rangle.
\label{eq28a}
\ee

The case when the gluon is soft but the photon is regulated  is described by the following formula
\be
\langle S_g (I - S_\gamma)  F_{\rm LM}(1_u,2_{\bar d} \; ; 4_g,5_\gamma) \rangle
= [\alpha_s] \frac{2 C_F}{\ep^2} \frac{\Gamma^2(1-\ep)}{\Gamma(1-2\ep)}
\left ( 2 E_{\rm max} \right )^{-2\ep}
\langle (I - S_\gamma) F_{\rm LM}(1_u,2_d \; ;5_\gamma) \rangle.
\label{eq12} 
\ee
The function $\langle (I - S_\gamma) F_{\rm LM}(1_u,2_d \; ; 5_\gamma) \rangle$ in Eq.(\ref{eq12}) contains collinear divergences that
arise when the photon is emitted along the directions of incoming quarks.  They can be extracted following the discussion
of NLO QED corrections to $u + \bar d \to W^+$ in the previous section. Note that at this stage we already benefit
from the fact that the energies of gluons and photons in soft limits are not correlated; compared to the QCD case, this simplifies
the calculation considerably.  We find the fully-regulated result
\be
\begin{split} 
& \langle S_g (I-S_\gamma)  F_{\rm LM}(1_u,2_d;5_\gamma) \rangle
= [\alpha_s] \frac{2 C_F}{\ep^2} \frac{\Gamma^2(1-\ep)}{\Gamma(1-2\ep)}
(2 E_{\rm max} )^{-2\ep}
\Bigg \{ \langle {\cal O}_{\rm NLO}^{\gamma} \left [ F_{\rm LM} ( 1_u,2_{\bar d} \; ; 5_\gamma) \right ] \rangle
\\
& -[\alpha] \frac{\Gamma^2(1-\ep)}{\ep \Gamma(1-2\ep)} 
 \int \limits_{0}^{1} {\rm d} z  \sum \limits_{i=1}^{2}
Q_i^2  (2 E_i)^{-2\ep} 
P_{qq}^{\rm NLO}(z,L_i)   \left \langle F^{(i)}_{\rm LM} \left (1_u,2_{\bar d} \; | \; z  \right ) 
 \right \rangle
\Bigg \}.
\end{split} 
\label{eq22}
\ee

Next, we discuss the soft-photon contribution. We write 
\be
\begin{split} 
& \langle S_\gamma ( I - S_g ) F_{\rm LM}(1_u,2_{\bar d};4_g,5_\gamma \rangle
= [\alpha] (2 E_{\rm max} )^{-2\ep} \; \frac{\Gamma^2(1-\ep)}{\Gamma(1-2\ep)}
\langle (I-S_g) J_{\gamma}(1,2,W)\; F_{\rm LM} \left (1_u,2_{\bar d}; 4_g \right ) \rangle
\\
& = [\alpha] (2 E_{\rm max} )^{-2\ep} \; \frac{\Gamma^2(1-\ep)}{\Gamma(1-2\ep)} \Bigg \{
\langle {\cal O}_{\rm NLO}^{g} \left [  J_\gamma(1,2,W)\; F_{\rm LM} \left (1_u,2_{\bar d}; 4_g \right ) \right ]  \rangle
\\
& \;\;\;\;\;\;\;\;\;\;\;\;\;\;\;\;\;\;\;\;\;\;\;\;\;\;\;\;\;\;\;\;\;\;
\;\;\;\;\;\;\;\;\;\;\;\;\;\;\;\;\;\;\;\;\;\;\;+ \sum \limits_{i=1}^{2}
\langle C_{gi} ( I - S_g)  J_\gamma(1,2,W) F_{\rm LM} \left (1_u,2_{\bar d} \; ; 4_g \right ) \rangle
\Bigg \},
\end{split} 
\ee
where  ${\cal O}^g_{\rm NLO} =(I-S_g) (I - C_{g1} - C_{g2} )$.

The calculation of the
soft-subtracted collinear gluon contribution proceeds in full analogy with the  NLO QED case discussed in Section~\ref{sect3}.
The only difference is the presence of  the  soft-photon factor $J_\gamma(1,2,W)$.  It is easy to see,
however, that this factor turns into ${\tilde J}_{\gamma}(z E_1,E_2)$ if
the collinear limit $p_4 || p_1$ is considered and 
 into ${\tilde J}_{\gamma}(E_1,z E_2)$  if the collinear limit $p_4 || p_2$ is considered. 
 We thus obtain 
 \be
\begin{split} 
& \langle S_\gamma ( I - S_g ) F_{\rm LM}(1_u,2_{\bar d} \; ; 4_g,5_\gamma \rangle
= [\alpha] (2 E_{\rm max} )^{-2\ep} \; \frac{\Gamma^2(1-\ep)}{\Gamma(1-2\ep)}
\langle (I-S_g) J_\gamma(1,2,W) \; F_{\rm LM} \left (1_u,2_{\bar d} \; ; 4_g \right ) \rangle
\\
& = [\alpha] (2 E_{\rm max} )^{-2\ep} \; \frac{\Gamma^2(1-\ep)}{\Gamma(1-2\ep)} \Bigg \{
\langle {\cal O}_{\rm NLO}^{g} \left [  J_{\gamma}(1,2,W) \; F_{\rm LM} \left (1_u,2_{\bar d} \; ; 4_g \right ) \right ] \rangle
\\
&  -\frac{[\alpha_s] C_F}{\ep} \frac{\Gamma^2(1-\ep)}{\Gamma(1-2\ep)}
\int \limits_{0}^{1} {\rm d} z \;
\sum \limits_{i=1}^{2} (2 E_i)^{-2\ep}
{P}^{\rm NLO}_{qq}(z,L_i) \; {\tilde J}^{(i)}_{\gamma}(E_1,E_2\; | \;  z  ) 
\left \langle  F^{(i)}_{\rm LM} \left (1_u,2_{\bar d} \; | \; z \right ) \right \rangle
\Bigg \},
  \label{eq24}
\end{split} 
\ee
where, similar to Eq.(\ref{eq22a}), we used the notation 
\be
   {\tilde J}^{(i)}_{\gamma}(E_1,E_2\; | \;  z  )
   =
   \left \{
   \begin{array}{cc}
  {\tilde J}_{\gamma}(z E_1,E_2  ),   &  i=1, \\
     {\tilde J}_{\gamma}(E_1,z E_2  ),   &  i = 2.
\end{array}
     \right. 
\ee

Eqs.(\ref{eq22},\ref{eq24}) provide fully-regulated results 
for single-soft gluon/photon
contributions. It remains to consider the term $\langle (I - S_g) (I-S_\gamma) F_{\rm LM}(1_u,2_{\bar d}; 4_g, 5_\gamma) \rangle$
in Eq.(\ref{eq2}) where both soft-photon and soft-gluon
singularities are regulated. 
This soft-regulated contribution possesses collinear singularities that need to be extracted. 
We do that following the NNLO QCD computations~\cite{Caola:2017dug,Caola:2019nzf}
but we make use of the peculiarities of mixed QCD-EW correction to simplify their treatment significantly. 

Indeed, similar to the QCD case, we deal with collinear singularities by introducing 
partition functions that select a subset of all the possible collinear configurations. 
We write
\be
1 = \omega^{\gamma 1, g1} + \omega^{\gamma2,g2} + \omega^{\gamma 1,g2}+ \omega^{\gamma 2,g1},
\label{eq26}
\ee
and note that partition functions are constructed in
such a way that   $\omega^{\gamma i, g j} F_{\rm LM}(1_u,2_{\bar d}; 4_g, 5_\gamma)$ develops collinear
singularities if and only if a photon is collinear
to parton $i$ and/or  a gluon is collinear to parton $j$. This is possible because no singularities appear 
when a photon is collinear to a gluon.

The first two contributions
in Eq.(\ref{eq26}) contain triple-collinear singularities and, for this reason, require further 
partitioning~\cite{Caola:2017dug,Caola:2019nzf}.
This is done by introducing sectors that order the gluon and photon emission angles relative to a particular
collinear
direction. We therefore write 
\be
\begin{split} 
  &  \omega^{\gamma 1, g1} = \omega^{\gamma 1, g1} \left ( \theta_A + \theta_B \right ),
  \\
  &  \omega^{\gamma 2, g2} = \omega^{\gamma 2, g2} \left ( \theta_A + \theta_B \right ).
  \end{split} 
\ee
The two sectors $A$ and $B$ are defined in the following way. In  the partition described by the function $\omega^{\gamma i,g i}$ the gluon and photon emission angles are ordered as 
\be
{\rm sector}~A:\;\;\; \theta^{gi} < \theta^{\gamma i},\;\;\;\;{\rm sector}~B:\;\;\; \theta^{\gamma i} < \theta^{g i}.
\ee
The full partitioning of the phase space becomes
\be
1 = \omega^{\gamma 1, g1} \left ( \theta_A + \theta_B \right )
+ \omega^{\gamma2,g2} \left ( \theta_A + \theta_B \right )
+ \omega^{\gamma 1,g2}+ \omega^{\gamma 2,g 1}.
\ee

We can now insert collinear projection operators  in relevant places taking into account the
ordering of angles in sectors
$A$ and $B$, see Refs.~\cite{Caola:2017dug,Caola:2019nzf} for details. We find the modified partition of unity 
\be
 1 = \Xi^{q \bar q} = \Xi^{q \bar q}_1 + \Xi^{q \bar q}_2 + \Xi^{q \bar q}_3 + \Xi^{q \bar q}_4,
\ee
where the different $\Xi^{q \bar q}$-operators read\footnote{As explained in Refs.~\cite{Caola:2017dug,Caola:2019nzf},
there is some freedom in the definition of the collinear operators $C_i$. In particular, one can decide whether
they should act only on the matrix element and momentum-conserving $\delta$ function or if they should also
modify the unresolved phase space. In this paper, we make the same choice we did in Ref.~\cite{Caola:2019nzf}:
all triple-collinear operators $C_{\gamma g,i}$ in Eq.(\ref{eq31}) {\it do not} modify the unresolved phase space,
while all the double collinear operators {\it do} act on it, see Ref.~\cite{Caola:2019nzf} for details.}
\be
\begin{split} 
\Xi^{q \bar q}_1 = {}& (I - C_{\gamma g, 1}) ( I - C_{g1}) \omega^{\gamma 1, g1} \theta_A
+  (I - C_{\gamma g, 1}) ( I - C_{ \gamma 1}) \omega^{\gamma 1, g1} \theta_B
+ ( I - C_{\gamma g,2}) ( I - C_{g2} ) \omega^{\gamma2,g2} \theta_A
\\
&
+ (I - C_{\gamma g,2}) ( I - C_{\gamma 2} ) \omega^{\gamma 2, g2} \theta_B
+ (I - C_{g2} ) ( I - C_{\gamma 1}) \omega^{\gamma 1, g2} + ( I - C_{g1}) (I - C_{\gamma 2}) \omega^{\gamma 2, g1},
\\
\Xi^{q \bar q}_2 = {}& 
C_{\gamma g,1} (I - C_{g1} ) \omega^{\gamma 1, g1} \theta_A 
+C_{\gamma g, 1} (I - C_{\gamma 1} ) \omega^{\gamma 1, g1} \theta_B 
+C_{\gamma g, 2} ( I - C_{g 2 }) \omega^{\gamma 2, g2} \theta_A
+C_{\gamma  g, 2} (I - C_{\gamma 2}) \omega^{\gamma 2, g 2} \theta_B,
\\
\Xi^{q \bar q}_3 = {}&  -C_{g2} C_{\gamma 1} \omega^{\gamma 1, g2} - C_{\gamma 2} C_{g1} \omega^{\gamma 2, g 1},
\\
\Xi^{q \bar q}_4 = {}& 
C_{g1} \left [ \omega^{\gamma 1, g1} \theta_A + \omega^{\gamma 2, g1} \right ]
+C_{\gamma 1} \left [ \omega^{\gamma 1, g1} \theta_B + \omega^{\gamma 1, g2} \right ]
+C_{g 2} \left [ \omega^{\gamma 2, g2} \theta_A + \omega^{\gamma 1, g2} \right ]
+C_{\gamma  2} \left [ \omega^{\gamma 2, g2} \theta_B + \omega^{\gamma 2, g1} \right ].
\end{split} 
\label{eq31}
\ee

We then re-write the soft-regulated contribution in Eq.(\ref{eq2}) in the following way
\be
\begin{split} 
\langle (I - S_g) (I-S_\gamma) F_{\rm LM}(1_u,2_{\bar d}; 4_g, 5_\gamma) \rangle
& = \langle (I - S_g) (I-S_\gamma) \; \Xi^{q \bar q} \; F_{\rm LM}(1_u,2_{\bar d} \; ; 4_g, 5_\gamma) \rangle
\\
& = \sum \limits_{i}^{4} \langle (I - S_g) (I-S_\gamma) \; 
\Xi^{q \bar q}_i \; F_{\rm LM}(1_u,2_{\bar d} \; ; 4_g, 5_\gamma) \rangle.
\end{split}
\label{eq32}
\ee
Among the four contributions that appear on the right hand side in Eq.(\ref{eq32}),
the one proportional to $\Xi^{q \bar q}_1$ is the  fully-regulated one. We  compute it numerically  in four dimensions.
The $\Xi^{q \bar q}_2$ term describes a triple-collinear singular contribution
which can be computed following the discussion in Ref.~\cite{Delto:2019asp}. We note that, because of
the different phase-space partition adopted here, the triple-collinear contribution required here is not identical to the one 
computed in Ref.~\cite{Delto:2019asp}. Further details about the computation 
are  given in Appendix~\ref{appendixtc}.  The result reads
\be
\langle ( I - S_\gamma) (I -S_g) \Xi^{q \bar q}_2 F_{\rm LM} (1_u,2_{\bar d}; 4_g, 5_\gamma ) \rangle
= -2 [\alpha] [\alpha_s] C_F  \int \limits_{0}^{1} {\rm d} z P^{\rm trc}_{qq}(z)
 \sum \limits_{i=1}^{2} \; (2 E_i)^{-4\ep} Q_i^2 \left \langle F^{(i)}_{\rm LM}(1_u, 2_{\bar d} \; | \; z  )
\right \rangle,  
\label{eq57}
\ee
where
\be
\begin{split} 
& P^{\rm trc}_{qq}(z)  = \frac{1}{\ep} \left ( 
\frac{3}{2}  (1 - z) + z \ln(z) + \frac{3+z^2}{4(1-z)}\ln^2(z)
\right )
+ \left ( \frac{11}{2} -6 \ln(1-z) \right ) (1-z) 
- \frac{2 \pi^2 z}{3} 
-  \frac{z}{2}  \ln^2(z)
\\
& \;\;\;\;
-\frac{(19 + 9 z^2)}{12(1-z)}\ln^3(z)
+ 4 z {\rm Li}_2(z)
-  \left ( z + \frac{\pi^2 (5 + 3 z^2)}{3 (1 - z)}
  + \frac{2 (1 + z^2)}{1-z} {\rm Li}_2(z) \right )  \ln(z) 
  + \frac{2 (5 + 3 z^2)}{1-z} \left ( {\rm Li}_3(z) - \zeta_3 \right ).
\end{split} 
\ee

The two contributions in Eq.(\ref{eq32}) that require further action  
are the ones proportional to $\Xi^{q \bar q}_{3,4}$.  These contributions are computed in a way which is similar
to the NNLO QCD case except for the required modifications of sectors and for the fact that in our case
the soft subtractions are done independently for a gluon and a photon. 

For definiteness, we now explicitly specify the partition functions
\be
\omega^{\gamma 1, g1} = \frac{\rho_{\gamma 2} \rho_{g2}}{4},\;\;\;\;
\omega^{\gamma 2, g2} = \frac{\rho_{\gamma 1} \rho_{g1} }{4},
\;\;\; \omega^{\gamma 1, g2} = \frac{\rho_{\gamma2} \rho_{g1}}{4},
\;\;\; \omega^{\gamma 2, g1} = \frac{\rho_{\gamma 1} \rho_{g2}}{4},
\label{eq33}
\ee
where $\rho_{ij} =  p_i\cdot p_j/(E_i E_j)$, 
and focus first on the  contribution to Eq.(\ref{eq32}) proportional to $\Xi_3^{q\bar q}$.
It  describes
singular emissions of  a photon and a gluon
collinear to  opposite directions, and reads 
\be
\begin{split}
& \langle (I - S_g) (I-S_\gamma)
\Xi_3^{q\bar q}  F_{\rm LM}(1_u,2_{\bar d} \; ; 4_g, 5_\gamma) \rangle  = \\
&
 - \langle (I - S_g) (I-S_\gamma)
\left ( C_{g2} C_{\gamma 1} \omega^{\gamma 1, g2} +  C_{\gamma 2} C_{g1} \omega^{\gamma 2, g 1}
\right ) F_{\rm LM}(1_u,2_{\bar d} \; ; 4_g, 5_\gamma) \rangle.
\end{split}
\ee
Proceeding as in the NNLO QCD case~\cite{Caola:2017dug,Caola:2019nzf}, we obtain
\be
\begin{split} 
& \langle ( I - S_\gamma) (I - S_g) \Xi^{q \bar q}_3  F_{\rm LM}(1_u,2_{\bar d};4_g,5_\gamma) \rangle =
\\
&   -[\alpha] [\alpha_s] \frac{(Q_u^2 + Q_d^2) C_F}{\ep^2} (2 E_1)^{-2\ep} (2E_2)^{-2\ep} 
\int \limits_{0}^{1} {\rm d} z_1 {\rm d} z_2 P^{\rm NLO}_{qq}(z_1,L_1)  P^{\rm NLO}_{qq}(z_2,L_2) 
\left \langle \frac{F_{\rm LM}(z_1\cdot 1_u, z_2 \cdot 2_{\bar d} )}{z_1 z_2}
\right \rangle,
\end{split}
\label{eq46a}
\ee
where the splitting function $P_{qq}^{\rm NLO}$ is defined in Eq.(\ref{eq21aa}), see also Appendix~\ref{app:split}.

Next, we consider single collinear limits described by the operator $\Xi^{q \bar q}_4$. The corresponding
contribution reads
\be
 \langle ( I - S_\gamma) (I - S_g) \Xi^{q \bar q}_4  F_{\rm LM}(1_u,2_{\bar d};4_g,5_\gamma) \rangle
= \sum \limits_{\alpha \in \{\gamma, g\},i \in \{1,2\}} \langle
( I - S_\gamma) (I - S_g) C_{\alpha i} \Omega_{\alpha i}
F_{\rm LM}(1_u,2_{\bar d} \; ; 4_g,5_\gamma) \rangle, 
\label{eq37}
\ee
where the various quantities $\Omega_{\alpha i}$ can be deduced  from  Eq.(\ref{eq31}).
Again, proceeding as in the NNLO QCD case~\cite{Caola:2017dug,Caola:2019nzf},  we obtain the following result for one of the four terms 
\be
\begin{split} 
& \left \langle
 C_{g1} \left (  \omega^{\gamma 1, g1} \theta_A + \omega^{\gamma 2, g1} \right )
(I-S_g)(I - S_\gamma) F_{\rm LM}(1_u,2_{\bar d} \; ; 4_g,5_\gamma)
\right \rangle 
=
\\
& -\frac{[\alpha_s] C_F}{\ep}
{\cal O}_{\rm NLO}^{\gamma} \left [ 
(2 E_1)^{-2\ep} \int \limits_{0}^{1} {\rm d} z\;
  P_{qq}^{\rm NLO}(z,L_1)
  \left \langle \frac{F_{\rm LM}( z \cdot 1_u, 2_{\bar d} \; ; 5_\gamma )}{z}
\left \{
\frac{\rho_{\gamma 2}}{2} \left ( \frac{\rho_{\gamma 1}}{2} \right )^{-\ep} 
+ \frac{\rho_{1\gamma}}{2}
\right \}
\right \rangle 
\right ]
\\
& +
\frac{ [\alpha_s] [\alpha] C_F Q_u^2}{2\ep^2}
\frac{\Gamma(1-2\ep) \Gamma(1-\ep)}{\Gamma(1-3\ep)}
(2 E_1)^{-4\ep} \int \limits_{0}^{1} {\rm d} z\; \left[P_{qq}^{\rm NLO} \otimes P_{qq}^{\rm NLO}\right](z,E_1)
\left \langle
\frac{F_{\rm LM}( z \cdot 1_u, 2_{\bar d})}{z }
\right  \rangle
\\
& +
\frac{ [\alpha_s] [\alpha] C_F Q_d^2}{2\ep^2}
\frac{\Gamma^2(1-\ep) }{\Gamma(1-2\ep)}
(2 E_1)^{-2\ep} (2 E_2)^{-2\ep}
\int \limits_{0}^{1} {\rm d} z_1\;{\rm d} z_2 \; P_{qq}^{\rm NLO}(z_1,L_1)\; P_{qq}^{\rm NLO}(z_2,L_2)
 \left \langle
\frac{F_{\rm LM}( z_1 \cdot 1_u, z_2 \cdot 2_{\bar d})}{z_1 z_2}
\right \rangle.
\label{eq53} 
\end{split} 
\ee
The  splitting function $\big[P_{qq}^{\rm NLO} \otimes P_{qq}^{\rm NLO}\big](z,E_1)$ that
appears in Eq.(\ref{eq53})  is defined
as follows
\be
\big[P_{qq}^{\rm NLO} \otimes P_{qq}^{\rm NLO}\big](z,E_1)
= \int\limits_0^1 {\rm d} z_1 {\rm d} z_2 \; z_1^{-2\ep} P_{qq}^{\rm NLO}(z_1,L_1)  P_{qq}^{\rm NLO}(z_2,L_{1z_1})
\delta(z - z_1 z_2),
\ee
where $L_{1z} = \ln (E_{\rm max}/(E_1 z_1))$.  It
can be written as
\be
\begin{split} 
\left [ P_{qq}^{\rm NLO} \otimes P_{qq}^{\rm NLO}\right ] (z,E_1) 
& = \left [ P_{qq} \otimes P_{qq} \right ]^{\rm ab}_{RR}(z)
- \frac{1}{\ep}  \left (1 - e^{-2 \ep L_1} \right ) \left (P_{qq}^{\rm NLO}(z) - \frac{1}{\ep}\delta(1 - z)e^{-2\ep L_1} \right )
\\
  & -  \frac{1}{\ep^2} e^{-2\ep L_1} \left (1 - e^{-2 \ep L_1} \right )\delta(1 - z),
\end{split}
  \ee
  where $ \left [ P_{qq} \otimes P_{qq} \right ]^{\rm ab}_{RR}$ can be obtained from
   Eq.(A.18) of Ref.~\cite{Caola:2017dug} by setting $C_F \to 1$.

The remaining three contributions to  $\Xi^{q \bar q}_4$ can be analyzed in a similar manner; the   results
can be obtained from Eq.(\ref{eq53}) with a few natural replacements.
We find 
\begin{align} 
& \langle (I - S_g) (I-S_\gamma)
  \Xi^{q \bar q}_4  F_{\rm LM}(1_u,2_{\bar d}; 4_g, 5_\gamma) \rangle = \nonumber
\\
&   -\frac{1}{\ep}  \sum \limits_{i=1}^{2} 
  \int \limits_{0}^{1} {\rm d} z\;
  (2 E_i)^{-2\ep} P_{qq}^{\rm NLO}(z,L_i)
   \left \langle 
[\alpha_s] C_F
  {\cal O}_{\rm NLO}^{\gamma} \bigg  [  F^{i}_{\rm LM}( 1_u, 2_{\bar d}; 5_\gamma |  z )\Delta_{\gamma i} \bigg ]
+ [\alpha] Q_i^2 {\cal O}_{\rm NLO}^{g} \bigg  [F^{i}_{\rm LM}( 1_u, 2_{\bar d}; 4_g |  z ) \Delta_{g i} \bigg ]
\right \rangle 
\nonumber\\
& +
\frac{ [\alpha_s] [\alpha] C_F }{\ep^2}\Bigg\{
\frac{\Gamma(1-2\ep) \Gamma(1-\ep)}{\Gamma(1-3\ep)}
\sum \limits_{i=1}^{2}
 \int \limits_{0}^{1} {\rm d} z\;
Q_i^2 (2 E_i)^{-4\ep} \;\left[ P_{qq}^{\rm NLO} \otimes P_{qq}^{\rm NLO}\right](z,E_i)
\left \langle
F^{(i)} _{\rm LM}(1_u, 2_{\bar d}\; | \; z )
\right  \rangle
\label{eq52a}
\\
& +
( Q_u^2 + Q_d^2)
\frac{\Gamma^2(1-\ep)}{\Gamma(1-2\ep)}
 \left \langle
(2 E_1)^{-2\ep} (2 E_2)^{-2\ep}
\int \limits_{0}^{1} {\rm d} z_1\; P_{qq}^{\rm NLO}(z_1,L_1)
\int \limits_{0}^{1} {\rm d} z_2\; P_{qq}^{\rm NLO}(z_2,L_2)
\frac{F_{\rm LM}( z_1\cdot 1_u, z_2 \cdot2_{\bar d})}{z_1 z_2}
\right \rangle
\Bigg\},
\nonumber
\end{align}
where the functions  $\Delta_{\gamma(g)i}$ 
are defined as follows 
\be
\Delta_{xi} =  \frac{\rho_{x j}}{2} \left ( \frac{\rho_{x i}}{2} \right )^{-\ep} 
+ \frac{\rho_{x i}}{2}, \;\;\; j \in \{1,2\} \ne i.
\ee

The final result for the
differential cross section of the process
$\bar u + d \to W^+  g\gamma$
is obtained
by summing Eqs.(\ref{eq28a},\ref{eq22}, \ref{eq24},\ref{eq57},\ref{eq46a},\ref{eq52a}) and
the fully regulated  contribution  $\langle (I - S_g) (I-S_\gamma)
\Xi^{q \bar q}_1 F_{\rm LM}(1_u,2_{\bar d}; 4_g, 5_\gamma) \rangle $ that is computed numerically.


  \subsection{The $q\bar q'$ channel: other contributions}
  \label{sect6a}
Apart from the double-real contribution discussed in the previous sub-section, the other contributions
that are required for the mixed QCD-EW calculation are the so-called double-virtual and 
real-virtual corrections, as well as collinear renormalization counterterms. The structure of virtual
corrections is discussed in Appendix~\ref{sectA1}. In this section, we consider the real-virtual and 
PDFs renormalization contributions. 

Real-virtual corrections to the $u + \bar d \to W^+$ process account for one-loop corrections
to processes with $W^+g$ and $W^+\gamma$ final states  produced in $u \bar d$ annihilation. 
It is straightforward to analyze these contributions since for both cases the structure of soft and collinear singularities
is very similar to that of a NLO calculation and, for this reason, the construction of subtraction terms is less complicated
than for the double-real case discussed in the previous sub-section.
For mixed QCD-EW corrections the situation is, in fact, simpler than in QCD because  soft limits for both the $u \bar d  \to W^+ + \gamma$
and $u \bar d  \to W^+ +g $ processes  are not affected by loop corrections. The regularization of soft divergences
is then identical to the NLO case discussed in Section~\ref{sect4} (soft-photon emission) and in Refs.~\cite{Caola:2017dug,
Caola:2019nzf} (soft-gluon emission). 
The regularization of collinear singularities is identical to the
NNLO QCD case. The only point that requires additional care is the abelianization 
of the one-loop QCD collinear splitting function and the replacement  $C_F \to Q_{u,d}^2$ where  appropriate.
The result for the $W\gamma$ final state reads
\be
\begin{split}
  &  2s \cdot {\rm d} \sigma_{RV}^\gamma =
  [\alpha] (2 E_{\rm max})^{-2\ep} \frac{\Gamma^2(1-\ep)}{\Gamma(1-2\ep)}\;
  \langle {\tilde J}_\gamma(E_1,E_2) F^{\rm QCD}_{\rm LV}(1_u,2_{\bar d}) \rangle
  +\left\langle {\cal O}_{\rm NLO}^{\gamma} \left [ 
  F^{\rm QCD}_{\rm LV} \left (1_u, 2_{\bar d} \; ; 4_\gamma \right ) \right ]\right\rangle
  \\
& -\frac{[\alpha]}{\ep}\frac{\Gamma^2(1 - \ep)}{\Gamma(1 - 2\ep)}
\int \limits_{0}^{1} {\rm d} z\;   \sum \limits_{i=1}^{2} \; (2 E_i)^{-2\ep}Q_i^2
P_{qq}^{\rm NLO}(z, L_i) \left \langle  F_{\rm LV}^{(i),\rm QCD}(1_u, 2_{\bar d}\; | \; z)
\right \rangle 
\\
& + \frac{[\alpha][\alpha_s] C_F}{\ep} \frac{\Gamma(1 - \ep)^4
  \Gamma(1 + \ep)}{\Gamma(1 - 3\ep)} 
\int\limits_0^1 {\rm d} z  \; P_{qq}^{\rm RV}(z) \; \sum \limits_{i=1}^{2} \;  (2E_i)^{-4\ep}  Q_i^2
\left \langle 
F_{\rm LM}^{(i)}(1_u, 2_{\bar d} \; | \; z)
\right \rangle, 
   \end{split}
\ee
with $P_{qq}^{\rm RV}$ defined in Appendix~\ref{app:split}, while the result  for the $Wg$ final state is
\be
\begin{split} 
  & 2s \cdot {\rm d}\sigma_{RV}^{g} =
  [\alpha_s] (2 E_{\rm max})^{-2\ep} \frac{\Gamma^2(1-\ep)}{\Gamma(1-2\ep)} \frac{2C_F}{\ep^2}\;
  \left\langle F^{\rm EW}_{\rm LV}(1_u,2_{\bar d}) \rangle
  + \langle {\cal O}_{\rm NLO}^{g} \left [ 
  F^{\rm EW}_{\rm LV} \left (1_u, 2_{\bar d} \; ; 4_g \right ) \right ] \right\rangle
  \\
& -\frac{[\alpha_s] C_F}{\ep}\frac{\Gamma^2(1 - \ep)}{\Gamma(1 - 2\ep)}
  \int\limits_0^1 {\rm d} z \; \sum \limits_{i=1}^{2} (2 E_i)^{-2\ep} P_{qq}^{\rm NLO}(z, L_i) \left \langle
F_{\rm LV}^{(i), \rm EW}(1_u, 2_{\bar d} \; | z \; ) \right \rangle 
\\
& + \frac{[\alpha][\alpha_s] C_F}{\ep} \frac{\Gamma(1 - \ep)^4
  \Gamma(1 + \ep)}{\Gamma(1 - 3\ep)}   \int\limits_0^1 {\rm d} z \; P_{qq}^{\rm RV}(z) \; 
\sum \limits_{i=1}^{2} 
(2E_i)^{-4\ep}  Q_i^2   \left \langle  F^{(i)}_{\rm LM}(1_u, 2_{\bar d}\; | z \; )
\right \rangle .
\end{split} 
\ee
The one-loop contributions  $F_{\rm LV}^{\rm QCD/EW}$ are defined in Appendix~\ref{sectA1}.

The PDFs renormalization contribution 
is obtained by computing convolutions of parton distribution functions
with lower-order cross sections. Following the steps described in Refs.~\cite{Caola:2017dug,Caola:2019nzf}, we obtain 
\be
\begin{split}
  2s\cdot {\rm d} \sigma_{\rm pdf} & = 
   2  [\alpha(\mu)] [\alpha_s(\mu)]
   \Bigg\{
     -\frac{(Q_u^2 + Q_d)^2 C_F}{2\ep^2}
     \int \limits_{0}^{1} {\rm d} z_1 {\rm d} z_2 \; \bar P^{{\rm AP},0}_{qq}(z_1) \; \bar P^{{\rm AP},0}_{qq}(z_2)
     \left\langle\frac{F_{\rm LM}(z_1\cdot 1_u , z_2 \cdot 2_{\bar d})}{z_1 z_2}\right\rangle
     \\
     &
     -\frac{C_F}{2\ep^2}\sum_{i=1}^{2} Q_i^2 \int\limits_0^1\d z
     \big[\bar P_{qq}^{{\rm AP},0}\otimes \bar P_{qq}^{{\rm AP},0}\big](z)
     \left\langle F^{(i)}_{\rm LM}(1_u, 2_{\bar d}\; | z \; )\right\rangle\Bigg\}
 + \frac{2s}{\ep}
\Bigg\{
C_F[\alpha_s(\mu)] \bigg (  \bar P_{qq}^{{\rm AP},0}  \otimes  {\rm d} \sigma_{{\rm NLO},q\bar q'}^{\rm EW}  
\\
&
+
{\rm d} \sigma_{{\rm NLO},q\bar q'}^{\rm EW} \otimes \bar P_{qq}^{{\rm AP},0} \bigg ) 
       + [\alpha(\mu)] \bigg ( Q_u^2\; \bar P_{qq}^{{\rm AP},0} \otimes  \d\sigma_{{\rm NLO},q\bar q'}^{\rm QCD} 
       + Q_d^2 \;\d\sigma_{{\rm NLO},q\bar q'}^{\rm QCD} \otimes \bar P_{qq}^{{\rm AP},0}
       \bigg ) 
        \Bigg\}
\\
&         + \frac{ [\alpha(\mu)][\alpha_s(\mu)] C_F    }{\ep} 
\int \limits_{0}^{1} {\rm d} z \;  \bar P_{qq}^{{\rm AP},1}(z) \;
\sum Q_i^2 \left\langle F^{(i)}_{\rm LM}(1_u, 2_{\bar d}\; | z \; )\right\rangle,
\end{split}
\label{eq59aaa} 
\ee
where $[\alpha(\mu)]$ and $[\alpha_s(\mu)]$ are the renormalized coupling constants defined in Eq.(\ref{eq26aaa}) 
and ``$\otimes$'' denotes the standard convolution product.
The splitting functions and their convolutions in Eq.(\ref{eq59aaa}) 
can be found in Appendices~\ref{app:split},\ref{app:AP}.
The NLO electroweak cross section $\d\sigma_{{\rm NLO},q\bar q'}^{\rm EW}$ is
given in Eq.(\ref{eqNLOns}), while its QCD equivalent $\d\sigma_{{\rm NLO},q\bar q'}^{\rm QCD}$ can be found in~\cite{Caola:2017dug,Caola:2019nzf}.

\subsection{The gluon-quark channel}
 \label{sect7a}

We now turn to the discussion of the gluon-quark channel. For definiteness, we focus on the 
process $g+ \bar d \to W^+ + \bar u$.
To compute  the mixed QCD-EW corrections to $pp  \to W^+$ arising from this partonic channel, we require 
 the real-emission contribution   $g(p_1) + \bar d(p_2) \to W^+ + \bar u(p_4) + \gamma(p_5)$, virtual electroweak corrections
to $g(p_1) + \bar d(p_2) \to W^+ + \bar u(p_4)$, as well as collinear renormalization. 

We begin with the real-emission process $g + \bar d \to W^+ + \bar u + \gamma$ and write its cross section as 
  \be
     2s\cdot {\rm d} \sigma_{gq}^{RR} =
     \langle F_{\rm LM} ( 1_g, 2_{\bar d} \; ; 4_{\bar u}, 5_\gamma ) \rangle. 
     \ee
     A soft singularity  can only be caused by a   photon. We write
     \be
     \begin{split} 
     \langle F_{\rm LM} ( 1_g, 2_{\bar d} \; ;  4_{\bar u} 5_\gamma ) \rangle & =
     \langle S_\gamma \; F_{\rm LM} ( 1_g, 2_{\bar d} \; ;  4_{\bar u}, 5_\gamma ) \rangle
     + \langle (I - S_\gamma) F_{\rm LM} ( 1_g, 2_{\bar d} \; ; 4_{\bar u}, 5_\gamma ) \rangle
     \\
     & =
      [\alpha] (2 E_{\rm max} )^{-2\ep}
     \frac{\Gamma^2(1-\ep)}{\Gamma(1-2\ep)} \langle J_\gamma(2,4,W) \; F_{\rm LM} ( 1_g, 2_{\bar d} \; ; 4_{\bar u} ) \rangle
     + \langle (I - S_\gamma) F_{\rm LM} ( 1_g, 2_{\bar d} \; ; 4_{\bar u}, 5_\gamma ) \rangle.
\label{eq83}
     \end{split} 
     \ee
     The integrated photon eikonal function $J_\gamma(2,4,W)$ is defined in Eq.(\ref{eq14a}). The single-soft
     piece in Eq.(\ref{eq83}) requires an additional collinear subtraction; the collinear singularity
     occurs when  the outgoing antiquark becomes collinear to the incoming gluon. We write
\be
\langle J_\gamma(2,4,W) \; F_{\rm LM} ( 1_g, 2_{\bar d}; 4_{\bar u} ) \rangle =
\langle {\cal O}^{\bar u}_{\rm NLO} \left [ J_\gamma(2,4,W) \; F_{\rm LM} ( 1_g, 2_{\bar d}; 4_{\bar u} ) \right ] \rangle
 + \langle C_{41} J_\gamma(2,4,W) \; F_{\rm LM} ( 1_g, 2_{\bar d}; 4_{\bar u} ) \rangle, 
\ee
where ${\cal O}^{\bar u}_{\rm NLO} = I - C_{41}$. To compute the contribution proportional to
$C_{41}$, we note that
$ C_{41} J_\gamma(2,4,W) = {\tilde J}_\gamma(z E_1, E_2)$ where $z = (E_1 - E_4)/E_1$ and
${\tilde J}_\gamma(E_1, E_2)$ is given in Eq.(\ref{eq11}).

Repeating the NLO QED  computation of Section~\ref{sect4}, we find that the soft-photon contribution is given by  
  \be
\begin{split} 
&   \langle S_\gamma \; F_{\rm LM} ( 1_g, 2_{\bar d}; 4_{\bar u}, 5_\gamma ) \rangle
  =
  [\alpha] (2 E_{\rm max} )^{-2\ep}
  \frac{\Gamma^2(1-\ep)}{\Gamma(1-2\ep)}
  \Bigg  [
    {\cal O}^{\bar u}_{\rm NLO}  
    \left [
      \left \langle  J_\gamma(2,4,W) \; F_{\rm LM} ( 1_g, 2_{\bar d}; 4_{\bar u} )
   \right \rangle
      \right ] 
    \\
& -
\frac{[\alpha_s] T_R}{\ep}
\frac{\Gamma^2(1-\ep)}{\Gamma(1-2\ep)}
 \left  \langle
(2 E_1)^{-2\ep} \int \limits_{0}^{1} \; {\rm d} z \; {P}_{qg}^{\rm NLO}(z) \; 
  {\tilde J}_\gamma( z E_1, E_2) \;  \frac{F_{\rm LM}( z\cdot 1_{u}, 2_{\bar d} )}{z}
\right \rangle 
    \Bigg ],
\end{split}
\label{eq61a}
\ee
where $P_{qg}^{\rm NLO}(z)$ is defined in Eq.(\ref{PqgNLOz}).

The soft-subtracted contribution in Eq.(\ref{eq83}) $\langle (I - S_\gamma) F_{\rm LM} ( 1_g, 2_{\bar d} \; ; 4_{\bar u}, 5_\gamma ) \rangle$
needs to
be further analyzed since it contains collinear divergences. Since the final state antiquark $\bar u$ can only develop
a collinear singularity when its momentum is along the momentum of the incoming gluon, we only need
to introduce partition functions  for the photon. We write
\be
1 = \omega^{2 \gamma} + \omega^{4 \gamma},
\label{eq91}
\ee
where
\be
\omega^{2 \gamma} = \frac{\rho_{4 \gamma}}{ \rho_{2 \gamma} + \rho_{4 \gamma} },
\;\;\;
\omega^{4 \gamma} = \frac{\rho_{2 \gamma}}{ \rho_{2 \gamma} + \rho_{4 \gamma} }.
\ee
We now rewrite Eq.(\ref{eq91}) by introducing different  collinear projection
operators for  different partition functions.
We note that  we have to  introduce the same four sectors as in the NNLO QCD case to order the 
angles of the photon and
of the up antiquark.  We find
\be
1 = \Xi_1^{gq} + \Xi_2^{gq} + \Xi^{gq}_3 + \Xi_4^{gq},
\label{eq65a}
\ee
where\footnote{As for the $q\bar q'$ channel, all double-collinear operators in
$\Xi_{i}^{gq}$ also act on the unresolved phase space, while the triple-collinear operators do not. See Ref.~\cite{Caola:2019nzf} for details.}
\be
\begin{split}
  &  \Xi_1^{gq} = (I-C_{2 \gamma} ) (I - C_{41}) \omega^{2 \gamma} + 
  \theta_C (I -C_{14 \gamma} ) (I - C_{41}) \omega^{4 \gamma} 
  + \theta_B (I - C_{14\gamma} ) ( I- C_{4\gamma}) \omega^{4 \gamma}
  \\
&    + \theta_A ( I - C_{14\gamma} ) \omega^{4 \gamma}
 + \theta_D ( I - C_{14\gamma}) (I - C_{4 \gamma} ) \omega^{4 \gamma},
  \\
  &  \Xi_2^{gq} = \theta_C C_{14\gamma} ( I - C_{41} ) \omega^{4 \gamma}
  + \theta_B C_{14\gamma} ( I - C_{4\gamma} ) \omega^{4 \gamma}
  + \theta_A C_{14\gamma} \omega^{4 \gamma} + \theta_D C_{14\gamma} ( I - C_{4\gamma} ) \omega^{4 \gamma},
  \\
  & \Xi_3^{gq} = - C_{2 \gamma} C_{41} \omega^{2 \gamma},
  \\
  &
  \Xi_4^{gq} = C_{2\gamma} \omega^{2\gamma}
+  C_{41} \left ( \theta_C  \omega^{4 \gamma} + \omega^{2\gamma} \right ) 
+ \theta_B C_{4\gamma} \omega^{4 \gamma} + \theta_D C_{4\gamma} \omega^{4\gamma}.
\end{split} 
\label{eq94}
\ee
As we mentioned, the four angular-ordered sectors $\theta_{A,...,D}$ are identical to the NNLO QCD case. 
We refer the reader to~\cite{Caola:2017dug,Caola:2019nzf} for their explicit definition. 

Using the partition of unity Eq.(\ref{eq65a}) in  Eq.(\ref{eq83}), we write the soft-subtracted terms as 
\be
\langle (I - S_\gamma) F_{\rm LM} ( 1_g, 2_{\bar d}; 4_{\bar u}, 5_\gamma ) \rangle
 = \sum \limits_{i=1}^{4} \langle (I -S_\gamma) \Xi_i^{gq}  F_{\rm LM} ( 1_g, 2_{\bar d}; 4_{\bar u}, 5_\gamma ) \rangle.
\ee

The different terms in Eq.(\ref{eq94}) have the following meaning.  The term proportional
to $\Xi_1^{gq}$ is fully regulated and  can be computed numerically in four dimensions. The term proportional
to  $\Xi_2^{gq}$ is the triple-collinear subtraction term;  $\Xi_3^{gq}$ describes collinear 
emissions of a photon and an up antiquark in opposite directions 
and $\Xi_4^{gq}$ describes the various single-collinear subtraction terms. 

We start by discussing the  triple-collinear contribution. Since the triple-collinear subtraction
term is independent of the partition and since for the $gq$ channel we use the same phase-space parametrization as
for the NNLO QCD case, the result for the integrated  triple-collinear 
subtraction term can be borrowed from the NNLO QCD results reported in Ref.~\cite{Delto:2019asp}.
We find 
\be
\begin{split}
  \langle (I -S_\gamma) \Xi_2^{gq}  F_{\rm LM} ( 1_g, 2_{\bar d}; 4_{\bar u}, 5_\gamma ) \rangle
  =  \frac{[\alpha_s][\alpha] T_R Q_u^2}{\ep}  (2 E_1)^{-4\ep}
  \int \limits_{0}^{1} {\rm d} z \; P_{qg}^{\rm trc}(z,E_1) \frac{F_{\rm LM}(z \cdot 1_u,2_{\bar d} )}{z},
\end{split}
\label{eq:xi2qg}
\ee
where the integrated triple-collinear subtraction term is given in Eq.(\ref{eq:PqgTRC}). 
We note that, similarly to what we did for $P_{qg}^{\rm NLO}$, we have included in $P_{qg}^{\rm trc}$ a 
factor $1/(1-\ep)$ to account for the different initial state in the $F_{\rm LM}$ structures in the 
left- and right- hand sides of Eq.(\ref{eq:xi2qg}).

    The double-collinear contribution is straightforward to compute. We find 
    \be
\begin{split} 
    &  \langle (I -S_\gamma) \Xi_3^{gq}  F_{\rm LM} ( 1_g, 2_{\bar d}\; ; 4_{\bar u}, 5_\gamma ) \rangle
    =   -\langle (I -S_\gamma) C_{2\gamma} C_{41} \omega^{2\gamma}  F_{\rm LM} ( 1_g, 2_{\bar d} \; ; 4_{\bar u}, 5_\gamma ) \rangle
\\
    & = -\frac{[\alpha][\alpha_s]   Q_d^2 T_R}{\ep^2}
   (2 E_1)^{-2\ep} (2 E_2)^{-2\ep}
    \int \limits_{0}^{1} \; {\rm d} z_1 {\rm d } z_2  P_{qg}^{\rm NLO}(z_1) P_{qq}^{\rm NLO}(z_2)
     \left \langle  \frac{F_{\rm LM} (z_1 \cdot 1_u, z_2 \cdot 2_{\bar d})}{z_1 z_2} \right \rangle.
    \end{split} 
        \ee

        The last contribution is the single-collinear one, proportional to $\Xi_4^{gq}$. It reads
        \be
         \begin{split} 
     \langle (I -S_\gamma) \Xi_4^{gq}  F_{\rm LM} ( 1_g, 2_{\bar d}; 4_{\bar u}, 5_\gamma ) \rangle &  =
           \langle (I-S_\gamma) \left (  C_{2\gamma} \omega^{2\gamma}
+  C_{41} \left ( \theta_C  \omega^{4 \gamma} + \omega^{2\gamma} \right ) \right )F_{\rm LM} ( 1_g,  2_{\bar d}; 4_{\bar u}, 5_{\gamma} )
  \rangle.
           \\
       +  &  \langle (I-S_\gamma) \left ( 
\theta_B C_{4\gamma} \omega^{4 \gamma} + \theta_D C_{4\gamma} \omega^{4\gamma} \right )
  \;  F_{\rm LM} ( 1_g,  2_{\bar d}; 4_{\bar u}, 5_{\gamma} )
  \rangle.
  \label{eq70a}
  \end{split} 
         \ee

         The computation proceeds in full analogy with the QCD case. For completeness, we present
         results for the various contributions to Eq.(\ref{eq70a}). 
                 We start with the term proportional to $C_{2\gamma}$ in Eq.(\ref{eq70a}). It reads
          \be
          \begin{split} 
& \langle (I-S_\gamma)   C_{2\gamma} \omega^{2\gamma} F_{\rm LM} ( 1_g,  2_{\bar d}; 4_{\bar u}, 5_{\gamma} )
\rangle
= - \frac{Q_d^2[\alpha] }{\ep}  (2 E_2)^{-2\ep}
\int\limits_0^1
 {\rm d} z  P_{qq}^{\rm NLO}(z) \left \langle\frac{F_{\rm LM}(1_g, z\cdot 2_{\bar d}; 4_{\bar u})}{z} \right \rangle =
\\
&  - \frac{Q_d^2 [\alpha]}{\ep}  (2 E_2)^{-2\ep}
\int \limits_{0}^{1} {\rm d} z  \; P_{qq}^{\rm NLO}(z,L_2)   \left \langle{\cal O}_{\rm NLO}^{\bar u} \left [
  \frac{F_{\rm LM}(1_g, z \cdot 2_{\bar d}; 4_{\bar u})}{z} \right ]
\right \rangle
\\
& + \frac{[\alpha][\alpha_s] Q_d^2 T_R}{\ep^2} \frac{\Gamma(^2(1-\ep)}{\Gamma(1-2\ep)}
(2 E_1)^{-2\ep}  (2 E_2)^{-2\ep}
 \int \limits_{0}^{1} {\rm d} z_1 \; {\rm d} z_2 \;  P_{qg}^{\rm NLO}(z_1) \;
 P_{qq}^{\rm NLO}(z_2,L_2)  \left \langle  \frac{F_{\rm LM}(z_1\cdot 1_u, z_2 \cdot 2_{\bar d})}{z_1 z_2 } \right \rangle.
 \end{split} 
           \ee

Next, we consider the term proportional to $C_{41}( \omega^{2 \gamma} + \theta_C \omega^{4\gamma})$.
           Since $C_{41} \omega^{2 \gamma} = \rho_{1\gamma}/2$, we find 
           \be
           \begin{split} 
& \langle (I-S_\gamma)   C_{41} \omega^{2\gamma} F_{\rm LM} ( 1_g,  2_{\bar d}; 4_{\bar u}, 5_{\gamma} )
             \rangle
             =
             \\
             &
             -\frac{[\alpha_s]T_R}{\ep}
           ( 2 E_1)^{-2\ep}
             \int \limits_{0}^{1}  {\rm d} z P_{qg}^{\rm NLO}(z) 
              \left \langle (I - S_\gamma)  \left[\frac{\rho_{1\gamma}}{2}
              \frac{ F_{\rm LM}(z\cdot 1_u,2_{\bar d};5_\gamma)}{z}
              \right] \right \rangle =
                      \\
           &
          - \frac{[\alpha_s]T_R}{\ep}
             ( 2 E_1)^{-2\ep}
             \int \limits_0^1{\rm d} z P_{qg}^{\rm NLO}(z) \; 
              \left \langle {\cal O}_{\rm NLO}^{\gamma} \left [ \frac{\rho_{1\gamma}}{2}
             \; \frac{ F_{\rm LM}(z_1 1_u,2_{\bar d};5_\gamma)}{z} \right ]\right \rangle
             \\
        &     +
             \frac{[\alpha_s][\alpha] T_R Q_d^2  }{\ep^2} \frac{\Gamma^2(1-2\ep)}{\Gamma(1-2\ep)}
             ( 2 E_1)^{-2\ep} ( 2 E_2)^{-2\ep}
             \int \limits_{0}^{1} {\rm d} z_1 {\rm d} z_2
             P_{qg}^{\rm NLO}(z_1) P_{qq}^{\rm NLO}(z_2,L_2)\; 
             \;  \left \langle \frac{ F_{\rm LM}(z_1\cdot 1_u, z_2 \cdot 2_{\bar d})}{z} \right \rangle.
           \end{split} 
           \ee

           The second contribution to
$C_{41}( \omega^{2 \gamma} + \theta_C \omega^{4\gamma})$
           is proportional to $C_{41} \omega^{4 \gamma} \theta_C = \theta_C \; \rho_{\gamma 2}/2\;  C_{41}$.
           We obtain 
           \be
           \begin{split} 
           & \langle (I-S_\gamma)   C_{41} \omega^{4 \gamma} \theta_C F_{\rm LM} ( 1_g,  2_{\bar d}\; ; 4_{\bar u}, 5_{\gamma} )
             \rangle =
\\
  &            
             - \frac{[\alpha_s] T_R}{\ep} 
            (2 E_1)^{-2\ep} \int\limits_0^1 {\rm d} z\;
             P^{\rm NLO}_{qg}(z) \;\left\langle 
               (I-S_\gamma)\left[ \frac{\rho_{2 \gamma}}{2} \; \left ( \frac{\rho_{1 \gamma}}{4} \right )^{-\ep}
             \frac{F_{\rm LM}(z\cdot 1_u, 2_{\bar d} \; ; 5_\gamma)}{z} \right]\right\rangle =
             \\
             &  
              - \frac{[\alpha_s] T_R}{\ep}
              (2 E_1)^{-2\ep} \int\limits_0^1 {\rm d} z\;
              P^{\rm NLO}_{qg}(z) \; 
              \left\langle {\cal O}_{\rm NLO}^{\gamma}
              \left [ \frac{\rho_{2 \gamma}}{2} \; \left ( \frac{\rho_{1 \gamma}}{4} \right )^{-\ep}
                \frac{F_{\rm LM}(z\cdot 1_u, 2_{\bar d} \; ; 5_\gamma)}{z} \right ] \right\rangle
              \\
              & 
              +  \frac{[\alpha_s] T_R Q_u^2 2^{\ep} [\alpha]}{2\ep^2}
              \frac{\Gamma(1-2\ep) \Gamma(1-\ep)}{\Gamma(1-3\ep)} 
              (2 E_1)^{-4\ep}
              \int\limits_0^1{\rm d} z \; \left[P_{qq}^{\rm NLO} \otimes P_{qg}^{\rm NLO}\right](z,E_1)
              \; 
              \left \langle 
             \frac{F_{\rm LM}(z \cdot 1_u, 2_{\bar d})}{z} \right \rangle.
                                    \end{split} 
                                    \label{eq:f83}
           \ee
           The convolution in Eq.(\ref{eq:f83}) is defined as
           \be
           \left[P_{qq}^{\rm NLO} \otimes P_{qg}^{\rm NLO}\right](z,E_1)
           = 
      \int\limits_0^1 {\rm d} z_1 {\rm d} z_\gamma \delta(z - z_1 z_\gamma) 
              P^{\rm NLO}_{qg}(z_1) \; z_1^{-2\ep} \; P^{\rm NLO}_{qq}(z_\gamma,L_{1\gamma}),  
           \ee
           where $L_{1\gamma} = \ln \left ( E_1/(E_{\rm max} z_\gamma) \right )$.  The result for this convolution
           is given in Eq.(\ref{eq:PqgNLOoPqqNLO}).

           The last contribution to $\Xi_4^{gq}$ is proportional to $C_{4\gamma}$;  it describes 
            collinear  splitting in the final state; the result can be borrowed from
          existing NNLO QCD computations. We find
           \be
           \begin{split} 
             & \left\langle (I-S_\gamma)   C_{4\gamma} \left ( \theta_B + \theta_D \right )
             \omega^{4 \gamma}  F_{\rm LM} ( 1_g,  2_{\bar d}; 4_{\bar u}, 5_{\gamma} )
             \right\rangle
             =
             -\frac{Q_u^2 [\alpha]}{\ep} \frac{2^{2\ep} \Gamma(1-\ep) \Gamma(1 + 2 \ep)}{\Gamma(1+\ep)}
             \times
             \\
             &
              \bigg \langle {\cal O}_{\rm NLO}^{\bar u}   \left [
               (2 E_4)^{-2\ep} \bar\gamma_{qg}(E_4,E_{\rm max})
               \eta_{41}^{-\ep}(1-\eta_{41})^{\ep}
              F_{\rm LM} ( 1_g,  2_{\bar d}; 4_{\bar u} ) \right ]
              \bigg \rangle
             + \frac{[\alpha] [\alpha_s] T_R 2^{2\ep} \Gamma^2(1-\ep) \Gamma(1+2\ep) \Gamma(1-2\ep)
             }{2 \ep^2 \Gamma(1+\ep)\Gamma(1-3\ep)}\times
             \\
             &             
 (2 E_1)^{-4\ep} Q_u^2
 \int \limits_{0}^{1} {\rm d} z \; P^{\rm NLO}_{qg}(z) (1-z)^{-2\ep}
       \bar\gamma_{qg}\big((1-z)E_1,E_{\rm max}\big)
        \Bigg  \langle
        \frac{F_{\rm LM} ( z\cdot 1_u,  2_{\bar d})}{z}
        \Bigg  \rangle,
           \end{split}
           \ee
           where  $\eta_{ij} = p_i \cdot p_j /(2 E_i E_j) = (1-\cos \theta_{ij})/2$, and
           \be
\bar\gamma_{qg}(E,E_{\rm max} ) =  
-\frac{(2-5\ep+\ep^2)}{2\ep(1-4\ep)} \frac{\Gamma^2(1-2\ep)}{\Gamma(1-4\ep)}
  + \frac{1}{\ep} \left ( \frac{ E_{\rm max}}{E} \right )^{-2\ep}.
  \ee

To complete the computation of mixed QCD-EW corrections to the  $g + \bar d \to W^+ + \bar u$ channel, 
we need to account
for one-loop virtual QED corrections. As for the $q\bar q'$ channel discussed earlier, the result can be 
easily obtained by abelianizing the NNLO QCD result~\cite{Caola:2017dug,Caola:2019nzf}. 
We do not go into further details and  just
quote  the result
           \be
           \begin{split} 
   & 2s\cdot {\rm d}\sigma_{RV,gq} = \langle F^{\rm EW}_{\rm LV}(1_g,2_{\bar d}; 4_u) \rangle
             = \bigg\langle {\cal O}_{\rm NLO}^{\bar u} \left [ F^{\rm EW}_{\rm LV}(1_g,2_{\bar d}; 4_{\bar u})\right ] \bigg\rangle
\\
&              -\frac{[\alpha_s]T_R}{\ep(1 - \ep)}\frac{\Gamma^2(1 - \ep)}{\Gamma(1 - 2 \ep)}
             (2 E_1)^{-2\ep} \int \limits_{0}^{1}  {\rm d} z\;  P_{qg}^{\rm NLO}(z)
                     \left \langle\frac{  F_{\rm LV}^{\rm EW} \left ( z \cdot 1_{u},2_{\bar d} \right ) }{z}
           \right \rangle 
           \\
           &            + \frac{[\alpha_s] [\alpha] T_R Q_u^2 \Gamma(1 - 2\ep)\Gamma(1-\ep)}{ 2\ep \Gamma(1-3\ep) }
           (2 E_1)^{-4\ep} \int\limits_0^1 {\rm d} z \; P_{qg}^{\rm RV}(z) 
	           \left \langle \frac{F_{\rm LM}(z \cdot 1_u,2_{\bar d})}{z}
           \right \rangle.
  \end{split} 
           \ee
           The splitting function  $P_{qg}^{\rm RV}(z)$ is given in Eq.(\ref{eq:PqgRV}), and the various one-loop
           contributions $F_{\rm LV}^{\rm EW}$ are discussed in Appendix~\ref{sectA1}.

           The last ingredient that is required for this channel is the PDFs renormalization. It can be easily obtained
           from the NNLO QCD result~\cite{Caola:2017dug,Caola:2019nzf} with obvious modifications. 
           The result is given by the following
           formula 
           \be
\begin{split}
2s\cdot   {\rm d} \sigma_{{\rm pdf},gq} & = 
     [\alpha(\mu)] [\alpha_s(\mu)]
   \Bigg\{
     - \frac{Q_d^2 T_R}{\ep^2}
     \int \limits_{0}^{1} {\rm d} z_1 {\rm d} z_2 \; \bar P^{{\rm AP},0}_{qg}(z_1) \; \bar P^{{\rm AP}s,0}_{qq}(z_2)
     \frac{F_{\rm LM}(z_1\cdot 1_u , z_2\cdot 2_{\bar d})}{z_1 z_2}
     \\
&  +  \frac{ Q_u^2 T_R}{2 \ep} \int \limits_{0}^{1} {\rm d} z \; \bar P_{qg}^{\rm AP,1}(z) \;
      \frac{F_{\rm LM}(z \cdot 1_u, 2_{\bar d})}{z}      
   - \frac{ Q_u^2 T_R}{2 \ep^2} \int \limits_{0}^{1} {\rm d} z \; \left[\bar P_{qq}^{{\rm AP},0} \otimes 
   \bar P_{qg}^{{\rm AP},0}\right](z) \;
     \frac{F_{\rm LM}(z \cdot p_1, p_2)}{z}
      \Bigg  \}
\\
&
+\frac{2s}{\ep}\bigg\{
       [\alpha(\mu)] Q_d^2 \; \d\sigma_{{\rm NLO},gq}^{\rm QCD} \otimes \bar P_{qq}^{{\rm AP},0}
+[\alpha_s(\mu)] T_R \; \bar P_{qg}^{{\rm AP},0} \otimes \d\sigma^{\rm EW}_{{\rm NLO},q\bar q'}
\bigg\},
\end{split}
\label{eq73} 
\ee
where ``$\otimes$'' stands for the standard convolution product and the various (color-stripped) Altarelli-Parisi
splitting functions $\bar P_{ij}^{{\rm AP},n}$ and their convolutions can be found in~\cite{Caola:2017dug,Caola:2019nzf} for the QCD part and in Appendices~\ref{app:split},\ref{app:AP} for the electroweak part.

\section{Mixed QCD-EW corrections at next-to-next-to-leading order: analytic results for the fully differential calculation
in all partonic channels}
\label{sect6}

To obtain mixed QCD-EW correction to the $u + \bar d \to W^+$ process, we need to combine the two-loop virtual, 
real-virtual, and 
real-real corrections as well as the collinear renormalization contributions for all the different partonic channels.
Each of these contributions is regulated by constructing subtraction terms
as we have explained in the previous sections. 
In this section, we present the finite remainders for all the different partonic channels. 

As we have explained in previous sections, our calculation is performed in a generic reference frame and
with arbitrary $E_{\rm max}$. 
We have explicitly checked that the cancellation of infra-red and ultraviolet poles occurs in an arbitrary reference frame and for generic  $E_{\rm max}$. 
However, for the sake
of simplicity in this section we present results in the center-of-mass frame of the colliding partons and
choose ${E}_{\rm max} = E_1 = E_2$. We will denote the
center-of-mass collision energy as $2 E_c$ so that  $E_1 = E_2 = E_c$. We also set $\mu=M_W$; using 
renormalization-group arguments, it is straightforward to obtain results for different choices of $\mu$. 

For convenience we  summarize the notation that we will use when presenting our results. 
We define
\be
L_c = \ln\lp{2 E_c}/{M_W}\rp,~~~
\beta = \sqrt{1-M_W^2/E_W^2},~~~
\kappa_{iw} = \frac{p_i\cdot p_W}{E_i E_W},
~~~
\eta_{ij} = \frac{\rho_{ij}}{2}=\frac{p_i\cdot p_j}{2E_i E_j} = \frac{1-\cos\theta_{ij}}{2},
~~~
s_{ij} = 2p_i\cdot p_j.
\ee
In the (generalized) splitting functions, we also use
\be
D_i(z) = \left[\frac{\ln^i(1-z)}{1-z}\right]_+.
\ee
We express our results in terms of the $\overline{\rm MS}$-renormalized strong coupling constant
$\alpha_s(\mu)$. We denote by $\alpha_{EW}$ the electromagnetic coupling constant in the $G_\mu$ scheme. 
The LO (color-stripped) Altarelli-Parisi splitting functions without the elastic piece $\bar P_{ij,R}^{{\rm AP},0}$ are
defined in Appendix~\ref{app:AP}. The splitting function describing the NLO finite remainders are defined as 
\be
\begin{split}
&\mathcal P_{qq}^{\rm NLO} = 4 D_1(z) -2(1+z)\ln(1-z)+(1-z)-
\frac{1+z^2}{1-z}\ln z,
\\
& \mathcal P_{qg}^{\rm NLO}(z) = \big[z^2+(1-z)^2\big]\ln\lp\frac{(1-z)^2}{z}\rp + 2z(1-z),
\end{split}
\ee
see Sect.~\ref{sect4}. We also find it convenient to introduce a slight generalization of these
equations
\be
\begin{split}
&
\widetilde{\cal P}_{qq}^{\rm NLO}(z,E_c) =  
4 D_1(z)
-2 (1+z)\ln(1-z)+(1-z)
+2 \ln \left ( \frac{2 E_c}{M_W} \right )\big[2 D_0(z)-(1+z)\big],
\\
&
\widetilde{\cal P}_{qg}^{\rm NLO}(z,E_c) =  
2\big[z^2+(1-z)^2\big]\ln(1-z) + 2z(1-z)
+2 \ln \left ( \frac{2 E_c}{M_W} \right )\big[z^2+(1-z)^2].
\end{split}
\ee
Finally, the one-loop finite remainders $F_{\rm LV}^{\rm fin, EW/QCD}$ and the two-loop remainder $F_{{\rm LVV+LV}^2}^{{\rm fin,QCD}\otimes{\rm EW}}$ are defined in Appendix~\ref{sectA1}. Their analytic and numerical 
expressions are given in Appendix~\ref{4a}.

\subsection{The $q\bar q'$ and $qq'$ channels}

\allowdisplaybreaks

We begin  by presenting formulas for mixed QCD-EW corrections in the $q\bar q'$ channel. As we have
explained in Section~\ref{sect5a}, this channel receives contributions from both $g\gamma$ and $q\bar q$
final states. We then write
\be
\d\sigma_{q\bar q'\to W(X)}^{{\rm QCD}\otimes{\rm EW}} = 
\d\sigma_{q\bar q'\to W(g\gamma)}^{{\rm QCD}\otimes{\rm EW}} +
\d\sigma_{q\bar q'\to W(q\bar q)}^{{\rm QCD}\otimes{\rm EW}},
\ee
where the terms in the brackets indicate the possible double-real contribution.
We consider the two cases separately. For definiteness, we present results for the $u\bar d$ initial state. 

We discuss the $u + \bar d \to W^++(g \gamma)$ case first.
We write the
two-loop contributions to the cross section in the following way 
\be
\d \sigma_{u \bar d\to W(g\gamma)}^{\rm QCD\otimes EW} = \d \sigma_{u \bar d \to W  (g \gamma)}^{\rm elastic}
+ \d \sigma_{u \bar d \to W  (g \gamma)}^{\rm boost}
+ \d \sigma_{u \bar d \to W  (g \gamma)}^{{\cal O}_{\rm NLO}}+ \d \sigma_{u \bar d\to W  (g \gamma)}^{\rm regulated}.
\ee
Below we present results for individual contributions. The elastic contribution reads
\be
\begin{split}
2s\cdot \d\sigma_{u \bar d \to W  (g \gamma)}^{\rm elastic} & =
\lp\frac{\alpha_s(\mu)}{2\pi}\;
\frac{\alpha_{EW}}{2\pi}\rp
C_F
\left[
\frac{8\pi^4}{45}\lp Q_u^2+Q_d^2\rp + \lp \frac{4\pi^2}{3}-\frac{\pi^4}{3}\rp Q_W^2
\right] 
\left\langle F_{\rm LM}(1_u,2_{\bar d} )  \right\rangle 
\\
&
+\lp\frac{\alpha_{EW}}{2\pi}\rp\left[\frac{\pi^2}{3}\lp Q_u^2 + Q_d^2\rp + \lp 2-\frac{\pi^2}{2}\rp Q_W^2\right]
\left\langle F_{\rm LV}^{\rm fin,QCD}(1_u,2_{\bar d}) \right\rangle
\\
&
+\lp\frac{\alpha_s(\mu)}{2\pi}\rp 
\left[C_F \; \frac{2 \pi^2}{3}\right]  \; \left\langle F_{\rm LV}^{\rm fin,EW}(1_u,2_{\bar d}) \right\rangle
+
\left\langle F_{{\rm LVV}+{\rm LV}^2}^{{\rm fin},{\rm QCD}\otimes {\rm EW}} (1_u,2_{\bar d})\right\rangle.
\end{split} 
\ee

The boosted contribution  $\sigma_{u \bar d\to W(g\gamma)}^{\rm boost}$  can be written as
\begin{align} 
& 2s\cdot \d\sigma_{u \bar d \to W(g\gamma)}^{\rm boost} =
\lp \frac{\alpha_{EW}}{2\pi}\;\frac{\alpha_s(\mu)}{2\pi}\rp \Bigg\{
C_F (Q_u^2 + Q_d^2)
  \int \limits_{0}^{1} {\rm d} z_1 {\rm d} z_2
\widetilde{\cal P}_{qq}^{\rm NLO}(z_1,E_c)
\frac{F_{\rm LM}(z_1 \cdot 1_{u}, z_2 \cdot 2_{\bar d} )}{z_1 z_2}
\widetilde{\cal P}_{qq}^{\rm NLO}(z_2,E_c)
\nonumber\\
&
\quad\quad
+
C_F \int \limits_{0}^{1} {\rm d} z \;  \sum \limits_{i=1}^{2}
                           P_{qq}^{\rm NNLO}(Q_i,Q_{j \ne i},z)
                            \left \langle F^{(i)}_{\rm LM}( 1_u,  2_{\bar d} \; | \; z ) \right \rangle                         
\Bigg\}
\\
&+
\int \limits_{0}^{1} {\rm d} z\;  {\cal P}_{qq}^{\rm NLO}(z) 
\sum \limits_{i=1}^{2} \left [ 
\lp\frac{\alpha_{EW}}{2\pi}\rp  
Q_i^2  \left \langle F_{\rm LV}^{(i), \rm fin, QCD}(1_u,2_{\bar d} \; | \; z ) \right \rangle 
+ C_F \lp\frac{\alpha_s(\mu)}{2\pi}\rp \left \langle F_{\rm LV}^{(i), \rm fin, EW}(1_u,2_{\bar d} \; | \; z ) \right \rangle
\right ].
\nonumber
\end{align}
Moreover, we write   
\be
P_{qq}^{\rm NNLO}(Q_u,Q_d,z) = Q_u^2 P^{\rm NNLO,u}_{qq}(z) + Q_u Q_d P^{\rm NNLO,ud}_{qq}(z)  + Q_d^2 P^{\rm NNLO,d}_{qq}(z).
\ee
The individual contributions read
\begin{align}
& P_{qq}^{\rm NNLO,u} = 
16 D_3(z)
+\left(8-\frac{10 \pi ^2}{3}\right) D_1(z)
+32 \zeta_3 D_0(z)
+\frac{5\pi^2}{2}(1-z)+2(8z-9) 
-8(1+z)\ln^3(1-z) 
\nonumber\\
&
+\left[
\lp\frac{15+23z^2}{1-z}\rp\zeta_2 - 7 -z
\right] \ln (z)+
\frac{(2-3z+4z^2+z^3)\ln^2(z)}{(1-z)^2} + 
\frac{(3+z^2)\ln^3(z)}{6(1-z)}
-\ln^2(1-z)
\times
\nonumber\\
&
\Bigg(
4(1-z) + \frac{(17+9z^2)\ln(z)}{1-z}
\Bigg)
+\frac{\ln(1-z)}{1-z}
\Bigg(
\frac{\pi^2(10-8z^2)}{3} + (4-21z+9z^2)+6(1+z^2)\ln^2(z)
\\
&
+
\frac{(2-10z+6z^2-6z^3)\ln(z)}{1-z}
\Bigg)
+
2\Bigg(
\frac{(5+z^2)\ln(z)}{1-z} +
\frac{(3z^2-5)\ln(1-z)}{1-z}-2(1-z)
\Bigg){\rm Li}_2(z)
\nonumber\\
&
+\frac{2(3z^2-5)}{1-z}{\rm Li}_{3}(1-z)
-\frac{4(3z^2+5)}{1-z}{\rm Li}_{3}(z)
+\frac{(4+28 z^2)\zeta_3}{1-z},
\nonumber\\
& P_{qq}^{\rm NNLO,ud} = 
(4\pi^2-16)D_1(z) + \pi^2(1-z)
+\lp\frac{16}{1-z}-2\pi^2(1+z)\rp\ln(1-z)+4z\ln(z)
\nonumber\\
&-\frac{\pi^2(1+z^2)}{1-z}\ln(z)
+\frac{4z(1+z^2)\big[2\ln(1-z)\ln(z)-\ln^2(z)\big]}{(1-z)^2},
\\
& P_{qq}^{\rm NNLO,d} = 
\lp 8 - \frac{2\pi^2}{3}\rp D_1(z) 
-\zeta_2(1-z)
+\lp 2\zeta_2(1+z)-\frac{8}{1-z}\rp\ln(1-z) - 2z\ln(z)
\nonumber\\
&
+\frac{2z(1+z^2)\big(\ln^2(z)-2\ln(z)\ln(1-z)\big)}{(1-z)^2}
+\frac{1+z^2}{1-z}\bigg( \zeta_2\ln(z) + 2\big[2\ln(1-z)-\ln(z)\big]{\rm Li}_2(1-z)\bigg)
\\
&+2(1-z) {\rm Li}_{2}(1-z).
\nonumber
\end{align}

We continue with the contribution  that involves NLO-like
processes,  $u \bar d \to W^+ + g$ and $u \bar d \to W^+ + \gamma$.
It reads
\begin{align}
&2s\cdot \d\sigma_{u\bar d\to W(g\gamma)}^{{\cal O}_{\rm NLO}} = 
\left\langle {\cal O}_{\rm NLO}^{\gamma} \left[F_{\rm LV}^{\rm fin, QCD}(1_u,2_{\bar d};4_\gamma)\right]\right\rangle
+
\left\langle {\cal O}_{\rm NLO}^{g} \left[F_{\rm LV}^{\rm fin,EW}(1_u,2_{\bar d};4_g)\right]\right\rangle
\nonumber\\
&
+ \lp\frac{\alpha_s(\mu)}{2\pi}\rp C_F \left[\frac{2\pi^2}{3}+6 L_c\right]
\big\langle {\cal O}_{\rm NLO}^{\gamma}\left[ F_{\rm LM}(1_u,2_{\bar d};4_\gamma)\right]\big\rangle
+\lp\frac{\alpha_{EW}}{2\pi}\rp\Bigg\{
\lp Q_u^2+  Q_d^2\rp \left[\frac{\pi^2}{3} + 3 L_c \right]
\nonumber\\
&
+Q_W^2 
\Bigg[
2  L_c^2
 -5  L_c - \frac{1}{2} \ln^2\lp\frac{1-\beta}{1+\beta}\rp
-\frac{1}{\beta}\ln\lp\frac{1-\beta}{1+\beta}\rp 
\Bigg]
+ Q_W\sum_{i=1}^{2} Q_i
\bigg[
4  L_c^2
-\frac{3}{2} \ln\lp\frac{2 p_i\cdot p_W}{M_W^2}\rp 
\nonumber\\
&
-4 L_c \ln\lp\frac{2 p_i\cdot p_W}{M_W^2}\rp 
+ \ln^2\lp\frac{2 p_i\cdot p_W}{M_W^2}\rp 
+2{\rm Li}_2\lp 1- \frac{1-\beta}{\kappa_{iW}}\rp
+2{\rm Li}_2\lp 1- \frac{1+\beta}{\kappa_{iW}}\rp
\bigg]
\Bigg\}
\big\langle {\cal O}_{\rm NLO}^{g}\left[ F_{\rm LM}(1_u,2_{\bar d};4_g)\right]\big\rangle
\\
&
+\int\limits_0^1\d z 
\sum_{\substack{i,j = 1\\i\ne j}}^2
\Bigg\{
\lp\frac{\alpha_s(\mu)}{2\pi}\rp
C_F \left\langle {\cal O}_{\rm NLO}^{\gamma}
\left[
\lp \widetilde{\cal P}_{qq}^{\rm NLO}(z,E_c) + \eta_{\gamma j}\ln(\eta_{\gamma i}) \bar P_{qq,R}^{{\rm AP},0}(z)
\rp
 F^{(i)}_{\rm LM}(1_u,2_{\bar d}; 4_\gamma \;|\; z)\right]\right\rangle
\nonumber\\
&+
\lp\frac{\alpha_{EW}}{2\pi}\rp
Q_i^2
 \left\langle {\cal O}_{\rm NLO}^{g}
\left[
\lp \widetilde{\cal P}_{qq}^{\rm NLO}(z,E_c) + \eta_{g j}\ln(\eta_{g i}) \bar P_{qq,R}^{{\rm AP},0}(z)
\rp
 F^{(i)}_{\rm LM}(1_u,2_{\bar d}; 4_g \;|\; z)\right]\right\rangle
\Bigg\}.
\nonumber
\end{align}

The fully-regulated contribution  has already been discussed. It reads
\be
2s\cdot \d\sigma_{u \bar d \to W(g\gamma) }^{\rm regulated} =
\langle (I - S_g) (I-S_\gamma) \Xi_1^{q\bar q} F_{\rm LM}(1_u,2_{\bar d}; 4_g, 5_\gamma ) \rangle, 
\ee
where the operator $\Xi_1$ is given in Eq.(\ref{eq31}). We compute it numerically. 

We now discuss the $u \bar d \to W^+ + (q\bar q)$ final state. The corresponding double-real matrix element
is only singular if the final-state $q\bar q$ pair is collinear to the initial-state $u$ or $\bar d$. We use the
same phase-space parametrization as for the $u \bar d \to W^+ + (g\gamma)$ case, and write
\be
\d\sigma_{u\bar d\to W(q\bar q)}^{{\rm QCD}\otimes{\rm EW}} = 
\d\sigma^{\rm boost}_{u\bar d \to W(q\bar q)} + 
\d\sigma^{\rm regulated}_{u\bar d \to W(q\bar q)}.
\ee
We write the boosted contribution as
\be
\begin{split}
2s\cdot \d\sigma^{\rm boost}_{u\bar d \to W(q\bar q)}  = 
\lp\frac{\alpha_s(\mu)}{2\pi}\;\frac{\alpha_{EW}}{2\pi}\rp
\int\limits_0^1 \d z \; P_{qq}^{{\rm NNLO},{\rm int}}(z)
\sum_{i=1}^2 C_F Q_i^2 \big\langle F_{\rm LM}^{(i)}(1_u,2_{\bar d}\; | \; z)\big\rangle,
\end{split}
\ee
with
\be
\begin{split}
&
P_{qq}^{{\rm NNLO},{\rm int}}(z) = 
\frac{2\pi^2}{3} (1+z) + 
\frac{2+12 z - 14 z^2 - (5-12z+4z^2)\ln^2(z)-8 \ln(1-z)(8-15z+7z^2)}{2(1-z)}
\\
&
-\frac{
4 \ln(1-z)\ln(z)(5-2z^2)+\ln(z)(6+11z-27z^2) - 8 \ln(z)\ln(1+z)(1-z^2)
+2(13-6z-z^2){\rm Li_2}(1-z)
}{1-z}
\\
&
+ 8 (1+z){\rm Li}_2(-z)
+\frac{1+z^2}{1-z}
\bigg\{
12 {\rm Li}_3(1-z) + 16 {\rm Li}_3(-z)+18{\rm Li}_{3}(z)-6 \zeta_3
-8 {\rm Li}_2(-z)\ln(z) + 2\big[5\ln(z)
\\
&
-4\ln(1-z)\big]{\rm Li}_{2}(1-z)
+5 \ln^2(z)\ln(1-z) - \frac{\ln^3(z)}{3}  - \frac{7\pi^2}{3}\ln(z)
\bigg\}.
\end{split}
\ee
The fully-regulated contribution reads
\be
2s\cdot\d\sigma^{\rm regulated}_{u\bar d\to W(q\bar q)} = 
\big\langle \lp I-C_{145}-C_{245}\rp
F_{\rm LM}(1_u,2_{\bar d};4_q,5_{\bar q})
\big\rangle.
\ee

Finally, we discuss the $q q'$ channel. At $\mathcal O(\alpha_s\alpha_{EW})$, it receives contributions
from interferences among two $t$-channel diagrams (see Fig.~\ref{fig:mixed_int}, right) with two identical quarks 
in the final state. As for the $u+\bar d \to W(q\bar q)$ case that we have just discussed, $qq'$ channels also only
have 
triple-collinear singularities. For definiteness, we present the results for the $u+d\to W(qq)$ channel. 
We employ the same phase-space parametrization as for the $q\bar q'$ channel, and write
\be
\d\sigma_{ud\to W(qq)}^{{\rm QCD}\otimes{\rm EW}} = 
\d\sigma^{\rm boost}_{ud\to W(qq)}+\d\sigma^{\rm regulated}_{ud\to W(qq)}.
\ee
The boosted contribution reads 
\be
2s\cdot \d\sigma^{\rm boost}_{u d \to W(q q)}  = 
\lp\frac{\alpha_s(\mu)}{2\pi}\;\frac{\alpha_{EW}}{2\pi}\rp
\int\limits_0^1 \d z \; C_F P_{q\bar q}^{{\rm NNLO},{\rm int}}(z)
\left\langle
\frac{ 
Q_d^2 F_{\rm LM}(1_u,z\cdot 2_{\bar d})}{z}\right\rangle,
\ee
\be
\begin{split}
P_{q\bar q}^{\rm NNLO,int}(z) &= -\pi^2(1+z)
+15(1-z)+4\ln^2(z)+16(1-z)\ln(1-z)+8(1+z)\ln(z)\ln(1-z)
\\
&
+
(11+19z)\ln(z)-12(1+z)\ln(z)\ln(1+z)
+4(3+z){\rm Li}_2(1-z)-12(1+z){\rm Li}_{2}(-z)
\\
&
+\frac{1+z^2}{1+z}
\Bigg\{
\frac{\ln^3(z)}{3}+6\ln^2(z)\ln(1+z)-12\ln(z)\ln^2(1+z)+4\ln^3(1+z)
\\
&
-\frac{2\pi^2}{3}\big[2\ln(1-z)-4\ln(z)+3\ln(1+z)\big]-4\ln(1-z)\ln(z)\big[\ln(z)+4\ln(1+z)\big]
\\
&
-4\ln(z){\rm Li}_2(1-z)+4\big[3\ln(z)-4\ln(1-z)\big]{\rm Li}_2(-z)+
16 {\rm Li}_3(1-z)-16 {\rm Li}_3(z) - 36{\rm Li}_{3}(-z)
\\
&
-24 {\rm Li}_{3}\lp\frac{z}{1+z}\rp - 8 {\rm Li}_{3}(1-z^2)+10\zeta_3
\Bigg\}.
\end{split}
\ee

The fully-regulated contribution is
\be
2s\cdot \d\sigma^{\rm regulated}_{ud\to W(qq)} = 
\big\langle \lp I-C_{245}\rp F_{\rm LM}(1_u,2_d;4_q,5_q)\big\rangle.
\ee


\subsection{The $qg$ channel}

For definiteness, we consider corrections to the $g+\bar d\to W^+ + \bar u$ annihilation process and write
\be
   {\rm d} \sigma_{g \bar d \to W (\bar u \gamma)}^{{\rm QCD}\otimes{\rm EW}} = {\rm d} \sigma_{g \bar d  \to W(\bar u \gamma) }^{\rm boost} +
   {\rm d} \sigma_{g \bar d \to W (\bar u \gamma)  }^{{\cal O}_{\rm NLO}} +{\rm d} \sigma_{g \bar d \to W(\bar u \gamma) }^{\rm regulated}.
\ee

The boosted contribution reads
\be
\begin{split}
& 2s\cdot \d\sigma_{g\bar d\to W(\bar u \gamma)}^{\rm boost} = 
\lp\frac{\alpha_s(\mu)}{2\pi}\;\frac{\alpha_{EW}}{2\pi}\rp
\Bigg\{
T_R Q_d^2 \int\limits_0^1\d z_1 \d z_2 
\widetilde{\cal P}_{qg}^{\rm NLO}(z_1,E_c)
\left\langle
\frac{F_{\rm LM}(z_1\cdot 1_u,z_2\cdot 2_{\bar d})}{z_1 z_2}
\right\rangle
\widetilde{\cal P}_{qq}^{\rm NLO}(z_2,E_c)
\\
&
+ T_R\int\limits_0^1 \d z\; P_{qg}^{\rm NNLO}(z) 
\left\langle
\frac{F_{\rm LM}(z\cdot 1_u,2_{\bar d})}{z}
\right\rangle
\Bigg\} + 
\lp\frac{\alpha_s(\mu)}{2\pi}\rp
T_R \int\limits_0^1\d z \;{\cal P}_{qg}^{\rm NLO}(z)
\left\langle
\frac{F_{\rm LV}^{\rm fin,EW}(z\cdot 1_u,2_{\bar d})}{z}
\right\rangle,
\end{split}
\ee
where
\be
P_{qg}^{\rm NNLO}(z) = Q_u^2 P_{qg}^{\rm NNLO,u}(z) + 
Q_d^2 P_{qg}^{\rm NNLO,d}(z) + 
Q_u Q_d P_{qg}^{\rm NNLO,ud}(z),
\ee
and
\begin{align}
&P_{qg}^{\rm NNLO,u}(z) = 
-\frac{69}{4}+\frac{255 z}{4}-49 z^2 + \frac{\pi^2(11-54z+46z^2)}{12}
+ \frac{11}{6}(1-2z+2z^2)\ln^3(1-z) 
\nonumber\\
&
+\frac{(3-6z+4z^2)\ln^3(z)}{12}
-\frac{(5-81z+140z^2-80z^3)\ln^2(z)}{8(1-z)}
-\bigg[
7-\frac{117z}{4}+27z^2-(11-22z+26z^2)\zeta_2
\bigg]\ln(z)
\nonumber\\
&
-\left[
7-21z+17z^2+
\lp\frac{3}{2}-3z-z^2\rp \ln(z)
\right]\ln^2(1-z)
+\bigg[
16-\frac{109}{2}z + 44 z^2 - \zeta_2(7-14z+22z^2)
\\
&
+\frac{(7-29z+40z^2-22z^3)\ln(z)}{1-z} - 
\lp\frac{3}{2}-3z+3z^2\rp \ln^2(z)\bigg]\ln(1-z)
+(18-36z+32z^2)\zeta_3
\nonumber\\
&
+\bigg[3+4z-4z^2 
- (1-2z-6z^2)\ln(1-z)+(1-2z-2z^2)\ln(z)\bigg]
{\rm Li}_{2}(z)
-(9-18z+10z^2){\rm Li}_3(1-z) 
\nonumber\\
&
-(9-18z+14z^2){\rm Li}_3(1-z),
\nonumber\\
&P_{qg}^{\rm NNLO,d}(z) = 
\big[z^2+(1-z)^2\big]
\bigg[
\zeta_2 \ln(z) - \frac{\pi^2}{3}\ln(1-z) + \frac{2z\big[\ln(z)-2\ln(1-z)\big]\ln(z)}{1-z}
\\
&
+2\big[2\ln(1-z)-\ln(z)\big]{\rm Li}_2(1-z)
\bigg]
+ z(1-z)\lp 4 {\rm Li}_2(1-z) - \frac{\pi^2}{3}\rp - 4z^2\ln(z),
\nonumber
\\
& P_{qg}^{\rm NNLO,ud}(z) = 
\lp\frac{4z\ln(z)}{1-z}+\pi^2\rp
\bigg[
(1-2z+2z^2)\big[2\ln(1-z)-\ln(z)\big]+2z(1-z)
\bigg].
\end{align}

The contribution involving NLO kinematics is given by
\begin{align}
& 2s\cdot \d\sigma^{{\cal O}_{\rm NLO}}_{g\bar d \to W(\bar u \gamma)} = 
\big\langle \mathcal O^{\bar u}_{\rm NLO} \big[ F_{\rm LV}^{\rm fin,EW}(1_g,2_{\bar d}; 4_{\bar u})
\big]\big\rangle
+\lp\frac{\alpha_{EW}}{2\pi}\rp \Bigg\{
\sum_{i\in\{2,4\}} Q_i^2\left[
2 L_c^2 - \zeta_2 - 2 \ln\lp\frac{E_c}{E_i}\rp
\ln\lp \frac{4 E_i E_c}{M_W^2}\rp
\right]
\nonumber\\
&
+ Q_u^2 \left[
\frac{13}{2}-\frac{2\pi^2}{3} - \left[\frac{3}{2}+2\ln\lp\frac{E_c}{E_4}\rp\right]\ln\lp\frac{\eta_{41}}{4(1-\eta_{41})}\rp
-3 \ln\lp\frac{2 E_4}{M_W}\rp
\right]
+ Q_W^2 \bigg[
\frac{1}{\beta}\ln\lp\frac{1+\beta}{1-\beta}\rp-\frac{1}{2}\ln^2\lp\frac{1+\beta}{1-\beta}\rp
\nonumber\\
&-2 L_c\bigg]
+ Q_u Q_d \left[
4 L_c \ln(\eta_{42}) + \ln^2(\eta_{42}) + 3 \ln\lp\frac{s_{24}}{M_W^2}\rp
-\ln^2\lp\frac{s_{24}}{M_W^2}\rp + 2 {\rm Li}_{2}(1-\eta_{42})
\right]
+\sum_{i\in\{2,4\}} Q_W Q_i
\label{ONLOqg}
\\
&
\times\bigg[
 4 L_c \ln\lp\frac{E_i M_W}{p_i\cdot p_W}\rp
 -\frac{3}{2}\ln\lp\frac{2 p_i\cdot p_W}{M_W^2}\rp
 +\ln^2\lp\frac{2 p_i\cdot p_W}{M_W^2}\rp
 +2 {\rm Li}_2\lp 1- \frac{1+\beta}{\kappa_{iW}}\rp
  +2 {\rm Li}_2\lp 1- \frac{1-\beta}{\kappa_{iW}}\rp
\bigg]
\nonumber\\
&
+\pi^2 Q_u Q_W 
\Bigg\} \big\langle {\cal O}^{\bar u}_{\rm NLO}\big[ F_{\rm LM}(1_g,2_{\bar d}; 4_{\bar u})\big]\big\rangle
+\lp\frac{\alpha_{EW}}{2\pi}\rp \int\limits_0^1\d z\;\widetilde {\cal P}_{qq}^{\rm NLO}(z)\; Q_d^2
\left\langle {\cal O}_{\rm NLO}^{\bar u} \left[\frac{F_{\rm LM}(1_g,z\cdot 2_{\bar d}; 4_{\bar u})}{z}
\right]\right\rangle
\nonumber\\
&
+\lp\frac{\alpha_s(\mu)}{2\pi}\rp 
\int\limits_0^1\d z\; T_R 
\left\langle {\cal O}^{\gamma}_{\rm NLO}
\left[
\left(\widetilde {\cal P}^{\rm NLO}_{qg}(z,E_c) + 
\eta_{42}\ln\lp\frac{\eta_{41}}{2}\rp \bar P^{{\rm AP},0}_{qg}(z)\right)
\frac{F_{\rm LM}(z\cdot 1_{u},2_{\bar d}; 4_\gamma)}{z}\right]\right\rangle,
\nonumber
\end{align}
with $Q_2 = Q_d$ and $Q_4 = -Q_u$.

  The fully-regulated gluon-quark contribution reads
  \be
2s\cdot   {\rm d}\sigma_{g\bar d \to W(\bar u \gamma)}^{\rm regulated}  = 
 \langle (I -S_\gamma) \Xi_1^{gq}  F_{\rm LM} ( 1_g, 2_{\bar d}; 4_{\bar u}, 5_\gamma ) \rangle,
\ee
where $\Xi_{1}^{gq}$ is defined in Eq.(\ref{eq94}). We compute it numerically.

\subsection{The $q\gamma$ channel}
The structure of the quark-photon channel is similar to the one of the quark-gluon channel. Actually, results in this
case are more compact because of the simplicity of soft-gluon limits.
For definiteness, we consider corrections to the $\gamma + \bar d \to W^+ + \bar u$ channel. We use the
same phase-space parametrization as for the quark-gluon channel and write
\be
   {\rm d} \sigma_{ \gamma \bar d \to W (\bar u g) }^{{\rm QCD}\otimes{\rm EW}} =
   {\rm d} \sigma_{ \gamma \bar d \to W (\bar u g) }^{\rm boost} +
   {\rm d} \sigma_{\gamma \bar d \to W (\bar u g) }^{{\cal O}_{\rm NLO}} +{\rm d} \sigma_{\gamma \bar d \to W (\bar u g)}^{\rm regulated}.
\ee

The boosted contribution reads
\be
\begin{split}
&
2s\cdot \d\sigma_{\gamma \bar d\to W(\bar u g)}^{\rm boost} = 
\lp\frac{\alpha_s(\mu)}{2\pi}\; \frac{\alpha_{EW}}{2\pi}\rp
N_c C_F Q_u^2\Bigg\{
\int\limits_0^1\d z_1 \d z_2 
\widetilde{\cal P}_{qg}^{\rm NLO}(z_1,E_c)
\left\langle
\frac{F_{\rm LM}(z_1\cdot 1_u,z_2\cdot 2_{\bar d})}{z_1 z_2}
\right\rangle
\widetilde{\cal P}_{qq}^{\rm NLO}(z_2,E_c)
\\
&+\int\limits_0^1\d z\; P_{q\gamma}^{\rm NNLO}(z)
\left\langle
\frac{F_{\rm LM}(z\cdot 1_u,2_{\bar d})}{z}
\right\rangle
\Bigg\}
+\lp\frac{\alpha_{EW}}{2\pi}\rp N_c Q_u^2
\int\limits_0^1\d z\;{\cal P}_{qg}^{\rm NLO}(z)
\left\langle
\frac{F_{\rm LV}^{\rm fin,QCD}(z\cdot 1_u,2_{\bar d})}{z}
\right\rangle,
\end{split}
\ee
where
\be
P_{q\gamma}^{\rm boost}(z) = P_{qg}^{\rm boost,u}(z) + P_{qg}^{\rm boost,d}(z) + P_{qg}^{\rm boost,ud}(z).
\ee

The $\d\sigma^{{\cal O}_{\rm NLO}}_{\gamma \bar d \to W(\bar u g)}$ term can be
obtained from the analogous result for the $qg$ channel Eq.(\ref{ONLOqg}) using the following replacements
\be
\begin{gathered}
\big. T_R \to N_c Q_u^2, ~~~ \{Q_u^2,Q_d^2,Q_u Q_d\} \to C_F,~~~ Q_W \to 0,
\\
\big. F_{\rm LM}(1_g,...) \to F_{\rm LM}(1_\gamma,...),
~~~
F_{\rm LM}(z\cdot 1_u,2_{\bar d};4_\gamma) \to F_{\rm LM}(z\cdot 1_u,2_{\bar d};4_g),
\\
\big.
\alpha_s(\mu)\leftrightarrow \alpha_{EW},
~~~
{\cal O}_{\rm NLO}^{\gamma} \to {\cal O}_{\rm NLO}^{g},
~~~
F_{\rm LV}^{\rm fin,EW}(1_g,2_{\bar d};4_{\bar u}) \to 
F_{\rm LV}^{\rm fin,QCD}(1_\gamma,2_{\bar d};4_{\bar u}).
\end{gathered}
\ee

The regulated contributions reads
  \be
2s\cdot   {\rm d}\sigma_{\gamma \bar d \to W (\bar u g)}^{\rm regulated}  = 
 \big\langle (I -S_g) \Xi_1^{\gamma q}  F_{\rm LM} ( 1_\gamma, 2_{\bar d}; 4_{\bar u}, 5_g ) \big\rangle,
\ee
where we define $\Xi_1^{\gamma q}$ in analogy to what we did for the $qg$ channel:\footnote{Similar to
what we did for $qg$ channel, all double-collinear operators in
$\Xi_{1}^{gq}$ also act on the unresolved phase space, while the triple-collinear operators do not. See Ref.~\cite{Caola:2019nzf} for details.}
\be
\begin{split}
  &  \Xi_1^{\gamma q} = (I-C_{2 g} ) (I - C_{41}) \omega^{2 g} + 
  \theta_C (I -C_{14 g} ) (I - C_{41}) \omega^{4 g} 
  + \theta_B (I - C_{14g} ) ( I- C_{4g}) \omega^{4 g}
  \\
&    + \theta_A ( I - C_{14g} ) \omega^{4 g}
 + \theta_D ( I - C_{14g}) (I - C_{4 g} ) \omega^{4 g},
\end{split}
\ee
with 
\be
\omega^{2 g} = \frac{\rho_{4g}}{ \rho_{2g} + \rho_{4 g} },
\;\;\;
\omega^{4g} = \frac{\rho_{2 g}}{ \rho_{2 g} + \rho_{4g} },
\ee
see Section~\ref{sect7a}.

\subsection{The $g\gamma$ channel}
This channel can be obtained straightforwardly by abelianizing the NNLO QCD $gg$ channel. Following Refs.~\cite{Caola:2017dug,Caola:2019nzf}, we do not order the final state partons either in energy
or in angle and we do not introduce any partitioning. 

For definiteness, we consider the partonic process $g+\gamma \to W + ({\bar u}+d)$ and write
\be
\d\sigma_{g\gamma \to W(\bar u d)}^{{\rm QC}\otimes{\rm EW}} = 
\d\sigma_{g\gamma \to W(\bar u d)}^{\rm boost} + 
\d\sigma_{g\gamma \to W(\bar u d)}^{{\cal O}_{\rm NLO}} + 
\d\sigma_{g\gamma \to W(\bar u d)}^{\rm regulated}.
\ee

The boosted contribution reads
\be
\begin{split}
2s\cdot \d\sigma_{g\gamma \to W(\bar u d)}^{\rm boost}  = &
\lp\frac{\alpha_s(\mu)}{2\pi}\; \frac{\alpha_{EW}}{2\pi}\rp
N_c T_R
\int\limits_0^1\d z_1 \d z_2 \; \widetilde{\cal P}_{qg}^{\rm NLO}(z_1,E_c)
\; \widetilde{\cal P}_{qg}^{\rm NLO}(z_2,E_c)\times
\\
&
\left\langle
\frac{Q_d^2 F_{\rm LM}(z_1\cdot 1_u,z_2\cdot 2_{\bar d})
+Q_u^2 F_{\rm LM}(z_1\cdot 1_{\bar u},z_2\cdot 2_{d})
}{z_1 z_2}
\right\rangle.
\end{split}
\ee

The term with NLO-like kinematics reads
\be
\begin{split}
\d\sigma_{g\gamma \to W(\bar u d)}^{{\cal O}_{\rm NLO}}  = 
\lp\frac{\alpha_s(\mu)}{2\pi}\rp
\int\limits_0^1 \d z \; T_R\widetilde{\cal P}_{qg}^{\rm NLO}(z)
\left\langle
{\cal O}_{\rm NLO}^{q_4}\left[
\frac{F_{\rm LM}(z\cdot 1_{u},2_\gamma;4_d)+
F_{\rm LM}(z\cdot 1_{\bar d},2_\gamma;4_{\bar u})}{z}\right]
\right\rangle
\\
+\lp\frac{\alpha_{EW}}{2\pi}\rp
\int\limits_0^1 \d z \; N_c\widetilde{\cal P}_{qg}^{\rm NLO}(z)
\left\langle
{\cal O}_{\rm NLO}^{q_4}\left[
\frac{Q_d^2 F_{\rm LM}(g,z\cdot 2_{\bar d}\;;4_{\bar u})+
Q_u^2 F_{\rm LM}(g,z\cdot 2_{u};4_{d})}{z}\right]
\right\rangle.
\end{split}
\ee

Finally, the regulated contribution reads
\be
\d\sigma_{g\gamma \to W(\bar u d)}^{\rm regulated} = 
\big\langle
\big(I-C_{41}-C_{42}-C_{51}-C_{52}+C_{42}C_{51}+C_{41}C_{52}\big)
F_{\rm LM}(1_g,2_\gamma;4_{\bar u},5_d)
\big\rangle.
\ee
In this case, the collinear operators always act on the unresolved phase space, see Refs.~\cite{Caola:2017dug,Caola:2019nzf} for
details. 


\begin{table}[t]
  \begin{tabular}{||c|r|c|c|c||}
    \hline \hline
 $\sigma[\mathrm{pb}]$ & channel & $\mu=M_W$ & $\mu=M_W/2$ & $\mu=M_W/4$ \\ \hline\hline
 $\sigma_{\rm LO}$ & & 6007.6 & 5195.0 & 4325.9  \\ \hline
 \hline
 $\Delta \sigma_{\rm NLO, \alpha_s}$ & all ch. & 508.8 & 1137.0 & 1782.2  \\ \hline
  & $q\bar{q}'$ & 1455.2 & 1126.7 & 839.2 \\ 
  & $qg/gq$ & -946.4 & 10.3 & 943.0 \\ \hline\hline
  $\Delta \sigma_{\rm NLO, \alpha}$ & all ch. & 2.1 & -1.0 & -2.6  \\ \hline
  & $q\bar{q}'$ & -2.2 & -5.2 & -6.7 \\ 
  & $q\gamma/\gamma q$ & 4.2 & 4.2 & 4.04 \\ \hline\hline
  $\Delta \sigma_{\rm NNLO, \alpha_s\alpha}$ & all ch. & -2.4 & -2.3 & -2.8  \\ \hline
  & $q\bar q'/qq'$ & -1.0 & -1.2 & -1.0 \\ 
  & $qg/gq$ & -1.4 & -1.2 & -2.1 \\ 
  & $q\gamma/\gamma q$ & 0.06 & 0.03 & -0.04 \\ 
  & $g\gamma/\gamma g$ & -0.12 & 0.04 & 0.30 \\ \hline \hline
  \end{tabular}

  \caption{Fiducial cross sections for $pp \to W^+(e^+ \nu_e)$ at the 13~{\rm TeV} LHC for three different values of the renormalization
    and factorization scales at different orders of perturbation theory. Contributions of different partonic
    channels are displayed separately. See text for details.}
    \label{table1}
\end{table}

\section{Numerical results}
\label{sect7}

We have implemented the above results for all the relevant partonic channels in a \texttt{Fortran} computer code that enables the computation
of mixed QCD-electroweak corrections to the  production of an on-shell $W^\pm$ boson in proton collisions at 
a fully-differential level.  Tree-level decays of the $W$ boson are included in the computation.
Note that  in this paper
we do not consider mixed corrections that originate from QCD corrections to $W$ production followed by electroweak corrections to
$W$ decay.  Such corrections are, essentially, of NLO-type and, for this reason,  are  much easier to deal
with; in fact, they have already been studied  in Ref.~\cite{ditt2}.\footnote{Similarly, we do not consider mixed QCD-EW corrections to the decay process. These are also very simple since they only come from the renormalization of the $W\to l\nu$ form factor.} 

We note that all the finite remainders
of one-loop electroweak and QCD corrections that we require are computed with  \texttt{OpenLoops} \cite{r56,r57,r58}. The calculation of the two-loop finite remainder of the mixed QCD-EW corrections to the $Wq\bar q'$ form factor is
presented in Appendix~\ref{4a}.

Before presenting selected results for the mixed QCD-electroweak corrections, we describe the various checks of the calculation
that we have performed to ensure its 
correctness.  First, we checked all  fully-resolved contributions by using our code to compute
cross sections and kinematic distributions for the process $pp \to W + \gamma + {\rm jet}$
and comparing the results with  \texttt{MADGRAPH}~\cite{Frederix:2018nkq} 
and \texttt{MCFM}~\cite{mcfm}.  Such a  comparison has been 
performed separately for  all the different partonic channels that contribute to the above process allowing for a thorough 
check of our code.

Second, we have used our code to compute
NLO QCD and NLO electroweak corrections to the processes $pp \to W+\gamma$ and $pp \to W+ {\rm jet}$ and checked the
results of the calculation against 
\texttt{MCFM} and \texttt{MADGRAPH}, respectively. In both cases excellent agreement for
these NLO contributions was found. 

Finally, we have  checked some  unresolved contributions by considering the limit of
equal up and down quark charges   $Q_u = Q_d$ and comparing the results with our earlier computation of
mixed QCD-electroweak corrections to $Z$ production in proton collisions \cite{Buccioni:2020cfi}.
This check is particularly useful  since,
compared to the case of $Z$ production, we have modified the   parametrization of the phase space and the
partitions for the  computation
reported in this paper.

We now turn to the presentation of numerical results. We renormalize weak corrections in the $G_\mu$ scheme and
use, as input parameters, $G_F = 1.16639 \times 10^{-5}~{\rm GeV}^{-2}$, $M_Z = 91.1876~{\rm GeV}$,
$M_W = 80.398~{\rm GeV}$, $M_t = 173.2~{\rm GeV}$ and $M_H = 125~{\rm GeV}$.  We also use
$\Gamma_W = 2.1054~{\rm GeV}$. The fine-structure
constant that is obtained with this setup is $\alpha_{EW} = 1/132.338$.  We use the NNLO \texttt{NNPDF3.1luxQED} parton distribution
functions~\cite{Bertone:2017bme,Manohar:2016nzj,Manohar:2017eqh} for {\it all} numerical computations reported in this paper.  The value of the strong coupling constant is provided
as part of the PDF set.  Numerically, it reads $\alpha_s(M_Z) = 0.118$. 

\begin{figure}[t]
\centering
\includegraphics[angle=0,width=0.49\textwidth]{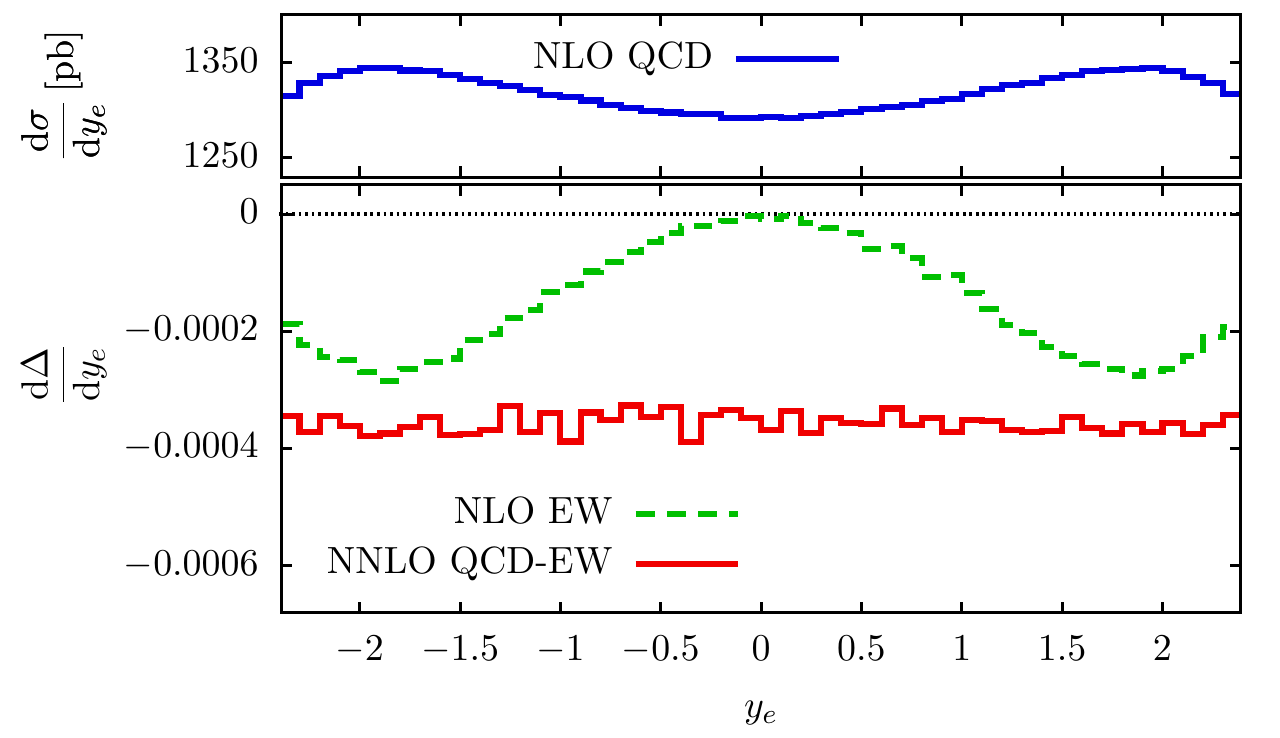}~~~~
\includegraphics[angle=0,width=0.49\textwidth]{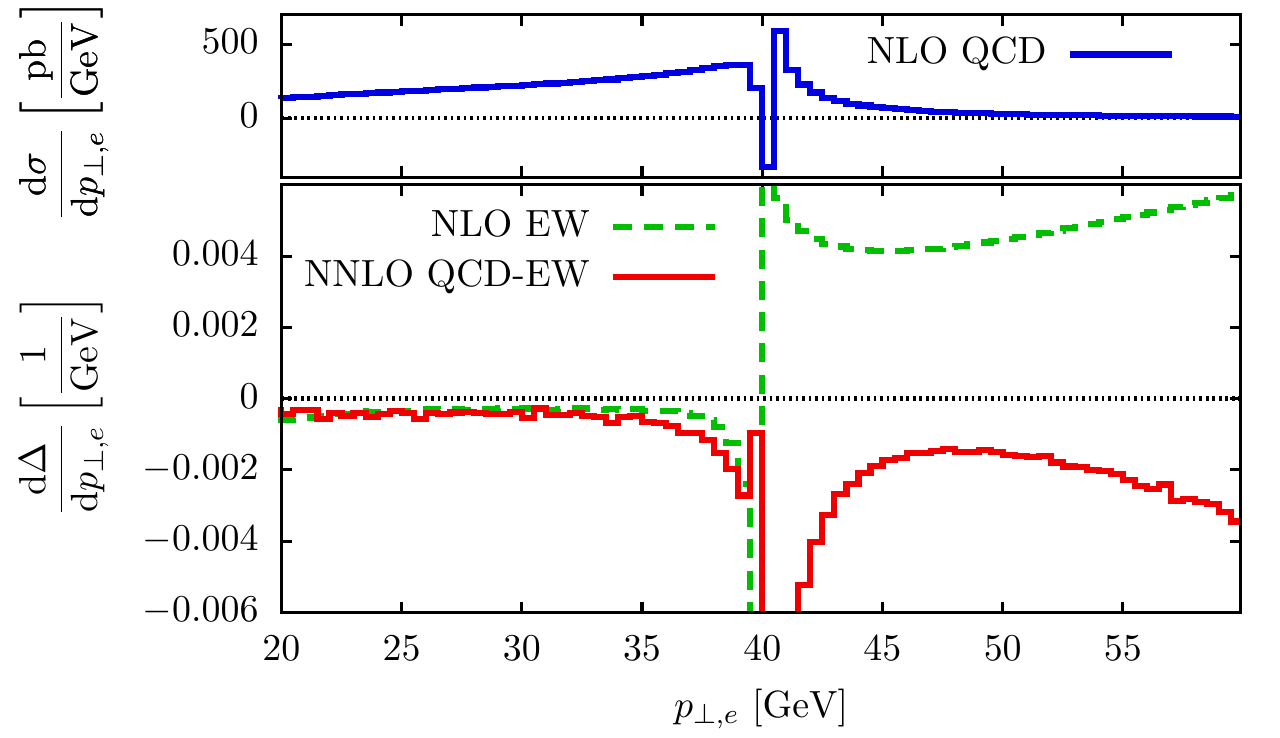} \\
\includegraphics[angle=0,width=0.49\textwidth]{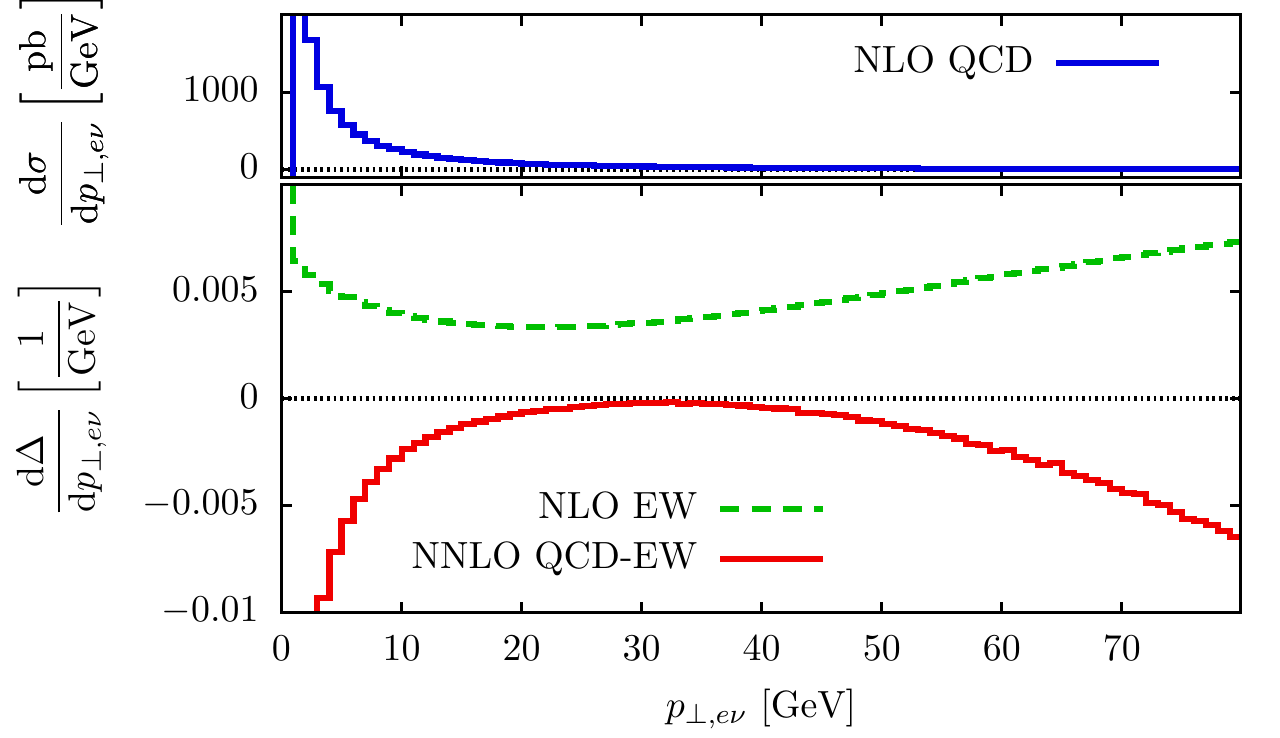}~~~~ 
\includegraphics[angle=0,width=0.49\textwidth]{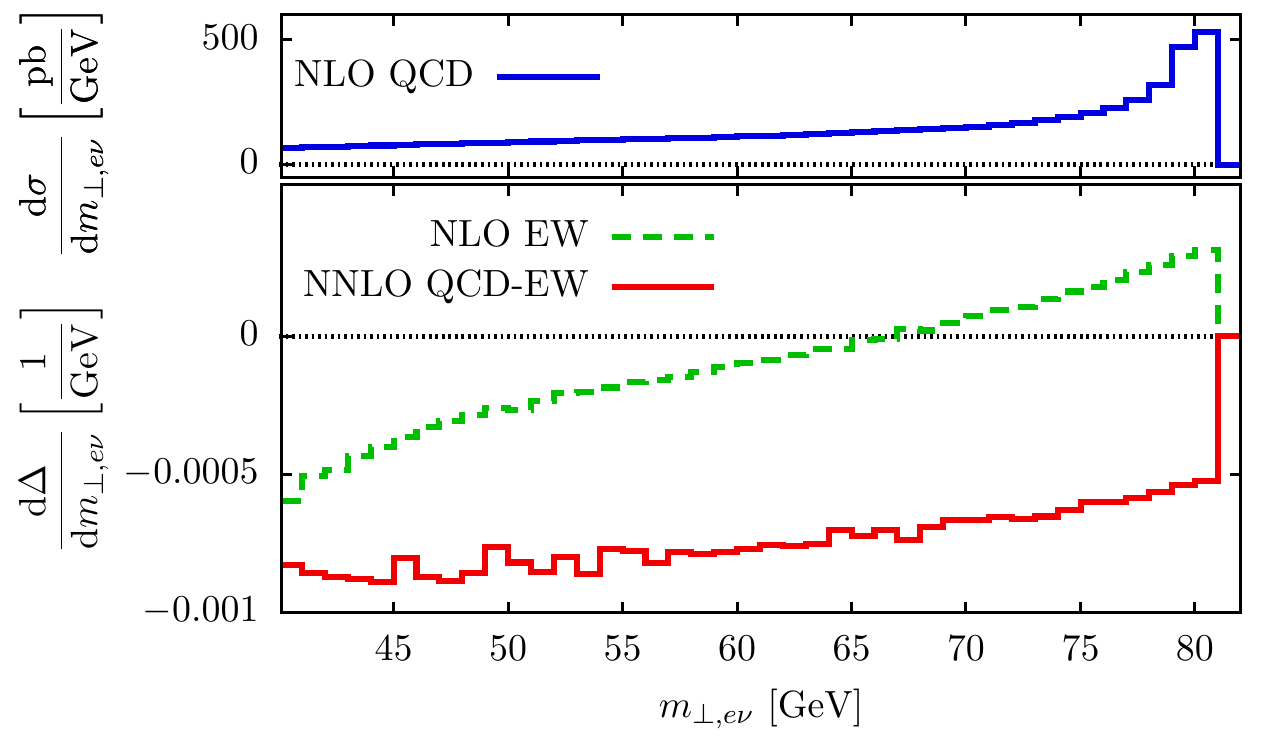} 
\caption{The impact of mixed QCD-electroweak  corrections to $pp \to W^+(e^+ \nu)$ production
  at $13~{\rm TeV}$ LHC on various kinematic distributions including
  lepton rapidity and transverse momentum,  the transverse momentum of the $W$-boson and the transverse mass. 
NLO electroweak corrections are also shown. See text for details.}
\label{fig:rr1}
\end{figure}

Since we do not aim at performing extensive  phenomenological studies in this paper, we apply very mild cuts
on the final state of the process $pp \to W^+(\bar e \nu) + X$. We require that the transverse momentum
of the positron $p_{\perp,e}$ and of the neutrino $p_{\perp,\rm miss}$ are larger than $15~{\rm GeV}$ and that
the absolute value of the positron rapidity does not exceed  $|y_e| < 2.4$.  We also set the
factorization and renormalization scales to be equal $\mu_R = \mu_F = \mu$ and
choose $\mu = M_W/2$ as the central scale for our computations.

To present the results, we write the fiducial cross section as
\be
\sigma_{pp \to W^+}  = \sigma_{\rm LO} + \Delta \sigma_{\rm NLO, \alpha_s}  + \Delta \sigma_{\rm NLO, \alpha}, 
+ \Delta \sigma_{\rm NNLO, \alpha \alpha_s} + ....
\label{eq111aaa}
\ee
where the first term on the right hand side is the leading order cross section, the second term is the NLO QCD contribution,
the third term is the NLO electroweak contribution and the last one is the mixed QCD-electroweak contribution. Ellipses in
Eq.(\ref{eq111aaa}) stand for other contributions to the cross section, e.g. NNLO QCD ones. 

We show the fiducial cross sections
$pp \to W+X$, using the cuts described above, in Table~\ref{table1}.
It follows from this table that
NLO electroweak contributions are tiny -- they
modify the leading order cross section by just about $-0.02$ percent. For comparison, we note that
NNLO QCD corrections
are of the order of a few percent. We note that the smallness of these corrections
is partially related to our choice of the $G_\mu$ renormalization scheme which appears to reduce the impact
of electroweak corrections significantly.  Although quite small as well,
mixed QCD-electroweak corrections turn out to be {\it larger}  than the NLO electroweak ones, at least for the setup
considered here. 

The relative  importance of mixed QCD-electroweak corrections, at least compared to NLO electroweak corrections,
is also apparent from the
  kinematic distributions  shown 
in Fig.~\ref{fig:rr1}. These distributions are computed with the fiducial cuts described above; results shown in Fig.~\ref{fig:rr1}
are obtained for $\mu = M_W/2$.  The $y$-axes in the lower panes correspond to bin-by-bin ratios of NLO electroweak and mixed QCD-electroweak contributions
to NLO QCD cross sections
\be
{\rm d} \Delta_i = \frac{{\rm d} \Delta\sigma_i }{ {\rm d} \sigma_{\rm LO} + {\rm d} \Delta\sigma_{\rm NLO,\alpha_s}}.
\ee
In Fig.~\ref{fig:rr1} we show the rapidity and transverse momentum distributions
of the charged lepton as well as  the transverse mass\footnote{We define the transverse mass as $m_{\perp,l\nu}=\sqrt{2 p_{\perp,l}\cdot p_{\perp,{\rm miss}}(1-\cos\Delta\phi_{l\nu})}$.} and the
transverse momentum distributions of the $W$ boson. 
It follows from Fig.~\ref{fig:rr1} that mixed QCD-electroweak corrections are often larger  than NLO electroweak
ones and that the two types of corrections often have different shapes.  It remains to be seen how these small effects
impact the extraction of the $W$-boson mass  from LHC data; we will investigate  this important question in a separate publication. 

\section{Conclusions}
\label{sect8}

A better understanding of  mixed QCD-electroweak  corrections to $W$-boson production in hadron collisions is  
important for the precision electroweak physics program at the LHC.
The calculation of these corrections is complicated by the fact that
they require two- and one-loop  virtual corrections with several
internal and external masses as well as control on infra-red and collinear singularities that
appear when photons and partons are radiated.

However, thanks to recent progress in developing subtraction
schemes for QCD computations at the LHC and in  technology for multi-loop computations, the
calculation  of mixed QCD-electroweak
corrections to on-shell vector-boson production becomes relatively straightforward.  To demonstrate this, in this paper we 
have presented   results for the two-loop QCD-EW corrections to the $q \bar q' \to W$ interaction vertex
and explained  how to construct 
a suitable  subtraction scheme for real-emission contributions.
We provided  relatively simple analytic results for fully- and partially-unresolved integrated subtraction terms
as well as analytic formulas and a numerical value
for the   two-loop form factor that describes mixed QCD-electroweak contributions to the $q \bar q' \to W$ on-shell interaction vertex.

We have implemented our calculation
in a flexible parton-level numerical code and used it to calculate  mixed QCD-electroweak corrections
to $pp \to W^+(e^+ \nu)$ at the LHC. We presented results for fiducial cross sections and selected kinematic distributions.
In the setup that we have considered, we have found that, in general, mixed QCD-electroweak corrections are rather small, often below a permille. However they appear
to be larger than one-loop electroweak corrections to $pp \to W^+$ at least when the latter are  computed in  the $G_\mu$ scheme. 

The calculation reported in this paper provides one of the last missing theoretical ingredients
whose understanding is considered to be essential for achieving  few MeV accuracy in  
the $W$-boson
mass measurement at the LHC. Needless to say that the actual impact of these  corrections on the $W$-mass measurement
is  unknown; we plan to study this question in a future publication.

{\bf Acknowledgment} This research is partially supported by BMBF grant
05H18VKCC1 and by the Deutsche Forschungsgemeinschaft (DFG, German Research Foundation) under grant
396021762 - TRR 257. The research of F.B. and F.C. was
partially supported by the ERC Starting Grant 804394 {\sc HipQCD}.

\appendix

\section{Infra-red structure of loop corrections}
\label{sectA1}

The computation of mixed QCD-EW corrections to $u  \bar d \to W^+$
requires   virtual corrections to a number of processes 
including {\it i}) two-loop mixed QCD-EW corrections to $ u \bar d \to W^+$,  {\it ii}) one-loop QCD corrections
to $ u \bar d \to W^+$,  {\it iii}) one-loop electroweak corrections to $u \bar d \to W^+$, 
{\it iv}) one-loop QCD corrections to $u \bar d \to W^+ + \gamma$, {\it v}) one-loop electroweak corrections
to $u \bar d \to W^+ + g$ as well as crossings of these processes.  
To demonstrate the cancellation of $1/\ep$ poles and identify the $\ep\to 0$ limit of the integrated subtraction terms, 
we need 
to isolate infra-red divergent contributions to these amplitudes. In case of QCD corrections, this can be accomplished
with the help of Catani's  formula \cite{Catani:1998bh}. In this appendix, we use the results of Ref.~\cite{Catani:1998bh} to explicitly extract the infra-red part of renormalized QCD amplitudes, and generalize them to deal with the electroweak case as well.

We use the following notation. We write a generic (renormalized) amplitude as
\be
\mathcal A = \mathcal A_0 + \lp\frac{\alpha_s(\mu)}{2\pi}\rp \mathcal A_1 + 
\lp\frac{\alpha_{EW}}{2\pi}\rp\bar{\mathcal A}_1 + 
\lp\frac{\alpha_s(\mu)}{2\pi}\; \frac{\alpha_{EW}}{2\pi}\rp \mathcal A_{\rm mix} + ...
\label{eq:A1}
\ee
and then define
\be
\begin{split}
F_{\rm LV}^{\rm QCD}(1,2;...) &= \mathcal N
\lp\frac{\alpha_s(\mu)}{2\pi}\rp\int \sum_{\rm col,pol} 2\Re\big[\mathcal A_0 \mathcal A_1^*\big]
(2\pi)^d\delta_d(P_i-P_f)\frac{\d^{d-1}p_W}{(2\pi)^{d-1}2E_W},
\\
F_{\rm LV}^{\rm EW}(1,2;...) &= \mathcal N
\lp\frac{\alpha_{EW}}{2\pi}\rp\int \sum_{\rm col,pol} 2\Re\big[\mathcal A_0 \bar{\mathcal A^*_1}\big]
(2\pi)^d\delta_d(P_i-P_f)\frac{\d^{d-1}p_W}{(2\pi)^{d-1}2E_W},
\\
F_{{\rm LVV}+{\rm LV}^2}^{{\rm QCD}\otimes\rm{EW}}(1,2) &= \mathcal N
\lp\frac{\alpha_s(\mu)}{2\pi}\;\frac{\alpha_{EW}}{2\pi}\rp
\int \sum_{\rm col,pol} 2\Re\big[\mathcal A_0 \mathcal A_2^*+\mathcal A_1\bar{\mathcal A_1^*}\big]
(2\pi)^d\delta_d(P_i-P_f)\frac{\d^{d-1}p_W}{(2\pi)^{d-1}2E_W},
\end{split}
\ee
where $P_{i(f)}$ stands for the sum of initial(final) momenta and $\mathcal N$ stands for all the required
($d$-dimensional) initial-state color and helicity averaging factors, see Eq.(\ref{eq:1}) and Refs.~\cite{Caola:2017dug,Caola:2019nzf}. In this appendix, we also use the notation 
\be
\bar \alpha = \left[(4\pi)^{\ep} e^{-\ep \gamma_E}\right]\alpha_{EW}.
\ee

Formulas provided in this appendix are used  in the main text to construct subtraction terms for
mixed QCD-EW corrections and to demonstrate 
cancellation of $1/\ep$ singularities analytically.

\subsection{The infra-red structure of real-virtual amplitudes}
We begin with  QCD corrections to the electroweak processes  
$u + \bar d \to W+\gamma$. The infra-red and collinear structure of the one-loop amplitude
directly follows from Catani's formula~\cite{Catani:1998bh}. We write 
\be
\begin{split}  
  \left\langle F^{\rm QCD}_{\rm LV}(1_u,2_{\bar d}; 4_\gamma ) \right\rangle  &=
  \lp\frac{\alpha_s(\mu)}{2\pi}\rp \left[\frac{e^{\ep\gamma_E}}{\Gamma(1-\ep)}\right]
   \lp\frac{\mu^2}{s_{12}}\rp^{\ep}
   \left[-2 C_F \cos ( \pi \ep )
  \left ( \frac{1}{\ep^2} + \frac{3}{2\ep} \right ) \right]\langle F_{\rm LM}(1_u,2_{\bar d};4_\gamma ) \rangle 
  \\&
   + \left\langle F_{\rm LV}^{\rm fin, QCD}(1_u,2_{\bar d} ; 4_\gamma ) \right\rangle,
\label{eq1a}
\end{split} 
\ee
where $s_{12} = 2 p_1\cdot p_2$.  
The required formula for QCD corrections to the photon-quark collision process  reads 
\be
\begin{split} 
  \left\langle F^{\rm QCD}_{\rm LV}(1_\gamma,2_{\bar d}\; ; 4_{\bar u} ) \right\rangle  &= 
  \lp\frac{\alpha_s(\mu)}{2\pi}\rp\left[\frac{e^{\ep \gamma_E}}{\Gamma(1-\ep)}\right]
  \lp\frac{\mu^2}{s_{24}}\rp^{\ep}
  \left[ -2 C_F 
  \left ( \frac{1}{\ep^2} + \frac{3}{2\ep} \right )\right] \langle F_{\rm LM}(1_\gamma,2_{\bar d}\; ; 4_{\bar u} ) \rangle 
  \\&
   + \left\langle F_{\rm LV}^{\rm fin,QCD}(1_\gamma,2_{\bar d} \; ; 4_{\bar u} ) \right\rangle,
\label{eq1b}
\end{split} 
\ee
with $s_{24} = 2 p_2\cdot p_4$. 

We also  require one-loop electroweak corrections to the process $u + \bar d \to W^++g$.
We parametrize them in the following
way 
\be
\begin{split} 
  \left\langle F^{\rm EW}_{\rm LV}(1_u,2_{\bar d};4_g ) \right\rangle &= 
  \lp\frac{\bar\alpha}{2\pi}\rp\left[\frac{e^{\ep \gamma_E}}{\Gamma(1-\ep)}\right]
  \lp\frac{\mu^2}{M_W^2}\rp^{\ep}
\big[ -Q_u Q_d f_1 - Q_u Q_W f_2 + Q_d Q_W f_3\big]
\langle F_{\rm LM}(1_u,2_{\bar d};4_g ) \rangle 
\\&
  + \left\langle F_{\rm LV}^{\rm fin,EW}(1_u,2_{\bar d};4_g ) \right\rangle,
\end{split} 
\label{eq:A6}
\ee
where $Q_W = Q_u - Q_d$ and 
\be
\begin{split} 
&   f_1 = \frac{2}{\ep^2} + \frac{3-2 L_s}{\ep} -3 L_s + L_s^2 - \pi^2,\\
& f_2 = \frac{1}{\ep^2} + \frac{5/2-2L_t}{\ep}-\frac{3}{2} L_t + L_t^2, \\
& f_3 = \frac{1}{\ep^2} + \frac{5/2-2L_u}{\ep}-\frac{3}{2} L_u + L_u^2,
\end{split} 
\ee
with $L_s = \ln(s/M_W^2)$, $L_t = \ln(1 - t/M_W^2)$, $L_u = \ln(1-u/M_W^2)$. The
Mandelstam invariants are defined as 
\be
s = (p_1+p_2)^2,\;\;\;\; t = (p_1 - p_W)^2,\;\;\;\; u = (p_2 - p_W)^2.
\ee

A related quantity is  the one-loop electroweak corrections to the gluon-initiated
process $g + \bar d \to W^+ + \bar u$. We   parametrize it  in the following
way 
\be
\begin{split} 
  \left\langle F^{\rm EW}_{\rm LV}(1_g,2_{\bar d};4_{\bar u} ) \right\rangle &= 
  \lp\frac{\bar\alpha}{2\pi}\rp\left[\frac{e^{\ep\gamma_E}}{\Gamma(1-\ep)}\right]
  \lp\frac{\mu^2}{M_W^2}\rp^{\ep}
\big[ -Q_u Q_d f^{gq}_1 - Q_u Q_W f^{gq}_2 + Q_d Q_W f^{gq}_3\big]
\langle F_{\rm LM}(1_g,2_{\bar d};4_{\bar u} ) \rangle 
\\&
  + \left\langle F_{\rm LV}^{\rm fin, EW}(1_g,2_{\bar d},4_{\bar u} ) \right\rangle,
\end{split} 
\ee
where $Q_W = Q_u - Q_d$ and 
\be
\begin{split} 
  &   f^{gq}_1 = \frac{2}{\ep^2}
  + \frac{3-2 L_{24}} {\ep} -3 L_{24} + L_{24}^2,\\
& f^{gq}_2 = \frac{1}{\ep^2} + \frac{5/2-2L_{4W}}{\ep}-\frac{3}{2} L_{4W} + L_{4W}^2-\pi^2, \\
  & f_3^{gq} = \frac{1}{\ep^2} + \frac{5/2-2L_{2W}}{\ep}-\frac{3}{2} L_{2W} + L_{2W}^2,
  \label{eq7a}
\end{split} 
\ee
with $L_{24} = \ln(2p_2\cdot p_4/M_W^2)$, $L_{4W} = \ln(2 p_4 \cdot p_W /M_W^2)$,
$L_{2W} = \ln(2 p_2 \cdot p_W/M_W^2)$.

In all the formulas above, the infra-red $1/\ep$ poles are explicitly extracted and $F_{\rm LV}^{\rm fin, QCD/EW}$
are finite remainders. 

\subsection{Infra-red structure of the $Wq\bar q'$ form factor}

The only two-loop amplitude that we require describes  mixed QCD-EW corrections to the $q + \bar q' \to W^+$ process. 
For definiteness, we present results for the $u+\bar d\to W^+$ channel. 

At one-loop, we parametrize QCD corrections as
\be
\begin{split}
&
\left\langle F_{\rm LV}^{\rm QCD}(1_u,2_{\bar d})\right\rangle = 
\lp\frac{\alpha_s(\mu)}{2\pi}\rp I_{12,{\rm QCD}} \left\langle  F_{\rm LM}(1_u,2_{\bar d})\right\rangle + 
\left\langle F_{\rm LM}^{\rm fin, QCD}(1_u,2_{\bar d})\right\rangle,
\\&
\left\langle F_{\rm LV}^{\rm EW}(1_u,2_{\bar d})\right\rangle = 
\lp\frac{\bar\alpha}{2\pi}\rp I_{12,{\rm EW}} \left\langle  F_{\rm LM}(1_u,2_{\bar d})\right\rangle + 
\left\langle F_{\rm LM}^{\rm fin,EW}(1_u,2_{\bar d})\right\rangle,
\end{split}
\ee
with
\be
\begin{split}
&I_{12,\rm QCD} = \left[\frac{e^{\ep \gamma_E}}{\Gamma(1-\ep)}\right]
\lp\frac{\mu^2}{M_W^2}\rp^{\ep}
\left[-2C_F\cos(\pi\ep)\lp\frac{1}{\ep^2}+\frac{3}{2\ep}\rp\right],
\\
&
I_{12,\rm EW} = \left[\frac{e^{\ep \gamma_E}}{\Gamma(1-\ep)}\right]
\lp\frac{\mu^2}{M_W^2}\rp^{\ep}
\big[-Q_u Q_d \bar f_1 - Q_u Q_W \bar f_2 + Q_d Q_W \bar f_3\big],
\end{split}
\ee
and
\be
\bar f_1 = \cos(\pi\ep)\left[\frac{2}{\ep^2}+\frac{3}{\ep}\right],
~~~\bar f_2 = \bar f_3 = \frac{1}{\ep^2}+\frac{5}{2\ep}.
\ee
We note that these formulas agree with the $s\to M_W^2$, $t,u\to 0$ limit of Eqs.(\ref{eq1a},\ref{eq:A6}).

We now discuss the infra-red structure of two-loop mixed QCD-EW corrections. As we have explained in the main text,
IR singularities in this case almost factorize into the product of two NLO-like structures. The only exceptions
are genuinely-NNLO hard triple-collinear configurations. As a consequence, we can write 
\be
\begin{split}
\left\langle F_{{\rm LVV}+{\rm LV}^2}^{{\rm QCD}\otimes{\rm EW}}(1_u,2_{\bar d})\right\rangle &= 
\lp\frac{\alpha_s(\mu)}{2\pi}\; \frac{\bar\alpha}{2\pi}\rp
\left[
I_{12,\rm QCD}\cdot I_{12,\rm EW} + \frac{e^{\ep\gamma_E}}{\Gamma(1-\ep)}
\frac{H^W_{{\rm QCD}\otimes{\rm EW}}}{\ep}\right]\left\langle F_{LM}(1_u,2_{\bar d})\right\rangle
\\
&
+\lp\frac{\alpha_s(\mu)}{2\pi}\rp I_{12,\rm QCD}\left\langle F_{\rm LV}^{\rm fin, EW}(1_u,2_{\bar d })\right\rangle
+\lp\frac{\bar\alpha}{2\pi}\rp I_{12,\rm EW}\left\langle F_{\rm LV}^{\rm fin, QCD}(1_u,2_{\bar d})\right\rangle
\\
&
+\left\langle F_{{\rm LVV}+{\rm LV}^2}^{{\rm fin, QCD}\otimes{\rm EW}}(1_u,2_{\bar d})\right\rangle.
\end{split}
\label{eq:A15}
\ee
In Eq.(\ref{eq:A15}), $F_{{\rm LVV}+{\rm LV}^2}^{{\rm fin, QCD}\otimes{\rm EW}}$ is the two-loop finite remainder.
The constant $H^W_{{\rm QCD}\otimes{\rm EW}}$ is related to the quark anomalous dimension and can
be extracted by abelianizing the corresponding contribution in Ref.~\cite{Catani:1998bh}. It reads
\be
H^{W}_{{\rm QCD}\otimes{\rm EW}} = \left ( \frac{\pi^2}{2} - 6 \zeta_3 - \frac{3}{8} \right )
C_F\big[Q_u^2+Q_d^2\big].
\label{eq:HwMix}
\ee

We present explicit formulas for the one- and two-loop finite remainders in the next appendix.

\section{Analytic expression for the mixed QCD-EW form factor}
\label{4a}

The double-virtual corrections to single on-shell $W$-boson production
require the form factor for the $q\bar q' \to W$ vertex at $\mathcal{O}(\alpha_s
\alpha_{EW})$. The on-shell condition simplifies the problem significantly;
in particular, we do not need complicated two-loop four-point functions
\cite{Bonciani:2016ypc,vonManteuffel:2017myy,Heller:2019gkq,%
Hasan:2020vwn} required to describe the process $pp \to l \nu$ with
${\cal O}(\alpha_s\alpha_{EW})$ accuracy in the off-shell case.
Moreover, if one assumes equal masses for internal $W$ and $Z$ bosons,
all necessary integrals are available in the literature and can be
extracted from Refs.~\cite{Aglietti:2003yc,Aglietti:2004tq,%
Bonciani:2016ypc,vonManteuffel:2017myy,Heller:2019gkq,Hasan:2020vwn}.
However, to the best of our knowledge, results for on-shell $W$
form factor that accommodate different masses of $W$ and $Z$ bosons are
not publicly available. We compute the relevant form factor in this
paper.

\begin{figure}[t]
  \centering
    \includegraphics[width=0.3\textwidth]{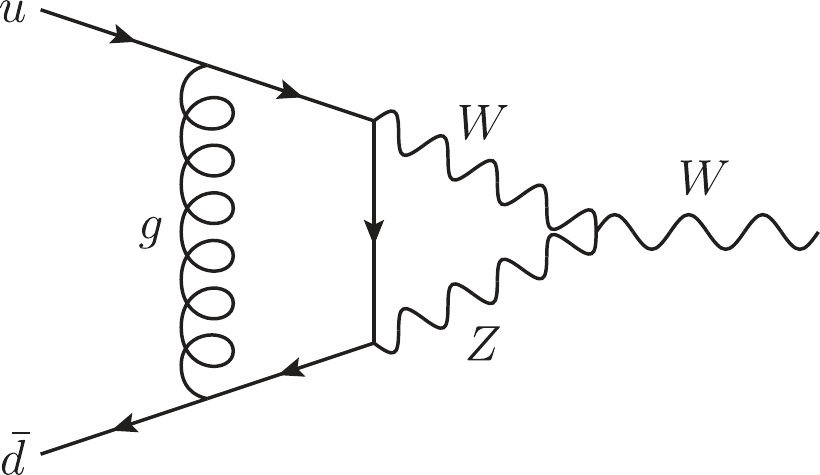}%
  \hspace*{2em}
    \includegraphics[width=0.3\textwidth]{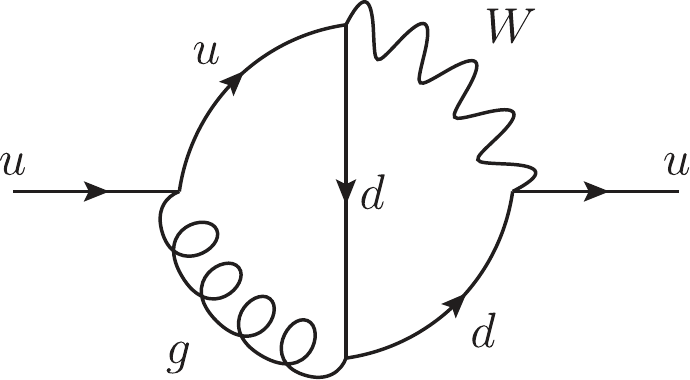}
  \caption{Examples for mixed QCD-electroweak two-loop diagrams.
           A form factor diagram with simultaneous internal $W$- and
           $Z$-boson propagators is shown on the left.
           A self-energy diagram which contributes to the wave function
           renormalization of the external quarks at $\mathcal{O}(
           \alpha_s\alpha_{EW})$ is shown on the right.}
  \label{fig:VV-diagrams}
\end{figure}

An example of a diagram that has to be computed is shown in
Fig.~\ref{fig:VV-diagrams}.
In order to calculate the form factor, we use 
\texttt{QGRAF} \cite{Nogueira:1991ex} to generate diagrams,
\texttt{FORM} \cite{Vermaseren:2000nd,Kuipers:2012rf,Kuipers:2013pba,%
Ruijl:2017dtg} to perform the Dirac and Lorentz algebra,
\texttt{color.h} \cite{vanRitbergen:1998pn} for the color algebra and
\texttt{Reduze2} \cite{vonManteuffel:2012np,Bauer:2000cp,Lewis:Fermat}
to reduce integrals that appear to master integrals using
integration-by-parts relations \cite{Chetyrkin:1980pr,Chetyrkin:1981qh,%
Tkachov:1981wb}. 
We work in the Feynman gauge and use Feynman rules from Ref.~\cite{Bohm:2001yx}.
Since we only require contributions of massless quarks and work at 
$\mathcal{O}(\alpha_s\alpha_{EW})$, the Dirac matrix $\gamma_5$ can
only appear on fermion lines that are connected to external lines. For
this reason, we consider $\gamma_5$ to be anti-commuting.

The form factor has to be renormalized in order to remove ultra-violet
divergences.
We choose to follow the procedure described in Ref.~\cite{Denner:1991kt}
and renormalize the wave functions and masses in the on-shell scheme.
We use the $\overline{\text{MS}}$ scheme to renormalize the strong
coupling constant $\alpha_s$ and the $G_\mu$ scheme\footnote{See Ref.~\cite{Denner:2019vbn} for a recent review.} for the electroweak input parameters. 
The weak mixing angle is defined as
$\cos \theta_W = M_W/M_Z$ in terms of the on-shell $W$ and $Z$ boson
masses.
The necessary renormalization constants at the one-loop order are given
explicitly in Ref.~\cite{Denner:1991kt}. The two-loop mixed QCD-electroweak
corrections to the self-energies of electroweak gauge bosons were
calculated in Ref.~\cite{Djouadi:1993ss}. 
In addition, we need the two-loop self-energies for massless fermions
which enters through the wave function renormalization of external
quarks.
A typical diagram that appears in this context is the self-energy
diagram shown in Fig.~\ref{fig:VV-diagrams}.
The required wave function renormalization has already been calculated
in Ref.~\cite{Buccioni:2020cfi}; it reads
\begin{align}
  Z_{f,L}
    ={}& 1
         -\left(\frac{\alpha}{2 \pi}
                \frac{(4\pi)^{\ep}}{\mu^{\ep}}\right)
          \frac{1 - \ep}{(2 - \ep) \ep} \Gamma(1 + \ep)
          \left(g_{f,L}^2 \left(\frac{M_Z^2}{\mu^2}\right)^{-\ep}
                +\frac{1}{2 \sin^2(\theta_W)}
                 \left(\frac{M_W^2}{\mu^2}\right)^{-\ep}
          \right)
  \notag \\ &
         +\left(\frac{\alpha}{2 \pi}
                \frac{\alpha_{s}}{2 \pi}
                \frac{(4\pi)^{2\ep}}{\mu^{2\ep}}
          \right) C_F
          \frac{(3 - 2 \ep) (1 - 3 \ep)}{4 \ep (2 - \ep) (1 - 2 \ep)}
          \Gamma(1 - \ep) \Gamma(1 + \ep) \Gamma(1 + 2 \ep)
\label{eq:B1}\\ & \hphantom{+~} 
        \times  \left(g_{f,L}^2 \left(\frac{M_Z^2}{\mu^2}\right)^{-2 \ep}
                +\frac{1}{2 \sin^2(\theta_W)}
                 \left(\frac{M_W^2}{\mu^2}\right)^{-2 \ep}
          \right)
  \,.\notag
\end{align}
The bare strong and electromagnetic coupling constants are
$\alpha_{s}$ and $\alpha$, respectively, and the subscript $f \in
\{u,d\}$ denotes the type of fermion. The $Z$-boson coupling is
defined as
\begin{align}
  g_{f,L} &= \frac{I_{3,f} - Q_f \sin^2(\theta_W)}{%
            \sin(\theta_W) \cos(\theta_W)},
\end{align}
where $I_{3,f}=\pm1/2$ and $Q_f$ are the third component of the weak isospin
and the electric charge of the fermion $f$.
We have checked  the renormalization constants by rederiving them using
the same set of programs as described above. We have also checked the
renormalization constants related to vector bosons by comparing
numerically against results of Ref.~\cite{ditt2}.

We note that the one-loop renormalization constants also enter the
two-loop renormalization where they are multiplied by infra-red
divergent quantities. Therefore, one would a priori also need
higher-order terms beyond $\mathcal{O}(\ep^0)$ for these renormalization
constants. However, once one-loop squared and genuine two-loop
contributions are combined, the higher-order terms cancel out and so
there is no need to compute them.

To check the correctness of our result, we performed two independent
calculations and found agreement. We have also checked that the $1/\ep$
infra-red poles of the renormalized form-factor agree with the general
structure discussed in the previous appendix. 

We now discuss some details of the calculation. 
After IBP reduction, we find ten master integrals with two different internal masses.
We compute them using  differential equations
\cite{Kotikov:1990kg,Caffo:1998du,Gehrmann:1999as}.
The integration constants are fixed by matching to the known results in
the equal mass limit which we take from 
Refs.~\cite{Aglietti:2003yc,Aglietti:2004tq,Bonciani:2016ypc}.%
\footnote{Partially, Loopedia \cite{Bogner:2017xhp} was used to identify
references for these integrals.}
To verify the computed master integrals,  we have numerically checked our results using
\texttt{pySecDec} \cite{Borowka:2017idc,Borowka:2018goh,%
Vermaseren:2000nd,Kuipers:2012rf,Kuipers:2013pba,Ruijl:2017dtg,%
Hahn:2004fe,Hahn:2014fua,GSL}.

To write down the differential equations that the master integrals
satisfy, we find it convenient to rationalize the square root present
in the alphabet by introducing the standard Landau variable $y$ defined as follows 
\begin{align}
  \frac{M_Z^2}{M_W^2} &= \frac{(1+y)^2}{y}.
\end{align}
When written in this variable, the differential equation 
for the vector of master integrals $I$
can be written as
\begin{align}
  \partial_y \vec{I} 
    &= \sum\limits_{a \in S}^{} \sum_{k=0}^5
       \frac{\hat A_{a,k}}{(y - a)^k} \vec{I},
  \label{eq31aaa}
\end{align}
where 
\begin{align}
  S = \left\{
    0,
    \pm 1,
    \pm i,
    e^{\pm \frac{2 i \pi}{3}},
    -\varphi^2,
    -\varphi^{-2}
  \right\}
  \,,
\end{align}
and  $\varphi = \frac{1 + \sqrt{5}}{2}$ is the golden ratio.

The differential equation Eq.(\ref{eq31aaa}) is solved in terms of
Goncharov Polylogarithms (GPLs).
Manipulations of GPLs were done using different tools, including
\texttt{HarmonicSums} \cite{Ablinger:2010kw,Ablinger:2013hcp,%
Vermaseren:1998uu,Remiddi:1999ew,Blumlein:2009ta,Ablinger:2011te,%
Ablinger:2013cf,Ablinger:2014rba,Ablinger:2016ll,Ablinger:2017rad,%
Ablinger:2019mkx} and \texttt{PolyLogTools}
\cite{Maitre:2005uu,Maitre:2007kp,Duhr:2019tlz}.
We have simplified GPLs that appear in the calculation using relations
from Refs.~\cite{Henn:2015sem,Ablinger:2011te}.
In the solutions of the master integrals the letters  $a = \pm i$ no
longer appear. Although the result expressed in terms of the $y$ variable
is straightforward to evaluate, its analytic form is somewhat unwieldy. 
Because of this, we decided to present our analytic result expressed
in terms of
\be
z = \frac{M_W^2}{M_Z^2} + i0.
\ee
Results in terms of $y$ can be obtained from the authors upon request. 

We express our results in terms of iterated integrals defined as
\begin{align}
  H_{\underbrace{0,\dots,0}_{k\text{ times}}}(z)
    &= \frac{1}{k!} \ln^k(z)
  \,, &
  H_a(z) &= \int_0^z \mathrm{d}x \, f_a(x)
  \,, &
  H_{a_1,\dots,a_n}(z)
    &= \int_0^z \mathrm{d}x \, f_{a_1}(x) H_{a_2,\dots,a_m}(x)
\end{align}
with the alphabet
\begin{align}
  f_0(x) &= \frac{1}{x}
  \,, &
  f_1(x) &= \frac{1}{1-x}
  \,, &
  f_{-1}(x) &= \frac{1}{1+x}
  \,, &
  f_r(x) &= \frac{1}{\sqrt{x(4-x)}}
  \,.
\end{align}
We also find it convenient to define the following combinations
\begin{align}
  \mathcal{H}_{1} &= H_r({z^{-1}})-\pi
  \,, \notag\\
  \mathcal{H}_{2} &= \pi H_0({z})+H_{0,r}({z^{-1}})
  \,, \notag\\
  \mathcal{H}_{3} &= i \pi^2-3 i \pi H_r({z^{-1}})-3 H_{r,1}({z^{-1}})
  \,, \notag\\
  \mathcal{H}_{4} &= H_{r,0}({z^{-1}})
  \,, \notag\\
  \mathcal{H}_{5} &= -\pi H_r({z^{-1}})+H_{r,r}({z^{-1}})
  \,, \\
  \mathcal{H}_{6} &= -\pi H_{0,r}({z^{-1}})+H_{0,r,r}({z^{-1}})
  \,, \notag\\
  \mathcal{H}_{7} &= -\pi H_{r,0}({z^{-1}})+H_{r,0,r}({z^{-1}})
  \,, \notag\\
  \mathcal{H}_{8} &= H_{r,r,0}({z^{-1}})
  \,, \notag\\
  \mathcal{H}_{9} &= -i\frac{\pi^3}{6} +i \pi^2 H_r({z^{-1}})
                     -3 i \pi H_{r,r}({z^{-1}})-3 H_{r,r,1}({z^{-1}})
  \,, \notag\\
  \mathcal{H}_{10} &= i\frac{\pi^3}{6}  H_0({z})+i \pi^2 H_{0,r}({z^{-1}})
                      -3 i \pi H_{0,r,r}({z^{-1}})-3 H_{0,r,r,0}({z^{-1}})
                      -3 H_{0,r,r,1}({z^{-1}})-4 i \pi \zeta_3
  \,.\notag
\end{align}
These combinations evaluate to real numbers in the relevant physical region
$M_W^2 \leq M_Z^2$, i.e. $0\leq z \leq 1$.
However, note that the iterated integrals where the square-root-valued
letter $f_r$ occurs are evaluated at argument $z^{-1} \geq 1$.
Therefore, the individual iterated integrals with simultaneous letters
$f_r$ and $f_1$ develop an imaginary part that cancels against the
explicit imaginary parts in $\mathcal{H}_3$, $\mathcal{H}_9$ and
$\mathcal{H}_{10}$.

We now present our results. For completeness, we first report
expressions for one-loop corrections. We write the finite remainders defined in Appendix~\ref{sectA1} as
\be
\left\langle F_{\rm LV}^{\rm fin,QCD}(1_u,2_{\bar d})\right\rangle= 
\lp\frac{\alpha_s(\mu)}{2\pi}\rp \mathcal F_{u \bar d}^{\rm QCD,fin} \left\langle F_{\rm LM}(1_u,2_{\bar d})\right\rangle,
~
\left\langle F_{\rm LV}^{\rm fin,EW}(1_u,2_{\bar d})\right\rangle= 
\lp\frac{\alpha_{EW}}{2\pi}\rp \mathcal F_{u \bar d}^{\rm EW,fin} \left\langle F_{\rm LM}(1_u,2_{\bar d})\right\rangle.
\label{eq:FF}
\ee
With this notation, the (renormalized) form factors read\footnote{Higher orders in the $\ep$-expansion
of the finite remainders are not needed for our calculation, so we do not report them here.}
$
\mathcal F_{u\bar d}^{\rm QCD,fin} = -8 C_F,
$
and
\begin{align}
   & \mathcal F_{u\bar d}^{\rm EW,fin} ={}
        (Q_u^2+Q_d^2) \Biggl[
            \frac{\big(z-1\big) (1+z)^2}{z^3} H_{-1,0}({z})
            -\frac{1}{2} \frac{\big(z-1\big) \big(3 z+2\big)}{z^2} H_0({z})
            -\frac{1}{4} \frac{9 z^2+3 z+4}{z^2}
        \Biggr]
    \notag \\ &
        +N_c \Biggl[
            \frac{1}{4} \frac{\big(z_t-1\big) \big(z_t+1\big)}{\big(z-1\big) z_t^3} H_1({z_t})
            -\frac{1}{8} \frac{4 z_t^2-z_t-2}{\big(z-1\big) z_t^2}
        \Biggr]
        -\frac{1}{2} \frac{\big(2 z^2-2 z+1\big) (1+z)^2}{\big(z-1\big) z^3} H_{-1,0}({z})
    \notag \\ &
        +\frac{1}{16} \frac{4 z_H^4-22 z_H^3+17 z_H^2-6 z_H+1}{\big(z-1\big) \big(z_H-1\big) z_H^3} H_0({z_H})
        -\frac{1}{16} \frac{16 z^5+20 z^4-118 z^3+79 z^2-2 z-1}{\big(z-1\big)^2 z^3} H_0({z})
\label{eq:B20}\\ &
        +\frac{1}{48} \frac{-6 z^2+21 z^2 z_H-30 z_H^2-69 z z_H^2+418 z^2 z_H^2}{\big(z-1\big) z^2 z_H^2}
        +\frac{1}{6} \frac{2 z+7}{z-1} \pi^2
        +\frac{z+2}{z-1} \Bigl(
            H_{0,0}({z})
            +\mathcal{H}_{5}(z)
        \Bigr)
    \notag \\ &
        -\frac{\sqrt{4 z_H-1}}{z_H} \frac{1}{16} \frac{28 z_H^3-20 z_H^2+7 z_H-1}{\big(z-1\big) \big(4 z_H-1\big) z_H^2} \mathcal{H}_{1}(z_H)
        +\frac{\sqrt{4 z-1}}{z} \frac{1}{16} \frac{28 z^3+52 z^2-13 z-1}{\big(z-1\big) z^2} \mathcal{H}_{1}(z).
        \notag
\end{align}
In Eq.(\ref{eq:B20}), we used
\begin{align}
  z &= \frac{M_W^2}{M_Z^2} + i0
  \,, &
  z_H &= \frac{M_W^2}{M_H^2} + i0
  \,, &
  z_t &= \frac{M_W^2}{M_t^2} + i0
  \,.
\end{align}

We now present results for the mixed QCD-EW corrections. We find it convenient to factor out the LO amplitude
and separate factorizable and non-factorizable contributions. We write
\be
\mathcal A_{\rm mix} = \mathcal A_0 \left[\mathcal M_1 \bar{\mathcal M}_1 + \mathcal M_{\rm mix,n.f.}\right],
\label{eq:B22}
\ee
where $ \mathcal A_i$ are defined in Eq.(\ref{eq:A1}) and we used analogous definitions for $\mathcal M_i$.
Following what was done in Ref.~\cite{kkv} for the $Z$ boson, we also separate the renormalization contribution coming from two-loop gauge-bosons self-energy
corrections, which is finite. We then write
\be
\mathcal M_{\rm mix,n.f.} = \widetilde{\mathcal M}_{\rm mix} 
+\delta Z_{{\rm mix},2},
~~~~
\widetilde{\mathcal M}_{\rm mix} = \mathcal M_{\rm mix, bare} + \delta Z_{{\rm mix},1},
\label{eq:B23}
\ee
where $\delta Z_{{\rm mix},1}$ contains the mixed fermion wave-function renormalization 
Eq.(\ref{eq:B1}) and $\delta Z_{{\rm mix},2}$ contains the remaining renormalization. In analogy
with Ref.~\cite{kkv}, we now present results for $\widetilde{\mathcal M}_{\rm mix}$. We obtain
\begin{align}
    \widetilde{\mathcal M}_{\rm mix}={}&
        \big(Q_u^2+Q_d^2\big) C_F \Biggl[
            \frac{1}{\ep}
            \biggl(
                -\frac{3}{16}
                +\frac{1}{4} \pi^2
                -3 \zeta_3
            \biggr)
            +\biggl(
                \frac{3}{8}
                -\frac{1}{2} \pi^2
                +6 \zeta_3
            \biggr) \ln\left(\frac{M_W^2}{\mu^2}\right)
            +\frac{1}{4} \frac{\big(27 z+13\big) (1-z)^2}{z^3} H_1
    \notag \\ &
            +\frac{(1-z)^2 (1+z)}{z^3} \biggl(
                \frac{3}{4} H_1 \pi^2
                -\frac{9}{2} H_{1,0,0}
                -\frac{9}{2} H_{1,0,1}
            \biggr)
            -\frac{1}{4} \frac{\big(5 z+3\big) (1-z) (1+z)}{z^3} H_{-1,0}
    \notag \\ &
            +\frac{(1-z) (1+z)^2}{z^3} \biggl(
                -\frac{3}{2} H_{-1,-1,0}
                +\frac{3}{2} H_{-1,0,0}
                +3 H_{-1,0,1}
                +2 H_{-1,-1,-1,0}
                -2 H_{-1,-1,0,0}
                -6 H_{-1,-1,0,1}
    \notag \\ &
                -2 H_{-1,0,-1,0}
                +H_{-1,0,0,1}
                +H_{0,-1,0,0}
                +4 H_{0,-1,0,1}
                +\Bigl(
                    -\frac{1}{4} H_{-1}
                    +\frac{1}{6} H_{-1,-1}
                    -\frac{1}{6} H_{0,-1}
                \Bigr) \pi^2
                -3 H_{-1} \zeta_3
            \biggr)
    \notag \\ &
            +\frac{1}{32} \frac{7 z^2-72 z+64}{z^2}
            +\frac{1}{24} \frac{50 z^2-5 z-16}{z^2} \pi^2
            -\frac{3}{2} \frac{8 z^2-z-2}{z^2} \zeta_3
            -\frac{11}{180} \pi^4
            +\frac{(1-z)}{z^2} \biggl(
                \frac{1}{2} \big(9 z+11\big) H_{0,1}
    \notag \\ &
                -\frac{1}{2} \big(3 z+4\big) H_{0,0,1}
                +\big(3 z+2\big) \Bigl(
                    -\frac{17}{8} H_0
                    +\frac{1}{2} H_{0,-1,0}
                \Bigr)
                +\frac{1}{4} \big(23 z+16\big) H_{0,0}
            \biggr)
    \notag \\ &
            +\frac{\big(z^2+3 z+1\big) (1-z)}{z^3} \biggl(
                \frac{1}{3} H_{0,1} \pi^2
                -2 H_{0,1,0,0}
                -2 H_{0,1,0,1}
            \biggr)
        \Biggr]
    \notag \\ &
        +C_F \Biggl[
            \frac{z+2}{1-z} \Bigl(
                -\frac{1}{6} H_{0,0} \pi^2
                +4 H_0 \zeta_3
            \Bigr)
            +\frac{1}{8} \frac{\big(5 z-2\big) \big(2 z^2+12 z+11\big)}{(1-z) z^2} H_{0,1}
            +\frac{1}{8} \frac{43 z^2+7 z-16}{(1-z) z^2} H_{0,0}
    \notag \\ &
            -\frac{1}{16} \frac{8 z^3+142 z^2+23 z-34}{(1-z) z^2} H_0
            -\frac{1}{48} \frac{10 z^3+5 z^2+20 z-16}{(1-z) z^2} \pi^2
            +\frac{1}{120} \frac{5 z-36}{1-z} \pi^4
            -\frac{1}{8} \frac{4 z^2-17 z+8}{(1-z) z^2}
    \notag \\ &
            +\frac{2 z^2-2 z+1}{(1-z) z^2} \biggl(
                \big(3 z+2\big) \Bigl(
                    -\frac{3}{4} \zeta_3
                    -\frac{1}{4} H_{0,-1,0}
                \Bigr)
                +\frac{1}{4} \big(3 z+4\big) H_{0,0,1}
            \biggr)
            +\frac{\big(2 z^2-6 z+3\big) (1+z)}{z^3} \biggl(
                \frac{3}{4} H_{1,0,0}
    \notag \\ &
                +\frac{3}{4} H_{1,0,1}
                -\frac{1}{8} H_1 \pi^2
            \biggr)
            -\frac{1}{(1-z) z} \biggl(
                \frac{1}{8} H_{0,0,0}
                +\frac{1}{2} \big(9 z^2-8 z-2\big) \zeta_3
                +\frac{5}{48} H_0 \pi^2
            \biggr)
    \notag \\ &
            +\frac{\big(2 z^2-2 z+1\big) (1+z)^2}{(1-z) z^3} \biggl(
                \frac{3}{4} H_{-1,-1,0}
                -\frac{3}{4} H_{-1,0,0}
                -\frac{3}{2} H_{-1,0,1}
                -H_{-1,-1,-1,0}
                +H_{-1,-1,0,0}
                +3 H_{-1,-1,0,1}
    \notag \\ &
                +H_{-1,0,-1,0}
                -\frac{1}{2} H_{-1,0,0,1}
                -\frac{1}{2} H_{0,-1,0,0}
                -2 H_{0,-1,0,1}
                +\Bigl(
                    \frac{1}{8} H_{-1}
                    -\frac{1}{12} H_{-1,-1}
                    +\frac{1}{12} H_{0,-1}
                \Bigr) \pi^2
                +\frac{3}{2} H_{-1} \zeta_3
            \biggr)
    \notag \\ &
            +\frac{1}{8} \frac{4 z^3+64 z^2-z-13}{z^3} H_1
            +\frac{1}{8} \frac{\big(5 z+3\big) \big(2 z^2-2 z+1\big) (1+z)}{(1-z) z^3} H_{-1,0}
            +\frac{z^4-4 z^2+z+1}{(1-z) z^3} \biggl(
                H_{0,1,0,0}
    \notag \\ &
                +H_{0,1,0,1}
                -\frac{1}{6} H_{0,1} \pi^2
            \biggr)
            +\biggl[
                \frac{\sqrt{4 z-1}}{z} \biggl(
                    -\frac{1}{8} \frac{10 z+3}{1-z} \mathcal{H}_{1}
                    -\frac{1}{8} \mathcal{H}_{2}
                    -\frac{1}{8} \frac{6 z+1}{1-z} \mathcal{H}_{3}
                    +\frac{1}{8} \frac{17 z+4}{1-z} \mathcal{H}_{4}
                \biggr)
    \notag \\ &
                -\frac{1}{8} \frac{3 z+2}{(1-z) z} \mathcal{H}_{5}
                +\frac{1}{8} \frac{1}{(1-z) z} \mathcal{H}_{6}
                -\frac{1}{8} \frac{6 z^2-4 z+1}{(1-z) z} \mathcal{H}_{7}
                -\frac{1}{8} \frac{30 z^2-20 z-1}{(1-z) z} \mathcal{H}_{8}
                +\frac{1}{2} \frac{3 z-2}{1-z} \mathcal{H}_{9}
                +\frac{z+2}{1-z} \mathcal{H}_{10}
            \biggr]
        \Biggr]
    \notag \\ &
        +i \pi \Biggl\{
             \big(Q_u^2+Q_d^2\big) C_F \Biggl[
                \frac{9}{2} \frac{(1-z)^2 (1+z)}{z^3} H_{1,0}
                +\frac{(1-z) (1+z)^2}{z^3} \biggl(
                    \frac{3}{2} H_{-1,-1}
                    -\frac{3}{2} H_{-1,0}
                    -2 H_{-1,-1,-1}
    \notag \\ &
                    +2 H_{-1,-1,0}
                    +2 H_{-1,0,-1}
                    -H_{0,-1,0}
                \biggr)
                +\frac{(1-z)}{z^2} \biggl(
                    -\frac{1}{2} \big(3 z+2\big) H_{0,-1}
                    -\frac{1}{4} \big(23 z+16\big) H_0
                \biggr)
    \notag \\ &
                +\frac{1}{4} \frac{\big(5 z+3\big) (1-z) (1+z)}{z^3} H_{-1}
                +\frac{\big(z^2+3 z+1\big) (1-z)}{z^3} 2 H_{0,1,0}
                -\frac{1}{8} \frac{54 z^2-17 z-34}{z^2}
                +\frac{1}{2} \pi^2
                -6 \zeta_3
            \Biggr]
    \notag \\ &
            +C_F \Biggl[
                \frac{1}{16} \frac{8 z^3+372 z^2+17 z-34}{(1-z) z^2}
                -\frac{3}{4} \frac{\big(2 z^2-6 z+3\big) (1+z)}{z^3} H_{1,0}
                -\frac{1}{8} \frac{\big(2 z-1\big) \big(23 z+16\big)}{(1-z) z^2} H_0
    \notag \\ &
                +\frac{\big(2 z^2-2 z+1\big) (1+z)^2}{(1-z) z^3} \biggl(
                    \frac{3}{4} H_{-1,0}
                    -\frac{3}{4} H_{-1,-1}
                    -H_{-1,-1,0}
                    +H_{-1,-1,-1}
                    -H_{-1,0,-1}
                    +\frac{1}{2} H_{0,-1,0}
                \biggr)
    \notag \\ &
                +\frac{1}{4} \frac{\big(3 z+2\big) \big(2 z^2-2 z+1\big)}{(1-z) z^2} H_{0,-1}
                -\frac{1}{8} \frac{\big(5 z+3\big) \big(2 z^2-2 z+1\big) (1+z)}{(1-z) z^3} H_{-1}
                -\frac{z^4-4 z^2+z+1}{(1-z) z^3} H_{0,1,0}
    \notag \\ &
                -\frac{1}{4} \pi^2
                -12 \frac{1}{(1-z)} \zeta_3
            \Biggr]
        \Biggr\}
    \,.
    \label{eq:amp2ldiff}
\end{align}
In Eq.~\eqref{eq:amp2ldiff} we omit writing down the argument $z$ of the
iterated integrals.
Real and imaginary parts in Eq.~\eqref{eq:amp2ldiff} are explicitly
separated.

For completeness, we also present results for the $\delta Z_{{\rm mix},2}$ term. We obtain
\begin{align}
&\delta Z_{{\rm mix},2} ={}
        \frac{N_c C_F}{4(1-z)} \Biggl\{
            \frac{\big(z_t+1\big) (1-z_t)}{z_t^3} \biggl[
                \frac{1}{6} H_1({z_t}) \pi^2
                -H_{0,1,1}({z_t})
                +H_{1,0,1}({z_t})
            \biggr]
            +\frac{1}{12} \frac{\big(17 z_t+15\big) (1-z_t)}{z_t^3} H_1({z_t})
    \notag \\ &
            +\frac{1}{3} \frac{2 z_t^2-3}{z_t^2} H_{0,1}({z_t})
            -\frac{1}{6} \frac{\big(4 z_t^2-5 z_t-17\big) (1-z_t)}{z_t^3} H_{1,1}({z_t})
            +\frac{1}{36} \frac{4 z_t^2-3 z_t-6}{z_t^2} \pi^2
            +\frac{1}{24} \frac{36 z_t^2-35 z_t-30}{z_t^2}
        \Biggr\}.
\end{align}

Starting from the definitions Eqs.(\ref{eq:B22},\ref{eq:B23}) it is straightforward to 
obtain the two-loop finite remainder $F_{{\rm LVV}+{\rm LV}^2}^{{\rm fin, QCD}\otimes{\rm EW}}$
Eq.(\ref{eq:A15}). It reads
\be
\begin{split}
&\left\langle F_{{\rm LVV}+{\rm LV}^2}^{{\rm fin, QCD}\otimes{\rm EW}}(1_u,2_{\bar d})\right\rangle  = {} 
\lp\frac{\alpha_s(\mu)}{2\pi}\;\frac{\alpha_{EW}}{2\pi}\rp\\
&\times
\bigg[
2\Re\big[\widetilde{\mathcal M}_{\rm mix}]+ 2 \delta Z_{{\rm mix},2} 
-\frac{H_{{\rm QCD}\otimes{\rm EW}}^W}{\ep}
+\mathcal F^{\rm fin,QCD}_{u\bar d}\mathcal F^{\rm fin,EW}_{u\bar d}
\bigg] \left\langle F_{\rm LM}(1_u,2_{\bar d})\right\rangle,
\end{split}
\ee
where $H_{{\rm QCD}\otimes{\rm EW}}^W$ is given in Eq.(\ref{eq:HwMix}) and $\mathcal F^{\rm fin,i}_{u\bar d}$
are defined in Eq.(\ref{eq:FF}).

We conclude this section by presenting numerical results for the finite remainders
of the one- and two-loop form factor.
Using the numerical values for the various input parameters reported in Section~\ref{sect7}, we obtain
\begin{align}
  \langle F_{\text{LV}}^\text{QCD,fin}(1_u,2_{\bar{d}}) \rangle
    &= \left(\frac{\alpha_s(\mu)}{2 \pi}\right)
       (-8 C_F)
       \langle F_{\text{LM}}(1_u,2_{\bar{d}}) \rangle
  \,, \\
  \langle F_{\text{LV}}^\text{EW,fin}(1_u,2_{\bar{d}}) \rangle
    &= \left(\frac{\alpha_{EW}}{2 \pi}\right)
       (-4.52495)
       \langle F_{\text{LM}}(1_u,2_{\bar{d}}) \rangle,
       \\
         \langle F_{\text{LVV}+\text{LV}^2}^{\text{fin,QCD}\otimes{\rm EW}}(1_u,2_{\bar{d}}) \rangle
   & = \left(\frac{\alpha_s(\mu)}{2 \pi}\;
\frac{\alpha_{EW}}{2 \pi}\right)
       \left[
         27.2702
         +3.92969 \ln\left(\frac{M_W^2}{\mu^2}\right)
       \right] \langle F_{\text{LM}}(1_u,2_{\bar{d}}) \rangle.
\end{align}


\section{Auxiliary splitting functions and their convolutions}
\label{app:split}

In this appendix, we collect the various splitting functions that we used in our derivations.

For the NLO calculation, we used
\be
P_{qq}^{\rm NLO}(z,L) = (1-z)^{-2\ep} {\bar P}_{qq}(z)
+ \frac{1}{\ep} \delta(1-z) e^{-2\ep L}.
\ee
Its expansion in powers of $\ep$ is given by
\be
\begin{split} 
P_{qq}^{\rm NLO}(z, L) & =  
- 2 L \delta(1 - z)  + 2 D_0(z)    - (1 + z)
\\
&   + (2 L^2 \delta(1 - z) - 4 D_1(z) + 2 (1 + z)\ln(1 - z) - (1 - z) )\ep
  \\
&    + \left (4 D_2(z) - \frac{4}{3} L^3 \delta(1 - z) + 2 (1 - z) \ln (1 - z) - 2 (1 + z)\ln^2(1 - z)
  \right )\ep^2
  \\
&    + \left ( \frac{2}{3}L^4\delta(1 - z) - \frac{8}{3}D_3(z) 
   -2 (1 - z) \ln^2(1 - z) + \frac{4}{3}(1 + z) \ln^3(1 - z) \right ) \ep^3.
   \end{split} 
\ee
The expansion of the analogous contribution for the $\gamma q$ channel 
\be
P_{qg}^{\rm NLO}(z) = (1-z)^{-2\ep} \big[ (1-z)^2 + z^2 - \ep \big]/(1-\ep)
\ee
is straightforward. 

When discussing real-virtual contributions, we introduced the following splitting functions
\begin{align} 
& P_{qq}^{\rm RV}(z) =  \frac{1}{\ep} \left[\frac{1+z^2}{1-z}\ln(z)\right]
-\frac{1+z^2}{1-z}\big[ {\rm Li}_2(1-z)+3 \ln(1-z) \ln (z)\big]-\frac{z}{2}-(1-z) \ln (z)
\nonumber\\
& + \ep \Bigg [
-\frac{\left(1+z^2\right) \text{Li}_3(1-z)}{1-z}
+\ln (1-z) \left(\frac{3 \left(1+z^2\right) \text{Li}_2(1-z)}{1-z}
+\frac{3 z}{2}+3 (1-z) \ln (z)\right)+(1-z) \text{Li}_2(1-z)
\\
& +\frac{9 \left(1+z^2\right) \ln (z) \ln ^2(1-z)}{2 (1-z)}-\frac{1}{2} (1+z)
    \Bigg ],
\nonumber
\end{align}
and
\be
    \begin{split} 
   P_{qg}^{\rm RV}(z) & =
    -\frac{\Gamma (1-\ep)^3 \Gamma (\ep+1)}{\Gamma(1-2\ep)}
    \Bigg \{  2 (1-z)^{-3 \ep}
    \left[1-\frac{2 (1-z) z}{1-\ep}\right]  \left[\frac{1}{\ep^2}
    +\ep \text{Li}_3(1-z)-\frac{\ln (z)}{\ep}-\text{Li}_2(1-z)\right]
\\
&    -\frac{2 (1-z)^{-4 \ep} \left(1-\frac{2 (1-z) z}{1-\ep}\right)
  \Gamma (1-\ep)^2 \Gamma (\ep+1)^2}{\ep^2 \Gamma (1-2 \ep) \Gamma (2 \ep+1)}
-\frac{(2 \ep+1) (z-\ep) (1-z)^{-3 \ep}}{1-\ep}
\Bigg \}.
\end{split}
\label{eq:PqgRV}
\ee

When discussing the double-real contribution in the $gq$ channel, we introduced the following convolution
\be
    \begin{split} 
& \left[P_{qq}^{\rm NLO} \otimes P_{qg}^{\rm NLO}\right](z,E)  =
-2+5z
      -3 z^2+2 \left(2 z^2-2 z+1\right) \ln (1-z)-\left(4 z^2-2 z+1\right) \ln (z)
\\
  &    +\ep \Bigg  \{
4 (1-2 z) \text{Li}_2(z)+\frac{1}{3} \big[4 \left(\pi ^2-12\right) z^2+57 z-9\big]-6 \left(2 z^2-2 z+1\right) \ln ^2(1-z)+\left(4 z^2-2 z+1\right) \ln ^2(z)
\\
& +\big[8 \left(2 z^2-3 z+1\right)+4 \left(2 z^2-2 z+1\right) \ln (z)\big] \ln (1-z)+\left(3+6z-4z^2\right) \ln (z) \Bigg \}
\\
& + \ep^2  \Bigg\{ 16 (2 z-1) \text{Li}_3(1-z)+8 (2 z-1) \text{Li}_3(z)+\text{Li}_2(z) \big[4 (2 z+3)+16 (2 z-1) \ln (1-z)\big]
\\
&+8(3-6z+4z^2)\zeta_3 - \frac{2\pi^2}{3}(1+9z-5z^2)-16+84z-68z^2
+\frac{28}{3} \left(2 z^2-2 z+1\right) \ln ^3(1-z)
\\
& -\frac{2}{3} \left(4 z^2-2 z+1\right) \ln ^3(z)-4\big[\left(9 z^2-13 z+4\right)+\left(2 z^2-6 z+3\right) \ln (z)\big] \ln ^2(1-z)
\\
& +\left[-\frac{4}{3} \big[\left(4 \pi ^2-51\right) z^2+60 z-9\big]
-4 \left(2 z^2-2 z+1\right) \ln ^2(z)-8 z(1-z) \ln (z)\right] \ln (1-z)
\\
& +\left(4 z^2-6 z-3\right) \ln ^2(z)
-2\left[1-z+2z^2-\frac{\pi^2}{3}(1-2z)\right]\ln(z)
 \Bigg \}
-\frac{1}{\ep} P_{qg}^{\rm NLO}(z) \left [z^{-2\ep} - \left ( \frac{E_{\rm max}}{E} \right )^{-2\ep}   \right ].
           \end{split} 
           \label{eq:PqgNLOoPqqNLO}
    \ee

For the collinear renormalization counterterms, we also need the convolution of Altarelli-Parisi
splitting functions (defined in Appendix~\ref{app:AP}) and the finite one-loop remainder $P_{\rm fin}(z,E)$
Eq.(\ref{eqPqqfin}). We obtain
\be
\begin{split} 
  \left[\bar P^{{\rm AP},0}_{qq} \otimes P_{qq}^{\rm fin}\right](z,E)
  &=
  -\frac{1}{\ep} \left (
  \frac{(2 E)^{-2 \ep} \Gamma^2 (1-\ep) \bar P_{qq}^{\rm NLO,CV}(z)}{\Gamma(1-2\ep )}
  -e^{-\ep \gamma_E}  \mu^{-2\ep} \Gamma(1-\ep) \left[\bar P_{qq}^{\rm AP,0} \otimes \bar P_{qq}^{\rm AP,0}\right](z)
  \right ),
  \\
\left[\bar P_{qg}^{{\rm AP},0} \otimes P_{qq}^{\rm fin}\right](z,E)
& =-\frac{1}{\ep}
\Bigg (
\frac{ \Gamma^2(1-\ep) (2 E)^{-2 \ep}}{\Gamma (1-2 \ep )}
\left(  \frac{3}{2} z^{-2\ep} \bar P_{qg}^{{\rm AP},0}(z) +\bar P_{qg}^{\rm NLO,CV}(z)\right)
\\
& -e^{-\gamma_E \epsilon} \mu^{-2 \ep}  \Gamma(1-\ep) \left[\bar P_{qq}^{{\rm AP},0} \otimes \bar P_{qg}^{{\rm AP},0}\right](z)
\Bigg  ),
  \end{split} 
\ee
where the convolutions of Altarelli-Parisi splitting functions are reported in Appendix~\ref{app:AP} and
\begin{align} 
& 
\bar P_{qq}^{\rm NLO,CV}(z) =
\left( 6 D_0(z)+8 D_1(z)+\left(\frac{9}{4}-\frac{2 \pi ^2}{3}\right) \delta(1-z)\right)
+\ep \Bigg ( \frac{4}{3} \pi^2 D_0(z)-6 D_1(z)-12 D_2(z)
\nonumber\\
& -8 \zeta_3 \delta(1-z) \Bigg ) + \ep^2 \left(16 \zeta_3 D_0(z)-\frac{1}{3} 8 \pi ^2 D_1(z)+6 D_2(z)
+\frac{32 D_3(z)}{3}-\frac{8}{45} \pi^4 \delta (1-z)\right)
\nonumber\\
&
+\left(-\frac{\left(3 z^2+1\right) \ln (z)}{1-z}-z-4 (z+1) \ln (1-z)-5\right)
+\ep \Bigg (   (z+1) \left(2 \text{Li}_2(z)+6 \ln ^2(1-z)-3 \ln ^2(z)-\pi ^2\right)
\nonumber\\
& -\frac{3}{2} (1-z)+\frac{4 \ln ^2(z)-6 \ln (z)}{1-z}
+(z+5) \ln (1-z)+2 (z+3) \ln (z) \Bigg )
\\
& +\ep^2 \Bigg (
(z+1) \left(-4 \text{Li}_3(1-z)-4 \text{Li}_3(z)+\left(2 \pi ^2-4 \text{Li}_2(z)\right) \ln (1-z)
-\frac{16}{3} \ln ^3(1-z)
\right. 
\nonumber\\
& \left.  -(2 \ln (z)+3) \ln ^2(1-z)+\frac{2}{3} \pi ^2 \ln (z)
-4 \zeta_3 \right )
-\frac{2(1+3z^2)}{3(1-z)} \ln ^3(z)
+(6-2 z) \text{Li}_2(z)+\pi ^2 \left(z-\frac{5}{3}\right)
\nonumber\\
& -2 (z+3) \ln ^2(z)
+\frac{6 \ln ^2(z)}{1-z}-2 \ln (z)+(1-z) (3 \ln (1-z)-2) \Bigg ),
\nonumber\\
\nonumber\\
 & 
 \bar P_{qg}^{\rm NLO,CV}(z) = -3 z^2+2 \left(2 z^2-2 z+1\right) \ln (1-z)+\left(-4 z^2+2 z-1\right) \ln (z)+5 z-2
\nonumber  \\
  & +\ep  \Bigg (2 (1-2 z) \text{Li}_2(z)+\frac{1}{3} \left(\pi ^2 (2 z-1)-3 \left(8 z^2-9 z+1\right)\right)-2 \left(2 z^2-2 z+1\right) \ln ^2(1-z)
\nonumber\\
&   +\left(4 z^2-2 z+1\right) \ln ^2(z)+2 \left(3 z^2-5 z+2\right) \ln (1-z)
+(2 z+3) \ln (z)\Bigg )
\nonumber\\
&
+\ep^2 \Bigg (
4 (2 z-1) \text{Li}_3(1-z)+4 (2 z-1) \text{Li}_3(z)+\text{Li}_2(z) (2 (2 z+3)
+4 (2 z-1) \ln (1-z))
\\
& +\frac{1}{3} \left(-12 \left(4 z^2+z (2 \zeta (3)-5)-\zeta (3)+1\right)-\pi ^2 (2 z+3)\right)
+\frac{4}{3} \left(2 z^2-2 z+1\right) \ln ^3(1-z)
\nonumber\\
& -\frac{2}{3} \left(4 z^2-2 z+1\right) \ln ^3(z)
+\left(2 (2 z-1) \ln (z)-2 \left(3 z^2-5 z+2\right)\right) \ln ^2(1-z)
\nonumber\\
& +\frac{2}{3} \left(24 z^2
 -2 \pi ^2 z-27 z+\pi ^2+3\right) \ln (1-z)-(2 z+3) \ln ^2(z)
+\frac{2}{3} \left(\pi ^2 (1-2 z)-3\right) \ln (z)
\Bigg ).
\nonumber
\end{align}

The triple-collinear splitting function for the $qg$ channel reads
\begin{align}
&P_{qg}^{\rm trc}(z,E) = \frac{1+5z}{4} + \frac{(5-11z+9z^2)\ln(2)}{2} + 
\left[ \frac{3}{4} -\ln(2)\lp\frac{1}{2}-z\rp\right]\ln(z)
-\frac{(3-6z+4z^2)\ln^2(z)}{4}
\nonumber\\
&
+\big[z^2+(1-z)^2\big]\left[{\rm Li}_2(z)+\ln(1-z)\ln(z) 
+ 3 \ln(2)\ln\lp\frac{E_{\rm max}}{E(1-z)}\rp - \frac{\pi^2}{3}  \right]
+\ep\;\Bigg\{
\frac{1+37z}{4} 
-\frac{\pi^2}{3} z(1-z)
\nonumber\\
&
 - \frac{2\pi^2}{3}(1-2z+3z^2)\ln(2)
+ \frac{(8-17z+15z^2)\ln^2(2)}{4} + 
\lp 8 - \frac{51z}{2}+24z^2\rp\ln(2)
-6z(1-z)\ln(2)\ln\lp\frac{E_{\rm max}}{E}\rp
\nonumber\\
&
+
\left[
\frac{7z}{4}-\lp\frac{3}{2}+z\rp\ln(2)-\frac{(1-2z)\ln^2(2)}{4} + 
\frac{\pi^2}{3}(5-10z+8z^2)
\right]\ln(z)
+ 
\frac{(19-38z+28z^2)\ln^3(z)}{12}
\label{eq:PqgTRC}\\
&
+\left[
\lp\frac{1}{2}-z\rp\ln(2)-
\frac{(15-4z+8z^2)}{8}
\right]\ln^2(z) 
-\big[1+5z + (10-28z+24z^2)\ln(2)
+2z(1-z)\ln(z)\big]\ln(1-z)
\nonumber\\
&
+\big[3-2z+2z^2-(2-4z)\ln(2)\big]{\rm Li}_2(z)
-\big[z^2+(1-z)^2\big]
\bigg[
3\ln^2\lp\frac{E_{\rm max}}{E}\rp\ln(2)
-\frac{5}{2}\ln\lp\frac{E_{\rm max}}{E(1-z)}\rp \ln^2(2)
\nonumber\\
&
+
\lp 6\ln(2)\ln(1-z)-\frac{2\pi^2}{3}\rp\ln\lp\frac{E_{\rm max}}{E}\rp
+\big[4\ln(z)-9\ln(2)\big]\ln^2(1-z) + 
\lp\frac{\ln^2(z)}{2}-\frac{2\pi^2}{3}\rp\ln(1-z)
\nonumber\\
&
+\big[4\ln(1-z)-\ln(z)\big]{\rm Li}_2(z)
+4{\rm Li}_3(1-z)\bigg]
-(9-18z+14z^2){\rm Li}_3(z)+2(1-2z)\zeta_3\Bigg\}.
\nonumber
\end{align}


\section{Altarelli-Parisi splitting functions and their convolutions}
\label{app:AP}

Here we list the Altarelli-Parisi splitting functions that we use in this paper. These functions do not contain
color factors because in many cases they are used to  describe QED {\it and} QCD radiation at the same time. 
At NLO, we need
\be
\begin{split} 
&\bar P_{qq}^{{\rm AP},0}(z)  = 2D_0(z) - (1+z) + \frac{3}{2} \delta(1-z),
\\
&\bar P_{qg}^{{\rm AP},0}(z) = (1-z)^2 + z^2.
\end{split}
\ee
At NNLO, we also used the following (abelianized) one-loop splitting functions
\be
\begin{split}
\bar P_{qq}^{{\rm AP},1} &(z) =
3 - 2 z + 2 \left ( 1 - \frac{1+z^2}{1-z} \ln(1-z) \right ) \ln(z) +
\frac{1+3z^2}{2(1-z)} \ln^2(z) 
\\
& + 2 \frac{1+z^2}{1-z} {\rm Li}_2(1-z)
+ \delta(1-z)  \left ( \frac{3}{8} -\frac{\pi^2}{2} + 6 \zeta_3 \right ),
\\
\bar P_{qg}^{{\rm AP},1} &(z) =
2- \frac{9}{2}z + 2 \ln(1-z)-
\frac{(1-4z)\ln(z)}{2}-\frac{(1-2z)\ln^2(z)}{2} 
\\
&+
\big[z^2+(1-z)^2\big]\left[ 5 - \frac{\pi^2}{3} - 2 \ln\lp\frac{1-z}{z}\rp + \ln^2\lp\frac{1-z}{z}\rp\right].
\end{split} 
\label{eq:C2}
\ee
We note that the $\bar P_{qq}^{{\rm AP},1}$ splitting function in Eq.(\ref{eq:C2}) subtracts collinear singularities
arising from the $g\gamma$ final state only (and not from the $q\bar q$ final state).
The equivalent result inclusive over all possible final states (i.e. $g\gamma$ and $q\bar q$) can be obtained by
abelianizing the standard NLO Altarelli-Parisi non-singlet splitting function.

We also need the convolution of two LO splitting functions:
\begin{align}
&
\left[\bar P_{qq}^{{\rm AP},0} \otimes \bar P_{qq}^{{\rm AP},0}\right](z)
 = 
6 D_0(z)+ 8 D_1(z)+\left( \frac{9}{4}
-\frac{2 \pi ^2}{3}\right) \delta(1-z)-\frac{\left(3 z^2+1\right) \ln (z)}{1-z}-z-4 (z+1) \ln (1-z)-5,
\notag\\
& 
\left[\bar P_{qq}^{{\rm AP},0} \otimes \bar P_{qg}^{{\rm AP},0}\right](z) = 
-2+5z-3z^2 -(1-2z+4z^2)\ln(z) + \lp 2\ln(1-z) + \frac{3}{2}\rp \bar P_{qg}^{{\rm AP},0}(z).
\end{align}

\section{Calculation of the triple-collinear integrated counterterm for the gluon-photon final state}
\label{appendixtc}

According to the discussion in the main text, the triple-collinear limit of the process $u + \bar d \to W^+ + g + \gamma$ is
described by the following formula 
\be
{\cal I}_{\rm TC} = \langle ( I -S_g) (I - S_\gamma) \Xi_2^{q\bar q} F_{\rm LM}(1_u,2_{\bar d}, 4_g, 5_\gamma) \rangle, 
\ee
where
\be
\Xi_2^{q\bar q} =
C_{\gamma g,1} ( I - C_{g1} ) \omega^{\gamma 1, g 1} \theta_A
+ C_{\gamma g,1} ( I - C_{\gamma 1} ) \omega^{\gamma 1, g 1} \theta_B
+
C_{\gamma g,2} ( I - C_{g2} ) \omega^{\gamma 2, g 2} \theta_A
+ C_{\gamma g,2} ( I - C_{\gamma 2} ) \omega^{\gamma 2, g 2} \theta_B,
\ee
and $\theta_A = \theta(\rho_{\gamma i}-\rho_{g i})$ and $\theta_B = \theta(\rho_{g i} - \rho_{\gamma i})$, 
see Eq.(\ref{eq31}). We remind
the reader that triple-collinear operators do not act on the unresolved phase space, while double-collinear ones
do, see Refs.~\cite{Caola:2017dug,Caola:2019nzf} for details. 

We write
\be
\Xi_2^{q\bar q} = \Xi_{2}^{(1)} + \Xi_{2}^{(2)}, 
\ee
to describe emissions off incoming $u$ and $\bar d$ quarks respectively and focus on $\Xi_{2}^{(1)}$. 
Taking into account that $ C_{\gamma g,1} \omega^{\gamma 1, g 1} = C_{\gamma g,1}$ and factoring out color factors, we find 
\be
   {\cal I}^{(1)}_{\rm TC} = Q_u^2 C_F e^2 g_s^2
   \left\langle ( I -S_g) (I - S_\gamma) \big [ \left ( I - C_{g1} \right ) \theta_A + \left (I - C_{\gamma 1} \right ) \theta_B
     \big] \frac{1}{s_{1g\gamma}} {\bar P}_{1g\gamma}(...) F_{\rm LM}(1-\gamma - g,... ) \right\rangle,
\ee
where $\bar{P}_{1g\gamma}$ is the \textit{abelian} part of the $ q \rightarrow gg q^*$ splitting function computed in Ref.~\cite{Catani:1999ss}. Using the fact that $1= \theta_A + \theta_B$, we write this contribution as
\begin{align}
\mathcal{I}^{(1)}_{\rm TC} = Q_u^2C_Fe^2g_s^2 \bigg< (1-S_g)(1-S_\gamma)\big[ 1 - ( C_{g1} \theta_A + C_{\gamma1} \theta_B )  \big] \frac{ \bar{P}_{1g\gamma}(...) }{s_{1g\gamma}}F_{LM}(1-\gamma-g,...) \bigg> \,,
\end{align}
where terms proportional to $C_{\alpha i}$ are referred to as \textit{strongly-ordered}. \\
We would like to rewrite the expression for $\mathcal{I}^{(1)}_{\rm TC}$ in such a way that the result in Ref.~\cite{Delto:2019asp} can be employed. We recall that our current
parametrization differs from the one considered in Ref.~\cite{Delto:2019asp} because \textit{i)} we do not order the energies of the gluon and the photon in the final state
and \textit{ii)} we only consider two angular sectors instead of four. We first consider the issue of energy ordering, introduce the partition of unity
\begin{align}
1 = \theta(E_g-E_\gamma) + \theta(E_\gamma-E_g) \,,
\end{align}
and write
\begin{align}
\label{eqn_itc1_energyordered}
\begin{split}
& \mathcal{I}^{(1)}_{\rm TC} = Q_u^2C_Fe^2g_s^2 \\
& \times\bigg< (1-S_g)(1-S_\gamma)\big[ 1 - ( C_{g1} \theta_A + C_{\gamma1} \theta_B )  \big] \big[ \theta(E_g-E_\gamma) + \theta(E_\gamma-E_g)  \big] \frac{ \bar{P}_{1g\gamma}(...) }{s_{1g\gamma}}F_{LM}(1-\gamma-g,...) \bigg> \,.
\end{split}
\end{align}
We note that this expression is symmetric upon exchanging $g \leftrightarrow \gamma$. This is because, upon replacing $g \leftrightarrow \gamma$, we find that  $C_{g1} \theta_A \leftrightarrow C_{\gamma1} \theta_B$. All other terms in Eq.~(\ref{eqn_itc1_energyordered}), including the triple-collinear splitting function $\bar{P}_{1g\gamma}$ are manifestly symmetric under $g \leftrightarrow \gamma$. This allows us to remove one of the energy orderings. Accounting for the extra factor of two we write 
\begin{align}
& \mathcal{I}^{(1)}_{\rm TC} = 2 \times Q_u^2C_Fe^2g_s^2 \notag\\
& \times\bigg< (1-S_g)(1-S_\gamma)\big[ 1 - ( C_{g1} \theta_A + C_{\gamma1} \theta_B )  \big] \theta(E_\gamma-E_g)  \frac{ \bar{P}_{1g\gamma}(...) }{s_{1g\gamma}}F_{LM}(1-\gamma-g,...) \bigg> \,.
\end{align}
This form is now energy-ordered and, except for a different definition of sectors, compatible with the integrals studied in Ref.~\cite{Delto:2019asp}.

It is very simple to adapt the calculation~\cite{Delto:2019asp} to the definition of sectors used in this paper. Indeed, the new sector definition only affects the strongly-ordered terms proportional to the double-collinear operators $C_{\alpha i}$, whereas the purely triple-collinear term remains unchanged and we can borrow it directly from Ref.~\cite{Delto:2019asp}. At this point, we recall that double-collinear operators act on the unresolved phase-space and that the corresponding integrand drastically simplifies upon taking the limit \cite{Delto:2019asp}. The integration with the new sector definition is again straightforward, which allows us to compute the required integrated triple-collinear contributions with minimal effort.

\end{document}